\documentclass[10pt,aps,prl,twocolumn,superscriptaddress]{revtex4-2}
\usepackage{braket}
\usepackage{enumitem}
\usepackage{amsfonts, mathtools}
\usepackage{booktabs}
\usepackage{xcolor}
\usepackage{siunitx}
\usepackage[breaklinks=true,colorlinks,allcolors=blue]{hyperref}
\usepackage{orcidlink}

\renewcommand\_{\leavevmode \kern.06em\vbox{\hrule width.3em}}

\newcommand{\diag}{\text{diag}}

\newcommand{\norm}[1]{\left\lVert#1\right\rVert}
\newcommand{\proj}[1]{\ket{#1}\!\bra{#1}}
\newcommand{\eqlabel}[1]{Eq.~\eqref{#1}}
\newcommand{\figlabel}[1]{Fig.~\ref{#1}}
\newcommand{\figslabel}[1]{Figs.~\ref{#1}}
\newcommand{\tablabel}[1]{Table~\ref{#1}}

\DeclareMathOperator{\cov}{cov}

\setlist{itemsep=1mm, topsep=1mm, partopsep=0mm, parsep=0mm}

\begin{document}

\title{Kicked-Ising Quantum Battery}
\author{Sebastián V. Romero$^{\orcidlink{0000-0002-4675-4452}}$}
\email{sebastian.v.romero@csic.es}
\affiliation{Quantum Advanced Research Center (QuARC), CSIC, 28049 Madrid, Spain}
\affiliation{Instituto de Ciencia de Materiales de Madrid (ICMM), CSIC, 28049 Madrid, Spain}
\affiliation{\mbox{Departamento de Física Teórica de la Materia Condensada, Universidad Autónoma de Madrid, 28049 Madrid, Spain}}
\affiliation{Department of Physical Chemistry, University of the Basque Country EHU, Apartado 644, 48080 Bilbao, Spain}

\author{Xi Chen$^{\orcidlink{0000-0003-4221-4288}}$}
\email{xi.chen@csic.es}
\affiliation{Quantum Advanced Research Center (QuARC), CSIC, 28049 Madrid, Spain}
\affiliation{Instituto de Ciencia de Materiales de Madrid (ICMM), CSIC, 28049 Madrid, Spain}

\author{Yue Ban$^{\orcidlink{0000-0003-1764-4470}}$}
\email{yue.ban@csic.es}
\affiliation{Quantum Advanced Research Center (QuARC), CSIC, 28049 Madrid, Spain}
\affiliation{Instituto de Ciencia de Materiales de Madrid (ICMM), CSIC, 28049 Madrid, Spain}
\date{\today}

\begin{abstract}
    Entanglement has been identified as a key resource for enhancing charging performance in quantum batteries. We show that the kicked-Ising model at the self-dual point provides an explicit charging mechanism, where maximal entanglement growth yields maximal energy injection. Identifying the Floquet dynamics as a Clifford quantum cellular automata and considering exact diagonalization in momentum space, we analytically characterize the charging process, featuring a stable performance while achieving maximal charging. We further propose a fixed time window protocol that accelerates charging toward the continuously driven transverse-field limit. Spin-correlator analysis reveals that scrambling and light-cone spreading govern charging performance. The protocols remain compatible with diverse platforms, underscoring their scalability and practical feasibility.
\end{abstract}

\maketitle

A quantum battery (QB) is a system designed to store extractable work in an ensemble of identical quantum cells~\cite{hovhannisyan2013entanglement}. Entangled unitary operators acting on quantum cells (\emph{collective charging}~\cite{campaioli2017enhancing}) can enhance work extraction compared to unentangled controls (\emph{parallel charging}~\cite{binder2015quantacell})~\cite{alicki2013entanglement}. 
A variety of QB models have been proposed to date, with experimental realizations in platforms such as superconductors, quantum dots, organic microcavities, and nuclear spins~\cite{campaioli2023colloquium}.
Despite these advances, their experimental realization still faces persistent challenges, highlighting the need for analytically tractable models that clarify the microscopic charging mechanism.

Spin chains have been widely used for the design and study of QBs~\cite{le2018spinchain}, offering a versatile playground with different physical regimes contributing distinct advantages, such as the many-body localized~\cite{MBL-QB} and the Anderson localized phases~\cite{AL-QB}. 
Yet a key question persists: how to identify physically transparent mechanisms that enable rapid entanglement propagation while achieving maximal energy injection. In this context, the kicked-Ising model provides an appealing candidate. As a paradigmatic spin chain model governed by a periodically driven transverse field, it exhibits rich dynamical behavior ranging from quantum chaos~\cite{rozenbaum2017lyapunov,lin2018out,joshi2020quantum,santhanam2022quantum,anand2024quantum,PhysRevB.111.054314} to many-body localization~\cite{waltner2021localization}. Remarkably, at the self-dual operator regime it renders a Floquet operator invariant under duality transformations~\cite{Akila_2016}, achieving maximal entanglement growth along the evolution~\cite{bertini2019entanglement}.%

Building on these features, in this Letter we propose an analytically tractable mechanism for quantum battery charging, leveraging the maximal entanglement growth that the kicked-Ising model features at the self-dual operator regime (see~\figlabel{fig:scheme}). 
By linking its Floquet evolution to Clifford quantum cellular automata (CQCA)~\cite{schlingemann2008structure} and using exact diagonalization in momentum space, we obtain an exact characterization of the charging dynamics, verified with tensor-network simulations and IBM experiments. 
The injected energy exhibits a structured dependence on the number of Floquet cycles and system size, with entanglement emerging as a key indicator of performance. Furthermore, we introduce a kicking protocol within a fixed time window, which efficiently approaches the continuously driven transverse-field Ising limit while enhancing experimental flexibility. 
Through spin correlators~\cite{colmenarez2020lieb-robinson}, we establish a connection between scrambling and battery performance, observing different light cones and energies depending on the kicking frequency. The compatibility with existing platforms like trapped ions~\cite{kim2010quantum,britton2012engineered,jurcevic2017direct,zhang2017observation}, ultracold atoms in optical lattices~\cite{johnson2008rabi,beguin2013direct,hermann2014long,labuhn2016tunable,guardado-sanchez2018probing,graham2019rydberg}, and transmon qubits~\cite{egorova2020analog,stehlik2021tunable,greenaway2024analoguequantum} demonstrates scalability under realistic conditions.
\begin{figure}[!tb]
    \centering
    \includegraphics[width=\linewidth]{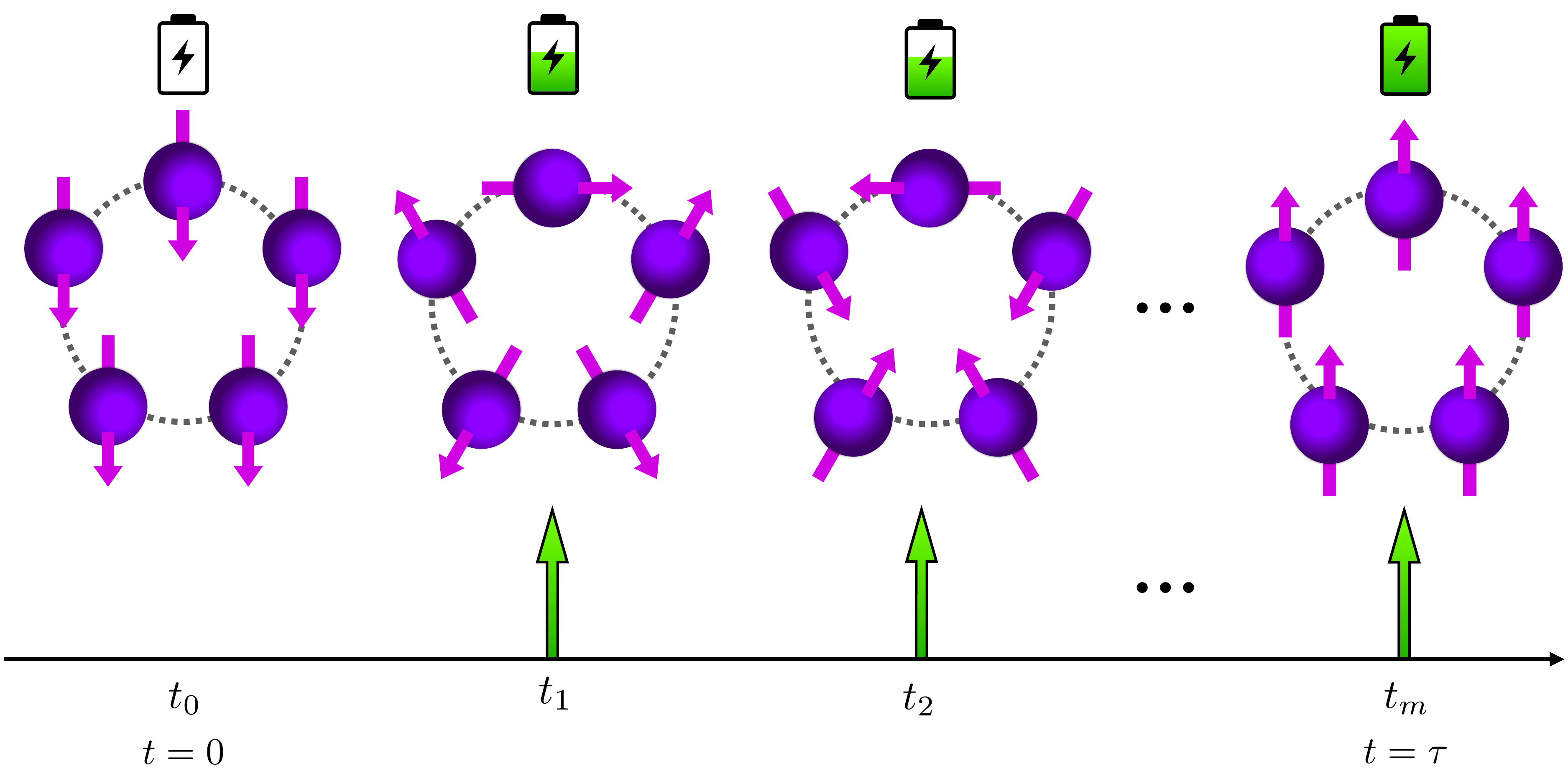}%
\caption{Kicked-Ising QB schematic. A set of coupled spins, driven by an external transverse field kicked at times $t_i\in[0,\tau]$ ($i=1,2,\dots,m$), is used to populate higher-excited states of an initially discharged QB. This model offers a stable and flexible protocol capable of maximally charging a QB.
}\label{fig:scheme}\vspace{-1mm}%
\end{figure}%

\emph{Formulation.}---Let $H(t)=H_0 + \lambda(t)[H_1(t) - H_0]$ be the charging protocol, with $H_0 = (\omega_0/2)\sum_{i=1}^N \sigma^\alpha_i$ the battery consists of $N$ quantum cells ($\alpha\in\{x,y,z\}$), $H_1$ the charger, and $\lambda(t)$ toggles $H_0$ and $H_1$. We use $\lambda(t)=1$ for $t\in[0,\tau]$ and zero otherwise. The state after charging completion is $\ket{\psi(\tau)}=\mathfrak{T}\exp[ -i\int_0^\tau \mathrm{d}t H(t)]\ket{\psi(0)}$, with the time-ordering operator $\mathfrak{T} [\cdot]$ and $\ket{\psi(0)}$ the ground state (GS) of $H_0$, which we shift to zero. The mean energy injected is then $E_N(\tau) = \braket{\psi(\tau) | H_0 | \psi(\tau)} = \braket{H_0}_\tau $. 

We study a kicked-Ising chain (KIC) as a charger,%
\begin{equation}\label{eq:charger}
    H_1(t) = H_I + H_K\sum_{t_i\in\mathcal{T}} \delta(t-t_i),
\end{equation}
where $\mathcal{T}$ denotes the chosen schedule, i.e., the sorted set of times $t_i\in[0,\tau]$ ($t_i<t_{i+1}$) where kicks are applied. We consider two scenarios: (i) uniform unit time interval ($t_i\in\mathbb{Z}$), and (ii) nonuniform protocol, where the spacing between pulses is relaxed. In~\eqlabel{eq:charger}, we consider $H^{\{xx,zz\}}_1(t)\coloneqq H^{\{xx,zz\}}_I + H^{\{xx,zz\}}_K\sum_{t_i\in\mathcal{T}} \delta(t-t_i)$ with Ising interactions and transverse-field terms
\begin{align}\label{eq:components_xx}
    H^{xx}_I &= \sum_{\braket{ij}} J_{ij}\sigma^x_i \sigma^x_j, \qquad H^{xx}_K = \sum_{i=1}^N b_i\sigma^z_i, \\ \label{eq:components_zz}%
    H^{zz}_I &= \sum_{\braket{ij}} J_{ij}\sigma^z_i \sigma^z_j, \qquad H^{zz}_K = \sum_{i=1}^N b_i\sigma^x_i.
\end{align}
For the $H^{xx}_1$ ($H^{zz}_1$) charger, cells are taken as 
$\sigma^\alpha_i=\sigma^z_i$ ($\sigma^\alpha_i=\sigma^y_i$). Here, $\braket{\cdot}$ denotes that enclosed qubit indices are nearest neighbors. Finally, since~\eqlabel{eq:charger} has a direct gate-based implementation on digital hardware, it can be used as a testbed given its exact solvability.

For both variants, $J_{ij}$ indicates the coupling strength between qubits $(i,j)$, and $b_i$ the transverse-field strength applied on qubit $i$. We consider the self-dual operator regime, $J_{ij}=J=\pi/4$, $b_i=b=-\pi/4$~\cite{bertini2019entanglement}. The absence of longitudinal fields makes the model integrable, thus nonergodic. Both open and periodic boundary conditions ($\sigma^{\{x,z\}}_{N+1}=\sigma^{\{x,z\}}_1$) are studied (OBC and PBC, respectively). For simplicity, we set $\hbar\equiv1$ and $\omega_0=1$ hereinafter, so that energies (times) are in units of $|J|$ ($1/|J|$). A noteworthy consequence is the trade-off between the coupling strength and the interval between subsequent kicks. Particularly, the constraints $|J\Delta t_i|\equiv|b\Delta t_i|=\pi/4 \pmod{2\pi}$ $\forall t_i\in\mathcal{T}$ with $\Delta t_i = t_i - t_{i-1}$ must be satisfied. Therefore, stronger couplings require shorter kick intervals, thereby accelerating the charging process.

\emph{Charging dynamics of KIC.}---For a uniform schedule, the Floquet operator after one kick with $\Delta t_i =1$ is
\begin{equation}\label{eq:floquet_kick}
    U(1,0) = \mathfrak{T}\exp\left[ -i\int_0^1 \mathrm{d}t H(t) \right] = e^{-iH_K}e^{-iH_I}.
\end{equation}
So the state evolves as $\ket{\psi(m)} = U(m,0)\ket{\psi(0)}$ after $m$ kicks, with the evolution operator $U(m,0)=U^m(1,0)$. 

In the self-dual regime,~\eqlabel{eq:floquet_kick} becomes a Clifford unitary, which renders the dynamics of a CQCA~\cite{schlingemann2008structure}, enabling to track exactly operator spreading and charging dynamics. This particular Clifford structure enforces ballistic operator spreading with maximal entanglement growth, establishing entanglement production as the microscopic mechanism behind charging. Alternatively, Eqs.~\eqref{eq:components_xx} and~\eqref{eq:components_zz} can also be \mbox{diagonalized} in momentum space via Fourier transform (End Matter). Choosing $H_1^{xx}$ as the charger, the energy injected simplifies to $E_N(m)=\omega_0\sum_k\sin^2 mk$, with $k$ the Fourier transform momenta. In~\figlabel{fig:kic} we see how the normalized injected energy, $E_N/N$, evolves with the number of kicks for OBC and PBC cases, verifying exact results with matrix product states (MPS)~\footnote{Notice that, as expected, the charging power scales extensively, defined as $P_N(\tau^*) = \max_\tau E_N(\tau)/\tau\sim\mathcal{O}[N]$.}. Using $q\in\mathbb{Z}$, we identify for PBC:%
\begin{itemize}%
    \item Even $N$ (periodicity $U(m+N,0)=U(m,0)$):%
    \begin{itemize}%
        \item At $m=(q+\frac{1}{2})N$ cells are maximally charged, $E_N(m)/N=1$, most excited energy of $H_0$.%
        \item At $m=qN$, cells are maximally discharged, $E_N(m)/N=0$, GS energy of $H_0$.%
        \item Otherwise, $E_N(m)/N=0.5$.%
    \end{itemize}%
    \item Odd $N$ (periodicity $U(m+N,0)=U(m,0)$):%
    \begin{itemize}%
        \item At $m=qN$, cells are maximally discharged, $E_N(m)/N=0$, GS energy of $H_0$.%
        \item Otherwise, $E_N(m)/N=0.5$.%
    \end{itemize}%
\end{itemize}%
For the $H^{zz}_1$ charger:%
\begin{itemize}%
    \item Even $N$ (periodicity $U(m+N,0)=U(m,0)$): same dynamics as the odd-$N$ case for $H^{xx}_1$.%
    \item Odd $N$ (periodicity $U(m+4N,0)=U(m,0)$):%
    \begin{itemize}%
        \item At $m=(2+4q)N$ cells are maximally charged, $E_N(m)/N=1$, most excited energy of $H_0$.%
        \item At $m=4qN$, cells are maximally discharged, $E_N(m)/N=0$, GS energy of $H_0$.%
        \item Otherwise, $E_N(m)/N=0.5$.%
    \end{itemize}%
\end{itemize}%
Interestingly, regardless of the parity of $N$, both chargers under OBC behave as their relatives under PBC with odd $N$. Moreover, we implement on the superconducting-qubit platform \textsc{ibm\_torino}~\cite{ibm} a KIC with $N=104$, the largest implementable closed chain (see End Matter). We use $H_1^{zz}$ under PBC up to $m=12$ kicks, which was the limit to obtain reliable results~\cite{sm}. Remarkably, IBM data greatly align with the analytical results, stressing the feasibility of our models on current platforms. The plateaus in~\figlabel{fig:kic} prove their stability, a key feature for storing and extracting energy. The injected energy may undergo fluctuations along charging, which can render a QB model impractical~\cite{santos2019stable}. Consequently it is of pivotal importance to suppress them, as our KIC model does~\footnote{As an example, when $J=0$ the energy oscillates as $(0,0.5,1,0.5,0)$ with kicks, fully charging without many-body interactions. However, this scheme is impractical due to its instability, with stability arising at the self-dual point, where entanglement is leveraged.}. Furthermore, in Sec. II of Supplemental Material, we show that the protocol remains robust against imperfect implementations by studying the effects of disorder, long-range interactions, narrow pulses (coined as \emph{quasikicks}), and slower quenches for $\lambda(t)$~\cite{sm}.%
\begin{figure*}[!tb]
    \centering
    \includegraphics[width=.92\linewidth]{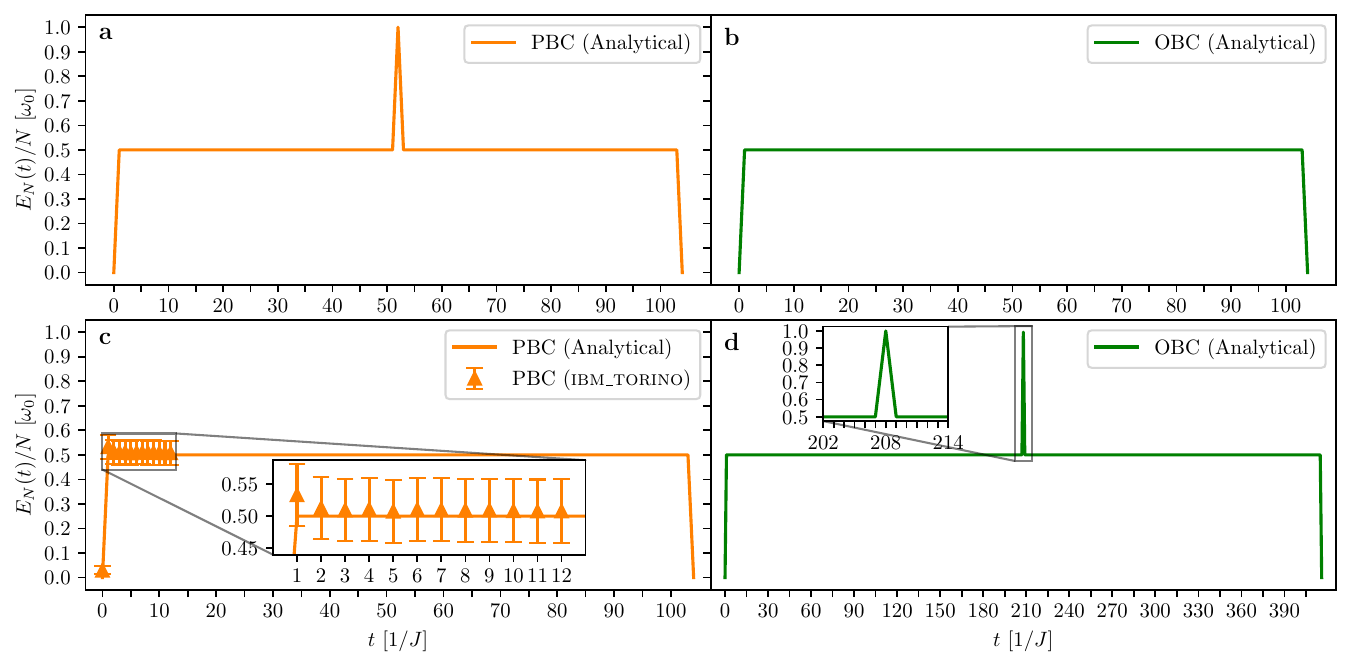}\vspace{-2mm}%
    \caption{Normalized injected energies for KIC with $N=104$. (a) Using $H_1^{xx}$ as charger [\eqlabel{eq:components_xx}], analytical results up to $m=N=104$ kicks under PBC. (b) Using $H_1^{xx}$, analytical results up to $m=N$ under OBC. (c) Using $H_1^{zz}$ as  [\eqlabel{eq:components_zz}], analytical (up to $m=N$) and \textsc{ibm\_torino} results (up to $m=12$) under PBC. (d) Using $H_1^{zz}$, analytical results up to $m=4N$ under OBC. In all panels, analytical (\textsc{ibm\_torino}) results are shown by solid lines (triangular marks).}\label{fig:kic}\vspace{-3mm}%
\end{figure*}%

To gain further insight, we study how the battery energy levels populate along the evolution. Its Hamiltonian decomposes as $H_0=\sum_{n=0}^N\varepsilon_n\sum_{i=1}^{\binom{N}{n}}\proj{\varepsilon_{n,i}}$ with $\ket{\varepsilon_{n,i}}$ the $i$th degenerate eigenstate of the $n$th energy level $\varepsilon_n=n\omega_0$ (multiplicity $\binom{N}{n}$) $\forall n\in[0,N]$~\cite{romero2024scramblingchargingquantumbatteries}, with each population evolving as $p_n(t)\coloneqq\sum_i|\braket{\varepsilon_{n,i}|\psi(t)}|^2$. In~\figlabel{fig:kic_populations} we replicate~\figlabel{fig:kic} for $N=14$ but tracking $p_n(t)$, observing the following for even $N$:
\begin{itemize}
    \item The populations are mirrored along $n=N/2$, i.e., $p_n(t)=p_{N-n}(t)$, except where the ground and most excited states are reached.
    \item Using $H_1^{xx}$ [\figslabel{fig:kic_populations}(a)-(b)], $p_n(\tau)=0$ for $n$ odd. As the charger conserves fermionic parity and both $\ket{\psi(\tau)}$ and $\ket{\varepsilon_{n,i}}$ are respectively in the orthogonal even and odd parity sectors, their overlap has to be zero.
    \item Using $H_1^{zz}$ under OBC [\figlabel{fig:kic_populations}(d)], at kicks $2(1+q)N$ and $2(1+3q)N$, the Y-basis states $\ket{\text{GHZ}^{\pm i}}\coloneqq (\ket{0}_\text{Y}^{\otimes N} \pm i\ket{1}_\text{Y}^{\otimes N})/\sqrt{2}$ are prepared, respectively.
\end{itemize}
Similar observations can be found for $N$ odd.
\begin{figure*}[!tb]
    \centering
    \includegraphics[width=.92\linewidth]{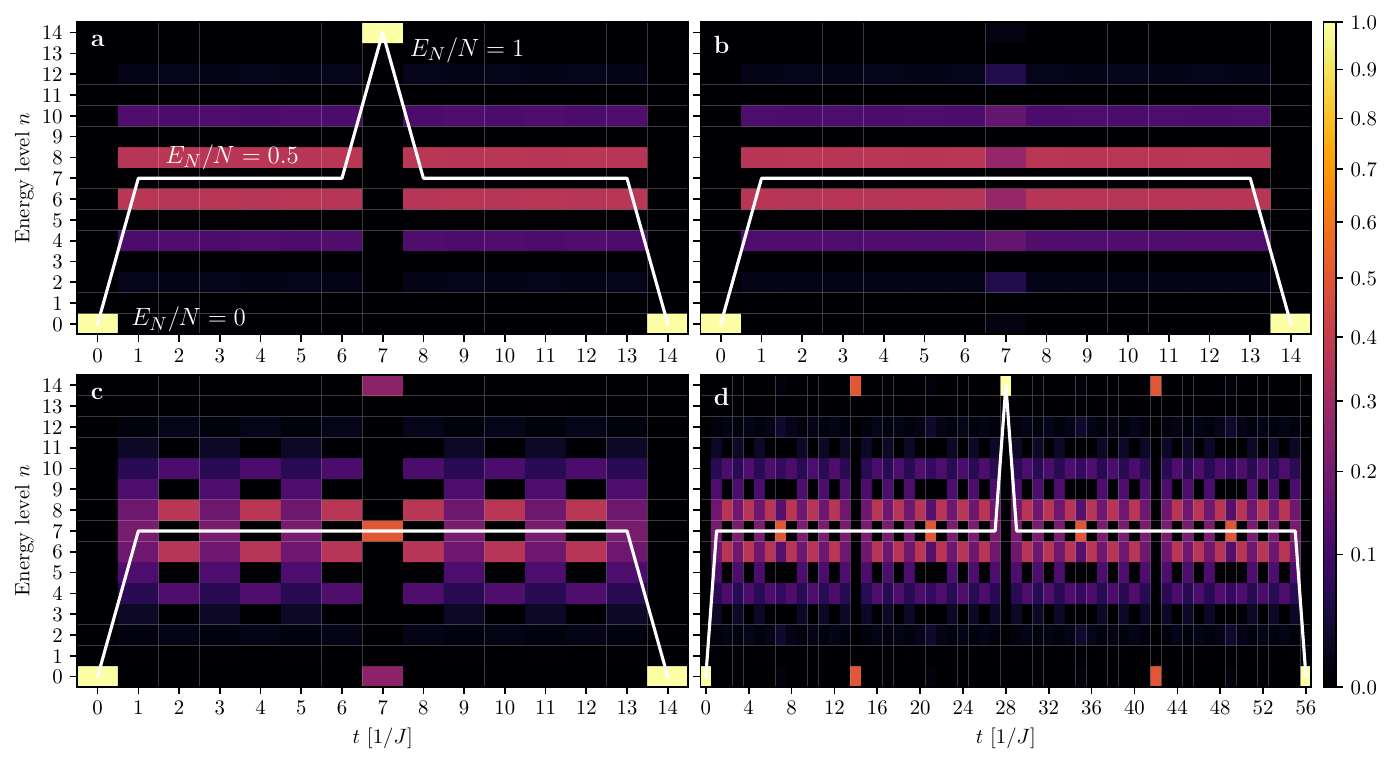}\vspace{-2mm}%
    \caption{Population dynamics of KIC QB with $N=14$. (a) Using $H_1^{xx}$  [\eqlabel{eq:components_xx}], populations up to $m=N=14$ kicks under PBC. (b) Same as (a) but under OBC. (c) Using $H_1^{zz}$ [\eqlabel{eq:components_zz}], populations up to $m=N=14$ kicks under PBC. (d) Same as (c) but evolving up to $m=4N=56$ kicks under OBC. White lines indicate the normalized injected energies [\figlabel{fig:kic}]. 
    }\label{fig:kic_populations}\vspace{-3mm}%
\end{figure*}%

\emph{Charging dynamics of nonuniform KIC.}---Studying nonuniform schedules is essential to make more realistic and flexible designs. Particularly, energy can be injected by applying kicks in a fixed time window. We consider as time-evolution operator after $m$ kicks
\begin{equation}\label{eq:time}
\begin{aligned}
    U(1,0) &= e^{-iH_I(1 - t_m)}\prod_{j=1}^m e^{-iH_K\Delta t_j}e^{-iH_I\Delta t_j},
\end{aligned}
\end{equation}
with $t_0=0$ and $\tau=1$. In the limit of $m\to\infty$, since $e^{A + B} = \lim_{m \to \infty} [ e^{A/m} e^{B/m} ]^m$, \eqlabel{eq:time} converges to $U_\infty=e^{-i(H_K+H_I)}$, thus a regular Ising chain.
Consequently, by applying increasingly dense kicking protocol within a unit time window, the battery can be charged to the value provided by a regular Ising QB, beyond $E_N/N=0.5$. Importantly, nonuniform kicks help to relax the stringent timing constraints of strictly periodic protocols, as seen for the self-dual operator constraints, thereby offering greater flexibility in practical implementations. From another perspective,  irregular kicks can be considered as introducing effective disorder, which breaks the periodicity of the KIC model. Nevertheless, the time window for applying kicks is crucial for efficient charging, as will be shown later.

In~\figlabel{fig:res} we can see how the normalized injected energy increases with the number of kicks for both OBC and PBC under $H^{zz}_1$ with $N=104$, averaged over $n_d=10$ realizations. We further implement on \textsc{ibm\_torino} a nonuniform KIC with $N=104$ under PBC up to $12$ kicks and $n_d=10$ realizations, where IBM data is again in good agreement with the analytical results.  Similar results hold for both boundary conditions, regardless of the parity of $N$, or whether kicks are uniform or nonuniform. While the regular KIC yields $E_N/N=0.5$ after a single kick [\figlabel{fig:kic}], applying multiple kicks in a random fashion can increase injected energy, and eventually reaches its maximal value of the regular Ising QB [inset of~\figlabel{fig:res}]. Notably, for $m\gtrsim10$ kicks the injected energies approach this maximal value, showing that, instead of a continuous transverse-field contribution, few kicks are needed for similar performance. Furthermore, while an Ising QB requires a continuous transverse-field applied for $\tau\sim0.6$ to reach $E_N/N\sim0.5$, our protocols can achieve it with a single kick, demonstrating its efficiency.
Since the protocol operates within a fixed time window, increasing the number of kicks yields an effective Ising evolution in the Lie-Trotter limit, which constrains the dynamics  and prevents full ergodic exploration of the Hilbert space.
\begin{figure}[!tb]
    \centering
    \includegraphics[width=.95\linewidth]{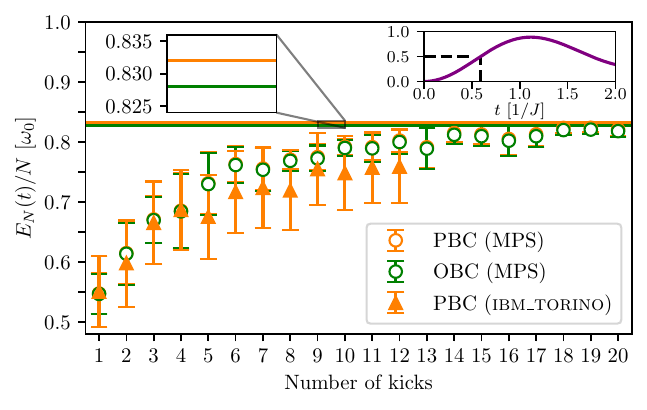}\vspace{-2.2mm}%
    \caption{Normalized injected energies for nonuniform KIC with $N=104$. Kicks are applied randomly with $t_i\in(0,1]$ up to $m=20$ ($m=12$) kicks for the MPS (\textsc{ibm\_torino}) results. Enlarged inset shows the saturation energies for PBC (orange) and OBC (green). PBC were considered for \textsc{ibm\_torino} data (triangular marks). Rightmost inset shows the same protocol but under an Ising chain (violet), with $E_N(\tau\sim0.6)/N=0.5$.%
    }\label{fig:res}\vspace{-3.5mm}%
\end{figure}%

\emph{Spin localization in time domain.}---The scrambling nature of the charging protocol can be analyzed by means of spin correlators under a butterfly effect perspective~\cite{colmenarez2020lieb-robinson}. Hereinafter, we use $H^{zz}_1$ as charger, but similar results are obtained for $H^{xx}_1$. The spin correlator is defined as $C_{ij}(t)=[\sigma^z_i(t), \sigma^z_j]$, with the Heisenberg picture of $\sigma^z_i(t)\coloneqq U^\dagger(t,0)\sigma^z_i U(t,0)$. When we choose $\sigma^z_i$ as operator, locality is preserved during evolution under the Ising interaction $H^{zz}_I$, since $[e^{iH^{zz}_It}\sigma^z_ie^{-iH^{zz}_It}, \sigma^z_j]=[\sigma^z_i, \sigma^z_j]=0$, and hence delocalization is solely induced by the kicks.%

We compute the light cones generated by a local operator initially placed at the chain center, tracking $\norm{C_{ij}(t)}$ for all $j \in [1,N]$, where $\norm{A}$ denotes the largest singular value. For a KIC with $N=13$ up to $m=4N$ kicks as shown in~\figlabel{fig:correlators}(a), we place the initial operator at $i=\lceil N/2\rceil=7$, and observe that spreading pattern. At kicks $2(1+q)N$ and $2(1+3q)N$ ($4qN$ and $2(1+2q)N$), GHZ$^{\pm i}$ states (the ground and most excited states) are prepared [\figlabel{fig:kic_populations}(d)], corresponding to the most delocalized (localized) configurations. Away from these regions, the light cones expand linearly in time~\cite{bertini2019entanglement}, consistent with linear growth of entanglement entropy under the self-dual operator regime~\cite{sm}, while the injected energy remains constant at $E_N/N=0.5$. This indicates the interplay between energy injection and entanglement growth that governs how the KIC dynamics lead to delocalization, whose spreading is well captured as a CQCA.

For nonuniform KIC [\figlabel{fig:correlators}(b)-(c)], we apply $10$ kicks across different time windows ($t_m=0.2$ and $1$). The time window has a decisive role: shorter time windows suppress entanglement spreading and yield low energy injection, whereas larger windows enable well-defined light cones and saturate the regular Ising model injected energy at $t=1$ [\figlabel{fig:res}]. High-frequency kicks in a small time window can induce a form of Anderson-like localization in time domain~\cite{lorenzo2018remnants,waltner2021localization,gyawali2025observationdisorderfreelocalizationusing}, hindering entanglement propagation, resulting in a lower injected energy. 
In contrast, lower-frequency kicks leave sufficient time between kicks for entanglement growth, allowing KIC to outperform regular Ising QBs in charging speed, by delivering energy in concentrated bursts to excite the system efficiently and exploit the coherent dynamics among kicks.%
\begin{figure}[!tb]
    \centering
    \includegraphics[width=.98\linewidth]{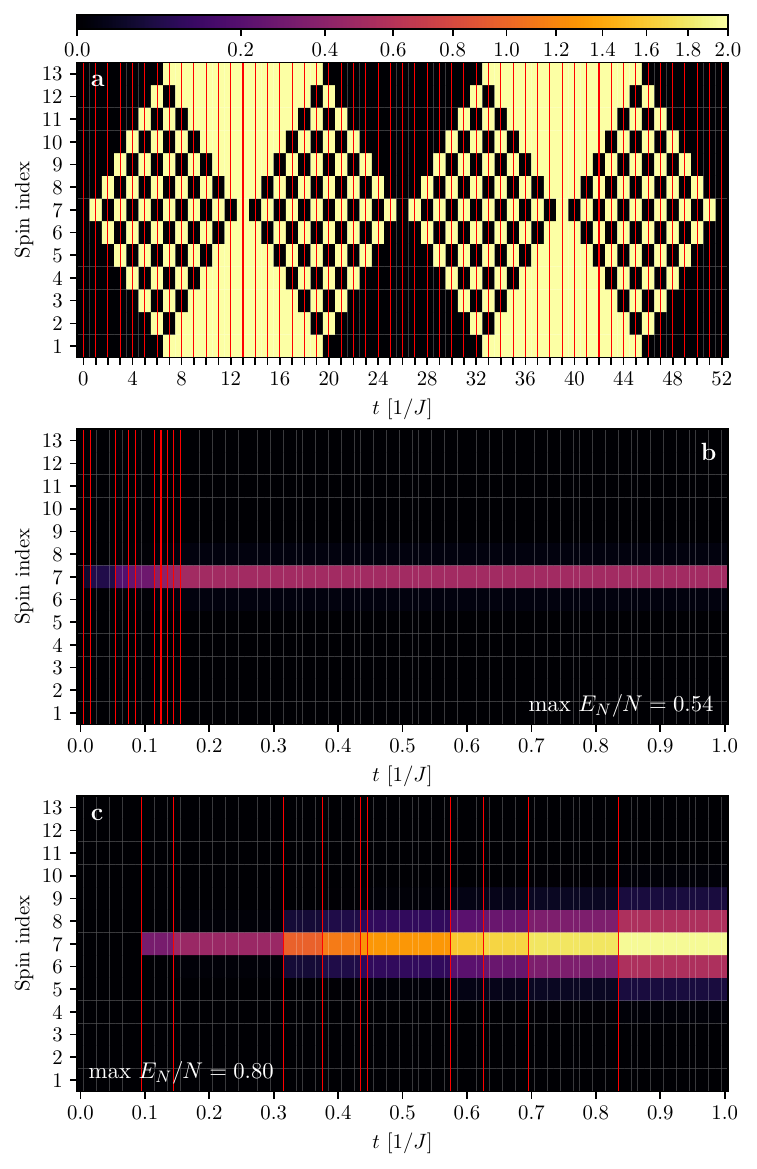}\vspace{-2mm}%
    \caption{Evolution of the spin-correlator norms. Using $H^{zz}_1$  [\eqlabel{eq:components_zz}], $N=13$ and OBC: (a) KIC results. (b) Non-uniform KIC results applying randomly 10 kicks within $(0,0.2]$. (c) Same as (b) but along $(0,1]$. For (b)-(c), the maximum energy injected shows the impact of the interval between kicks and the time window. Kicks are applied at red lines.}\label{fig:correlators}\vspace{-3mm}
\end{figure}%

\emph{Experimental implementation.}---We conclude by discussing potential platforms where our models can be realized. Trapped ions are a natural setup~\cite{kim2010quantum,britton2012engineered,jurcevic2017direct,zhang2017observation}, where Eqs.~\eqref{eq:components_xx} and~\eqref{eq:components_zz} can be engineered in current setups beyond 50 spins with $J_{ij}\sim 0.1-\SI{10}{kHz}$, $b_i\sim0.1-\SI{50}{kHz}$ and dephasing times $T^*_2\sim1-\SI{1000}{ms}$. Taking representative values $J_{ij},b_i\simeq\SI{10}{kHz}$ and $T^*_2\simeq\SI{100}{ms}$, we estimate a rate of $\SI{0.1}{ms/kick}$ to meet \mbox{self-dual} constraints, allowing for $1000$ kicks within coherence time.%

Ultracold atoms in optical lattices or Rydberg arrays~\cite{johnson2008rabi,beguin2013direct,hermann2014long,labuhn2016tunable,guardado-sanchez2018probing,graham2019rydberg} provide programmable Ising interactions with MHz-scale Rabi frequencies and coherence times up to hundreds of microseconds. Since long-range interactions are not expected to affect performance significantly~\cite{sm}, using as typical van der Waals interaction strengths $C_6\simeq\SI{10}{GHz\,\micro\meter^6}$, blockade radii 
$R_b\simeq\SI{5}{\micro\meter}$, Rabi frequencies $\Omega\simeq\SI{1}{MHz}$, $T_2^*\simeq\SI{100}{\micro\second}$ on the order of $100$ kicks can be implemented at a rate of $\SI{1}{\micro\second}$ per kick.%

Superconducting transmons, scalable to hundreds of qubits, naturally realize transverse-field Ising models in the hardcore boson limit~\cite{egorova2020analog,stehlik2021tunable,greenaway2024analoguequantum}. Using as representative qubit-qubit coupling $J_{ij}\simeq\SI{50}{MHz}$, anharmonicity $\alpha_i\simeq\SI{200}{MHz}$, qubit frequency $\omega_i\simeq\SI{5}{GHz}$ and relaxation times $T^*_2\simeq\SI{100}{\micro\second}$, a submicrosecond kick rate would allow up to $1000$ kicks within coherence time.

Waveguide quantum electrodynamics~\cite{douglas2015quantum,hung2016quantum} devices may be also suitable and Andreev spin qubits~\cite{pitavidal2024strong,pitavidal2025blueprint} emerge as solid candidates too despite present coherence times limited to tens of nanoseconds.

Given the feasibility of spin chains across different platforms, kicked-Ising QBs pose a compelling candidate for realizations in existing devices. Moreover, simulations on hardware confirm their potential as a programmable testbed, supporting near-term experimental realization. The impacts of finite-temperature initialization and realistic dissipative effects, including ergotropy and coherence properties, are further explored in subsequent work~\cite{romero2026impact}.


\emph{Acknowledgments.}---We thank Yongcheng Ding, David Guéry-Odelin, Anne-Maria Visuri, Javier Mas, Alfonso V. Ramallo, Juan Santos-Suárez and Grace M. Sommers for fruitful discussions. This work is supported by the project grants PID2024-157842OA-I00 and PID2021-126273NB-I00 funded by MCIN/AEI/10.13039/501100011033 and by ``ERDF A way of making Europe'' and ``ERDF Invest in your Future'', Spanish national project in the field of Artificial Intelligence (AIA2025-163435-C44), the Severo Ochoa Centres of Excellence program through Grant CEX2024-001445-S, the Spanish Ministry of Economic Affairs and Digital Transformation through the QUANTUM ENIA project call-Quantum Spain project. Y.B. acknowledges the Ayudas para contratos Ramón y Cajal (RYC2023-042699-I). We acknowledge the use of IBM Quantum services for this work. The views expressed are those of the authors and do not reflect the official policy or position of IBM or the IBM Quantum team.

\emph{Data availability.}---The data that support the findings of this article are openly available~\cite{repository}.

\bibliography{bibfile}


\onecolumngrid
\section*{---End Matter---}
\twocolumngrid

Classical simulations were performed using MPS via the ITensor library~\cite{itensor} with a matrix product operator cutoff of $10^{-5}$. The time-evolving block decimation (TEBD) method was used, as Eqs.~\eqref{eq:floquet_kick} and~\eqref{eq:time} suggest, which provide an efficient protocol for computing large-scale dynamics in short-range interacting systems. 

\emph{Analytical framework.}---Here we summarize two separate methods used to obtain the exact charging dynamics, referring to Supplemental Material for a step-by-step derivation~\cite{sm}. The first one is based on Clifford quantum cellular automata, using that any Clifford gate acting on a Pauli operator returns another Pauli operator, which holds for~\eqlabel{eq:floquet_kick} at the self-dual point. So, we can reconstruct the charging dynamics by evolving each quantum cell as:%
\begin{enumerate}%
    \item Using $U_{\{I,K\}} = \exp(-i H_{\{I,K\}})$, evolve each quantum cell $\sigma^\alpha_i(m) \coloneqq (U^\dagger_IU^\dagger_K)^m \sigma^\alpha_i (U_KU_I)^m$ after $m$ kicks by applying known conjugation rules.
    \item Evaluate the energy $E_N(m) = \braket{\psi(m)|H_0| \psi(m)} = \braket{\psi(0) | (U^\dagger_IU^\dagger_K)^m H_0 (U_KU_I)^m | \psi(0)}$.
\end{enumerate}%

To obtain the exact dynamics for arbitrary coupling strengths $J$ and $b$, we derive the charging dynamics from momentum space as the second method. The steps to consider are:
\begin{enumerate}%
    \item Apply the Jordan-Wigner transformation to work under a spinless fermion basis followed by a Fourier transform consistent with boundary conditions.%
    \item Integrate the Bogoliubov-de Gennes equations to obtain the Floquet operator $U_k(1,0)$ per mode.%
    \item Since $U_k(m,0)=U^m_k(1,0)$ after $m$ kicks, compute its $m$th power using Chebyshev polynomials.%
    \item Evolve the initial state $\ket{\psi(0)}=\prod_k \ket{\psi_k(0)}$ as $\ket{\psi(m)}=\prod_k U^m_k(1,0)\ket{\psi_k(0)}$ to obtain $E_N(m)$.%
\end{enumerate}%
\begin{figure*}[!tb]
    \centering
    \includegraphics[width=.9\linewidth]{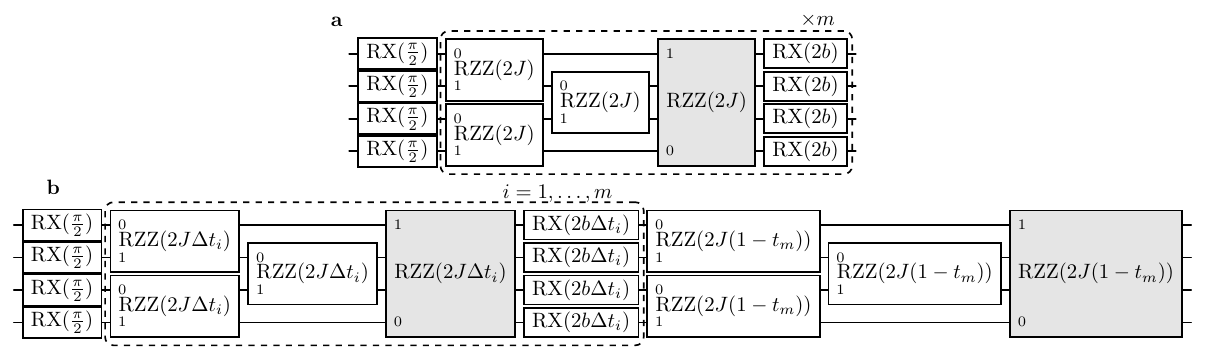}\vspace*{-1mm}%
    \caption{Gate decomposition of the circuits used in our experiments. (a) Four-qubit circuit implementation of the KIC QB Floquet operator $U(m,0)$ up to $m$ kicks following the decomposition presented in~\eqlabel{eq:rzz_rx_decom}. The first layer builds the GS $\ket{\psi(0)} = [\text{RX}(\frac{\pi}{2})\ket{0}]^{\otimes N}$. Inside the dashed line, effective implementation of $U(1,0)$. The shaded RZZ gate is applied if PBC are considered. (b) Same implementation as (a) but for the nonuniform KIC QB up to $m$ kicks, see~\eqlabel{eq:time}.}\label{fig:circuit}
\end{figure*}%

Despite addressing actual implementations of our proposals being desirable, in the following lines we indicate how we verified our findings using a quantum digital approach. The Floquet operator of~\eqlabel{eq:floquet_kick} has an efficient implementation on current IBM quantum platforms when $H^{zz}_1$ is considered as charger [\eqlabel{eq:components_zz}]. In our Letter we use~\textsc{ibm\_torino}, whose native gate set is given by
\begin{equation}
    X=\sigma^x,\text{ }\sqrt{X}=\frac{1}{2}\!\begin{bmatrix}1+i & 1-i \\ 1-i & 1+i\end{bmatrix}\!,\text{ } \text{RZ}(\theta)=e^{-i\theta\sigma^z/2},
\end{equation}
and $\text{CZ}=\diag(1,1,1,-1)$ as entangling gate. Moreover, IBM has recently added the fractional gates $\text{RZZ}(\theta)=e^{-i\theta\sigma_0^z\sigma_1^z/2}$ ($0<\theta\le\frac{\pi}{2}$) and $\text{RX}(\theta)=e^{-i\theta\sigma^x/2}$~\cite{frac}, which are suitable for the KIC model, since the Ising contribution can be encoded in a depth-two circuit as
\begin{equation}\label{eq:rzz_rx_decom}
    e^{-iH^{zz}_I} = \prod_{j\text{ odd}}\text{RZZ}_{j,j+1}(2J)\prod_{j\text{ even}}\text{RZZ}_{j,j+1}(2J),
\end{equation}
and $e^{-iH^{zz}_K}=\prod_{j=1}^N \text{RX}_j(2b)$, where subindices indicate the qubits where gates are applied. A similar decomposition is used for the nonuniform KIC case after the maps $2J\mapsto 2J\Delta t_i$ and $2b\mapsto 2b\Delta t_i$. Moreover, it is possible to write the initial state, GS of $H_0=(\omega_0/2)\sum_{i=1}^N \sigma^y_i$, as $\ket{\psi(0)} = [\text{RX}(\frac{\pi}{2})\ket{0}]^{\otimes N}$. So, both uniform and nonuniform models can be exactly implemented on current IBM platforms and compare the results obtained with the exact ones for benchmarking purposes. See~\figlabel{fig:circuit} for the gate decompositions of our experiments and~\figlabel{fig:experiment} for a schematic of the \textsc{ibm\_torino} qubit coupling map, where the $N=104$ qubits used are marked. \tablabel{tab:resources} includes the resources needed in terms of the system size, number of kicks and boundary conditions considered.
\begin{figure}[!tb]
    \centering
    \includegraphics[width=.9\linewidth]{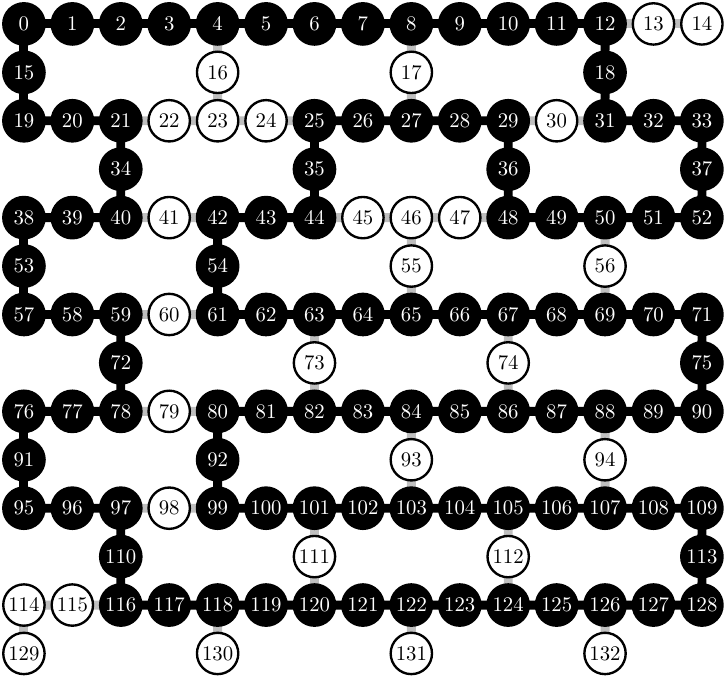}%
    \caption{Coupling map of \textsc{ibm\_torino}. In black, qubits used for the KIC QB experiments of $N=104$ cells under PBC.}\label{fig:experiment}
\end{figure}%
\begin{table}[!tb]
    \caption{Resources needed to embed $m$ kicks of KIC QB for $H_1^{zz}$ on \textsc{ibm\_torino} for different system sizes. See~\figlabel{fig:circuit}.}\label{tab:resources}
    \begin{ruledtabular}\begin{tabular}{lrrr}
        \multicolumn{1}{c}{Model} & \multicolumn{1}{c}{$\text{RX}(\theta)$} & \multicolumn{1}{c}{$\text{RZZ}(\theta)$} & \multicolumn{1}{c}{Depth} \\ \midrule
        \eqlabel{eq:floquet_kick} (PBC) & $(m+1)N$ & $mN$ & $3m+1$ \\
        \eqlabel{eq:floquet_kick} (OBC) & $(m+1)N$ & $m(N-1)$ & $3m+1$ \\
        \eqlabel{eq:time} (PBC) & $(m+1)N$ & $(m+1)N$ & $3(m+1)$ \\
        \eqlabel{eq:time} (OBC) & $(m+1)N$ & $(m+1)(N-1)$ & $3(m+1)$
    \end{tabular}\end{ruledtabular}
\end{table}%

One of the quantities to monitor along the evolution is the injected energy $E_N(\tau) = (\omega_0/2)\sum_{i=1}^N \braket{\sigma^\alpha_i}_\tau$. In order to measure Pauli expectation values, we can sample our circuit after rotating into their corresponding basis. While for Pauli-Z values no rotation is needed, for Pauli-Y the gate product $\text{RZ}(\frac{\pi}{2})\text{RX}(\frac{\pi}{2})$ is required right before measuring. To estimate the energy uncertainty from the sampled bitstrings, we use the propagation of uncertainties as a function of the Pauli expectation values. Each bit position can be treated as a Bernoulli variable, where if a bit $b$ has a probability $p$ of being 1, its variance is given by $\mathbb{V}(b)=p(1-p)$. For KIC, its variance reads as
\begin{equation}\label{eq:var}
    \mathbb{V}[E_N] = \frac{\omega_0^2}{4} \sum_{i=1}^N \Big[ \mathbb{V}[ \braket{\sigma^\alpha_i}_\tau ] + \sum_{i\neq j} \cov[\braket{\sigma^\alpha_i}_\tau, \braket{\sigma^\alpha_j}_\tau] \Big],
\end{equation}
where we considered $100 000$ samples per circuit. For the nonuniform KIC, the variance yields
\begin{equation}
    \mathbb{V}[E_N] = \frac{1}{n_\text{d}}\sum_{d=1}^{n_\text{d}}\left[\mathbb{V}[E_{N,d}] + \left( \mathbb{E}[E_{N,d}] - \bar{E}_N \right)^2 \right]
\end{equation}
with $E_{N,d}$ the energy injected at the $d$th disorder realization and $\bar{E}_N\coloneqq (1/n_\text{d})\sum_{d=1}^{n_\text{d}} E_{N,d}$, where we considered $n_\text{d}=10$ and $1 000$ samples per circuit. We have included covariances because, although $E_N$ is a sum of local Pauli expectation values, both the quantum state (e.g., via entanglement) and hardware effects (e.g., readout crosstalk) can induce interqubit correlations, thus ignoring covariance would miss these contributions.

\end{document}


\title{Supplemental Material for: ``Kicked-Ising Quantum Battery''}
\author{Sebastián V. Romero$^{\orcidlink{0000-0002-4675-4452}}$}
\email{sebastian.v.romero@csic.es}
\affiliation{Quantum Advanced Research Center (QuARC), CSIC, 28049 Madrid, Spain}
\affiliation{Instituto de Ciencia de Materiales de Madrid (ICMM), CSIC, 28049 Madrid, Spain}
\affiliation{\mbox{Departamento de Física Teórica de la Materia Condensada, Universidad Autónoma de Madrid, 28049 Madrid, Spain}}
\affiliation{Department of Physical Chemistry, University of the Basque Country EHU, Apartado 644, 48080 Bilbao, Spain}

\author{Xi Chen$^{\orcidlink{0000-0003-4221-4288}}$}
\email{xi.chen@csic.es}
\affiliation{Quantum Advanced Research Center (QuARC), CSIC, 28049 Madrid, Spain}
\affiliation{Instituto de Ciencia de Materiales de Madrid (ICMM), CSIC, 28049 Madrid, Spain}

\author{Yue Ban$^{\orcidlink{0000-0003-1764-4470}}$}
\email{yue.ban@csic.es}
\affiliation{Quantum Advanced Research Center (QuARC), CSIC, 28049 Madrid, Spain}
\affiliation{Instituto de Ciencia de Materiales de Madrid (ICMM), CSIC, 28049 Madrid, Spain}
\date{\today}

\begin{abstract}
    In this~\supinf, we provide additional notes and extended results supporting the findings in the main text. Specifically, we present a step-by-step analytical derivation of the energy injection dynamics under our kicked charging protocol using Clifford quantum cellular automata principles and momentum space, analyzing its dependence on the number of kicks and spins. Given that interactions beyond nearest neighbors naturally arise in many quantum platforms, such as neutral atoms in optical lattices, we examine the robustness of our models under power-law-decaying long-range interactions, demonstrating that their essential features persist. We also explore the effect of relaxing the kick duration, introducing a novel concept coined as \emph{quasikick}, where Dirac delta functions are replaced by narrow pulses; as well as the impact of slowly introducing the charger Hamiltonian instead of suddenly. Additionally, we derive the exact time evolution of the entanglement entropy, showing its linear growth with the number of kicks as reported in prior studies. Finally, we include a detailed analysis of the experimental data obtained on IBM hardware.
\end{abstract}%

\maketitle%

\renewcommand{\thetable}{S\arabic{table}}
\renewcommand{\theequation}{S\arabic{equation}}
\renewcommand{\thefigure}{S\arabic{figure}}
\renewcommand{\bibnumfmt}[1]{[S#1]}
\renewcommand{\citenumfont}[1]{S#1}

\vspace{-6mm}%
\tableofcontents

\section{Analytical solution of the kicked-Ising chain charging dynamics}

Before beginning with a detailed step-by-step derivation, first we summarize here two independent analytical methods to obtain the charging dynamics of the kicked-Ising chain (KIC), see Eqs.~(2) and~(3) in the main text, for arbitrary system sizes and number of kicks. These provide key tools to unveil the charging dynamics in our KIC quantum battery (QB) proposal. 

The first method considered is based on Clifford quantum cellular automata (CQCA), using the fact that any Clifford gate acting on a Pauli operator returns another Pauli operator, which holds for the Floquet operator [Eq.~(4) in the main text] under the self-dual operator regime, thereby establishing a connection between the kicked-Ising model and CQCA. As a result, this allows to track how each quantum cell of the battery Hamiltonian $H_0$ evolves after any number of kicks, thereby reconstructing the dynamics of energy injection.
The procedure is as follows:%
\begin{enumerate}%
    \item Under the self-dual operator regime, the Floquet operator becomes a Clifford unitary, allowing for an analytic description of the dynamics.
    \item Using $U_{\{I,K\}} = \exp(-i H_{\{I,K\}})$ [Eq.~(4) in the main text], evolve each quantum cell $\sigma^\alpha_i(m) \coloneqq (U^\dagger_IU^\dagger_K)^m \sigma^\alpha_i (U_KU_I)^m$ after $m$ kicks by applying known conjugation rules.
    \item Evaluate the energy $E_N(m) = \braket{\psi(m)|H_0| \psi(m)} = \braket{\psi(0) | (U^\dagger_IU^\dagger_K)^m H_0 (U_KU_I)^m | \psi(0)}$.
\end{enumerate}%
\begin{figure*}[!b]
    \centering%
    \includegraphics[width=.94\linewidth]{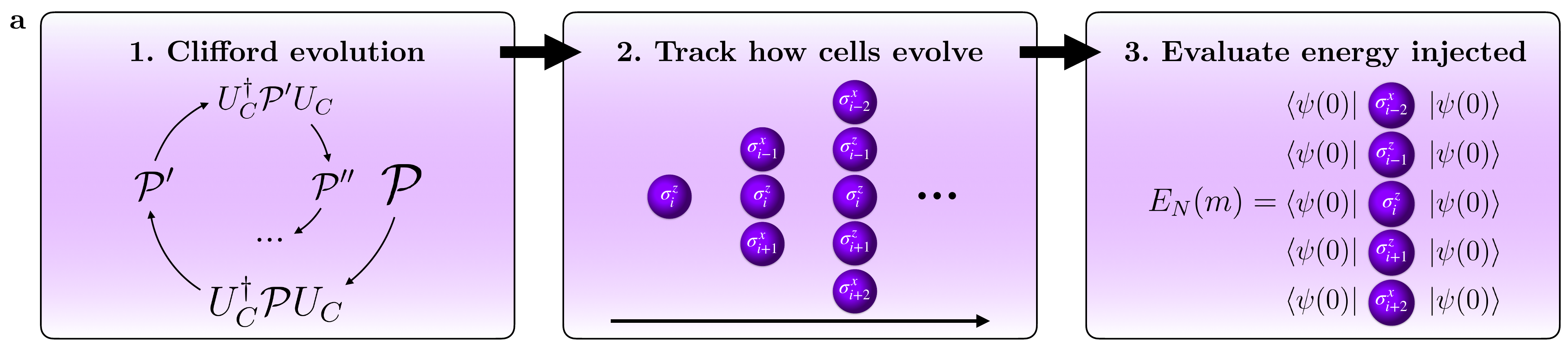}\\%
    \includegraphics[width=.94\linewidth]{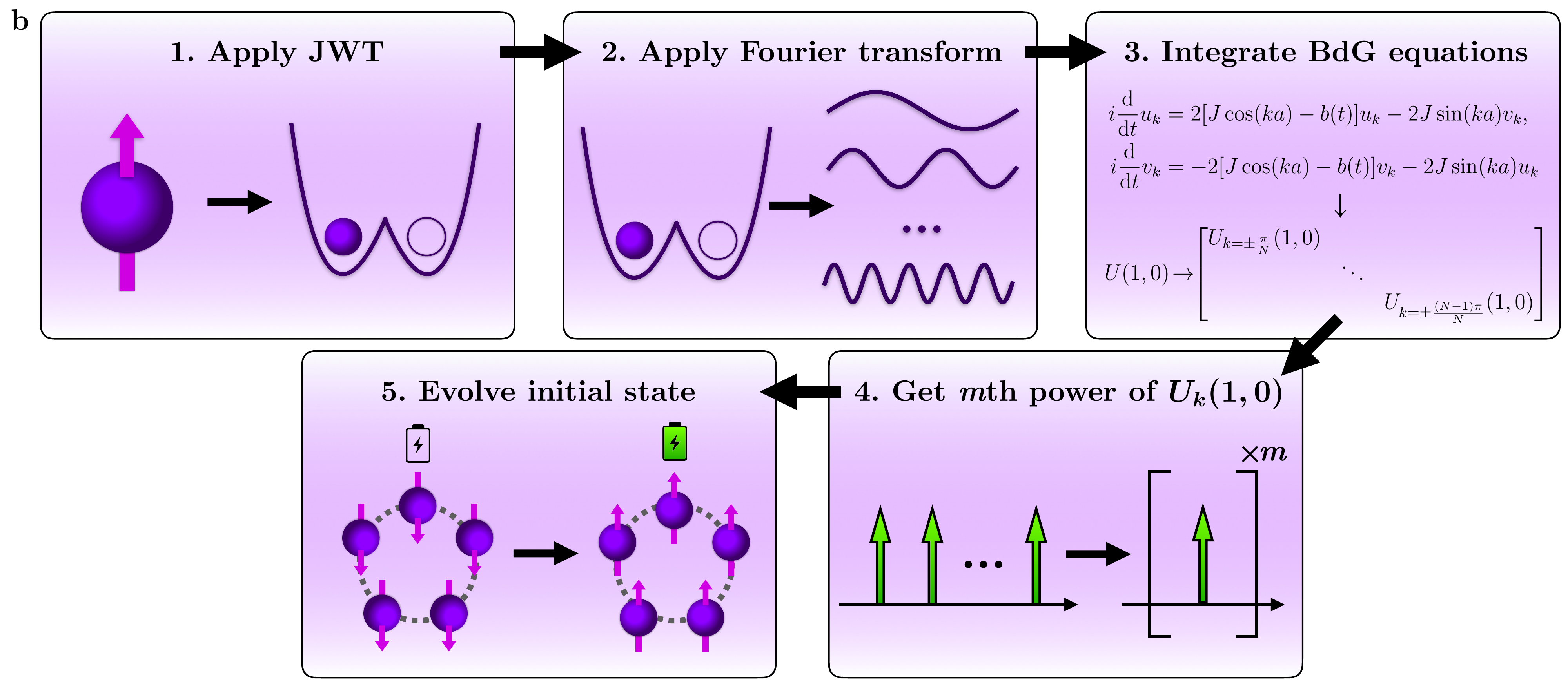}%
    \vspace{-1mm}%
    \caption{Exact charging dynamics of KIC. (a) Using CQCA principles: 1. in the self-dual operator regime, the system undergoes a Clifford evolution, preserving Pauli structure; 2. track how each cell operator ballistically evolves under repeated conjugation of $U(1,0)=U_KU_I$; 3. as the initial state is a product state, we compute $E_N(m) = \braket{\psi(0) | (U^\dagger_IU^\dagger_K)^m H_0 (U_KU_I)^m | \psi(0)}$ by measuring at every site each time-evolved operator. (b) Exact diagonalization from momentum space: 1. spins are mapped to spinless fermions via Jordan-Wigner transformation; 2. Fourier transforms are applied according to boundary conditions; 3. mode-wise Floquet operators $U_k(1,0)$ are obtained by applying the Bogoliubov transformation and integrating the Bogoliubov–de Gennes equations; 4. evolution after $m$ kicks, $U_k(m,0)=U^m_k(1,0)$, is computed using the Cayley-Hamilton theorem; 5. the state is reconstructed as $\ket{\psi(m)}=\prod_k U^m_k(1,0)\ket{\psi_k(0)}$, required to compute the injected energy $E_N(m) = \braket{\psi(m)|H_0| \psi(m)}$.}\label{fig:momentum_schematic}%
\end{figure*}

To obtain the exact dynamics for arbitrary coupling strengths, we derive the charging dynamics from momentum space. The steps to consider are (see~\figlabel{fig:momentum_schematic}):
\begin{enumerate}%
    \item Apply the Jordan-Wigner transformation~\cite{jordan1928uber} to work under a spinless fermion basis.%
    \item Depending on the parity of the initial state and number of spins, apply a Fourier transform consistent with boundary conditions~\cite{dziarmaga2005dynamics,damski2014exact}.%
    \item Apply the Bogoliubov transformation and integrate the Bogoliubov-de Gennes equations to obtain the Floquet operator $U_k(1,0)$ per mode.%
    \item Since $U_k(m,0)=U^m_k(1,0)$ after $m$ kicks, compute the $m$th power of the Floquet operator relying on the Cayley-Hamilton theorem~\cite{griffiths2001waves,das2024insights}.%
    \item Evolve the initial state $\ket{\psi(0)}=\prod_k \ket{\psi_k(0)}$ as $\ket{\psi(m)}=\prod_k U^m_k(1,0)\ket{\psi_k(0)}$ to obtain $E_N(m)$.%
\end{enumerate}%

\subsection{Spectrum of the Floquet operator}

To begin with the exact diagonalization of the KIC, we first derive the spectrum of its corresponding Floquet operator, unitary time-evolved operator obtained after applying one kick, which can be used to unveil the periodicity of the model under different system sizes and boundary conditions. For our study we consider as charger Hamiltonian $H_1^{xx}(t)=H^{xx}_I+H^{xx}_K(t)$ [Eq.~(2) in the main text], with Ising and kicked-transverse field contributions
\begin{equation}\label{eq:ref_ham1}
    H^{xx}_I = J\sum_{\braket{ij}}\sigma^x_i\sigma^x_j,\qquad H^{xx}_K(t) = \underbrace{b\sum_{j=1}^N\sigma^z_j}_{\coloneqq H^{xx}_K}\sum_{t_i\in\mathbb{Z}}\delta(t-t_i).
\end{equation}
Hereinafter, we work under the self-dual operator regime, where we consider as parameters $J=\pi/4$ and $b=-\pi/4$. Since this model is equivalent to the $H^{zz}_1(t)$ charger [Eq.~(3) in the main text] up to a global $\frac{\pi}{2}-$rotation around the Y-axis, the eigenvalues obtained for~\eqlabel{eq:ref_ham1} will be equivalent for those obtained for $H^{zz}_1(t)$. The resulting Floquet operator has the form
\begin{equation}\label{eq:floquet}
    U(1,0) = \mathfrak{T}\exp\left[ -i\int^1_0 \text{d}t H^{xx}_1(t) \right] = e^{-iH^{xx}_K}e^{-iH^{xx}_I}.
\end{equation}

We will obtain the spectrum by connecting in the Majorana basis the monodromy of the system, defined as the minimal-period net action of the Floquet evolution, i.e., the smallest $m$ for which $U^m(1,0)$ (or its adjoint conjugation) returns every state or operator to itself up to a global phase and parity operator $P\coloneqq \prod_{i=1}^N \sigma^z_i$, if present, followed by a block diagonalization of the Floquet operator. We set as Majorana operators
\begin{equation}
    \gamma_{2j-1}\coloneqq \Bigg( \prod_{l<j} \sigma^z_l \Bigg) \sigma^x_j, \qquad \gamma_{2j}\coloneqq \Bigg( \prod_{l<j} \sigma^z_l \Bigg) \sigma^y_j, \qquad (j=1,\dots,N)
\end{equation}
with $\gamma^\dagger_j=\gamma_j$, $\{\gamma_j,\gamma_k\}=2\delta_{jk}$, with $\sigma^z_j=-i\gamma_{2j-1}\gamma_{2j}$ and $\sigma^x_j\sigma^x_{j+1}=i\gamma_{2j}\gamma_{2j+1}$. In the bulk, the effect of conjugation for the kicked and Ising parts of the KIC are individually
\begin{equation}
\begin{aligned}
    e^{+iH^{xx}_K} \gamma_{2j-1} e^{-iH^{xx}_K} &= \gamma_{2j},  &\qquad e^{+iH^{xx}_K} \gamma_{2j} e^{-iH^{xx}_K} &= -\gamma_{2j-1}, \\
    e^{+iH^{xx}_I} \gamma_{2j-1} e^{-iH^{xx}_I} &= \gamma_{2j-2}, &\qquad e^{+iH^{xx}_I} \gamma_{2j} e^{-iH^{xx}_I} &= -\gamma_{2j+1}.
\end{aligned}
\end{equation}
Therefore, the conjugation of the Floquet operator in~\eqlabel{eq:floquet} returns
\begin{equation}\label{eq:conj_bulk}
    U^\dagger(1,0)\gamma_{2j-1}U(1,0) = e^{+iH^{xx}_I} \gamma_{2j} e^{-iH^{xx}_I} = -\gamma_{2j+1}, \qquad
    U^\dagger(1,0)\gamma_{2j}U(1,0) = e^{+iH^{xx}_I} (-\gamma_{2j-1}) e^{-iH^{xx}_I} = -\gamma_{2j-2}.
\end{equation}
While for periodic boundary conditions (PBC), one simply needs to take into account that $\sigma^\alpha_{N+1}=\sigma^\alpha_1$ with $\alpha\in\{x,y,z\}$, for open boundary conditions (OBC) the conjugation of the Floquet operator at the edges returns
\begin{equation}\label{eq:conj_edges}
    U^\dagger(1,0)\gamma_{2N-1}U(1,0) = e^{+iH^{xx}_I} \gamma_{2N} e^{-iH^{xx}_I} = \gamma_{2N}, \qquad
    U^\dagger(1,0)\gamma_2 U(1,0) = e^{+iH^{xx}_I} (-\gamma_1) e^{-iH^{xx}_I} = -\gamma_1.
\end{equation}
From the relationships in Eqs.~\eqref{eq:conj_bulk} and~\eqref{eq:conj_edges}, we can see that the monodromy under OBC (PBC) is $2N$ ($N$). 

We begin with the exact diagonalization of the OBC case. We can see that after $2N$ conjugations of the Floquet operator over every Majorana operator $\gamma_j$, we obtain the exact same operator but with a minus sign. Defining $\phi$ as the form of the eigenphases of~\eqlabel{eq:floquet}, we obtain that 
\begin{equation}\label{eq:phi_obc}
    e^{i2N\phi}=-1 \implies \phi = \frac{(2j-1)\pi}{2N}, \quad (\text{mod }2\pi)
\end{equation}
which we can use to obtain the set of different eigenvalues lying on the unit circle.

The next step is to connect $\phi$ with the eigenvalues obtained after a block-diagonalization of~\eqlabel{eq:floquet} under Majorana operators. Let $\Gamma\coloneqq[\gamma_1,\dots,\gamma_{2N}]^\text{T}$ be the column vector of $2N$ Majorana operators. Any quadratic Gaussian Floquet operator can be written as $U(1,0)=\exp(\Gamma^\text{T}A\Gamma/4)$ with $A^\text{T}=-A$. In the Heisenberg picture, this acts linearly on $\Gamma$ as $U^\dagger(1,0)\Gamma U(1,0)=O\Gamma$, with $O=e^A\in\text{SO}(2N)$~\cite{bravyi2005lagrangian}. There exists an orthogonal matrix $W\in\text{SO}(2N)$, with $W^\text{T}W=WW^\text{T}=I$, that block-diagonalizes $A$ as
\begin{equation}\label{eq:block_obc}
    W^\text{T}AW = \bigoplus_{j=1}^N \begin{bmatrix}
        0 & \epsilon_j \\ -\epsilon_j & 0
    \end{bmatrix} \implies W^\text{T}OW = \bigoplus_{j=1}^N \begin{bmatrix}
        \cos\epsilon_j & \sin\epsilon_j \\ -\sin\epsilon_j & \cos\epsilon_j
    \end{bmatrix},
\end{equation}
with $\epsilon_j\ge 0$. Defining as block Majoranas $\tilde{\Gamma}\coloneqq W^\text{T}\Gamma = [\tilde{\gamma}_1,\dots,\tilde{\gamma}_{2N}]^\text{T}$, the unitary factorizes as
\begin{equation}\label{eq:gaussian}
    U(1,0) = \exp\left( \frac{1}{4}\Gamma^\text{T} A \Gamma \right) = \exp\left[ \frac{1}{4} \tilde{\Gamma}^\text{T} (W^\text{T}AW) \tilde{\Gamma} \right] = \prod_{j=1}^N \exp\left( \frac{\epsilon_j}{2}\tilde{\gamma}_{2j-1}\tilde{\gamma}_{2j} \right).
\end{equation}
Let $f_j\coloneqq (\tilde{\gamma}_{2j-1} + i\tilde{\gamma}_{2j})/2$, satisfying $U^\dagger(1,0)f_jU(1,0)=e^{-i\epsilon_j}f_j$, and define $n_j\coloneqq f^\dagger_j f_j$ as the number operator. We can rewrite~\eqlabel{eq:gaussian} and factorize the Floquet operator as
\begin{equation}\label{eq:epsilon_obc}
    U(1,0) = \prod_{j=1}^N \exp\left[ -i \frac{\epsilon_j}{2}( 2n_j - 1 ) \right] = \exp\bigg[ -i \sum_{j=1}^N \frac{\epsilon_j}{2}( 2n_j - 1 ) \bigg].
\end{equation}
Connecting Eqs.~\eqref{eq:phi_obc} and~\eqref{eq:epsilon_obc} and defining $s_j\coloneqq 2n_j-1\in\{-1,+1\}$, each of the eigenphases of~\eqlabel{eq:floquet} can be written
\begin{equation}\label{eq:phi_obc_final}
    \Phi = \frac{\pi}{4N}\underbrace{\sum_{j=1}^N s_j(2j-1)}_{\coloneqq S(N)} = \frac{\pi}{4N}\Bigg[ \sum_{j\in G_+} (2j-1) - \sum_{j\in G_-} (2j-1) \Bigg] = \frac{\pi}{4N}\Bigg[ N^2 - 2\sum_{j\in G_-} (2j-1) \Bigg],
\end{equation}
with $G_+\coloneqq \{j=1,\dots,N \,|\, s_j=+1\}$ and $G_-\coloneqq \{j=1,\dots,N \,|\, s_j=-1\}$. In particular, for OBC we obtain $e^{i\Phi(N)}$ as distinct eigenvalues of~\eqlabel{eq:floquet}, with eigenphases 
\begin{equation}\label{eq:eigenvalues_obc}
    \Phi(N) = \begin{dcases}
        \left\{\frac{2\pi k}{8} \,|\, k\in\{1,2,6,7\}\right\} & \text{if }N=2 \\
        \left\{\frac{2\pi k}{24} \,|\, k\in \{1,3,7,9,15,17,21,23 \} \right\} & \text{if }N=3 \\
        \left\{\frac{2\pi k}{16} \,|\, k\in\mathbb{Z}_{16}-\{6,10\}\right\} & \text{if }N=4 \\
        \left\{ \frac{2\pi(2k+1)}{8N} \,|\, k\in\mathbb{Z}_{4N} \right\} & \text{if }N\ge 5\text{ and }N\text{ odd} \\
        \left\{ \frac{2\pi k}{4N} \,|\, k\in\mathbb{Z}_{4N} \right\} & \text{if }N\ge 6\text{ and }N\text{ even}
    \end{dcases},
\end{equation}
where we plot each of these sets in~\figlabel{fig:eigenvalues}(a)-(b). To observe these patterns, we can directly obtain them from~\eqlabel{eq:phi_obc_final} using brute force for $N\in[2,7]$ and, for $N\ge 8$, it is possible to prove them by using the following two lemmas to control $S(N)$ moduli:
\begin{enumerate}
    \item A single-site flip $s_j\mapsto -s_j$ changes $S(N)$ by $\Delta S=-2(2j-1)$, hence $\Delta S \bmod 8=2$ when $j$ is even and $\Delta S\bmod 8=6$ when $j$ is odd. This fixes $S(N)$ modulo $8$.
    \item An adjacent-pair flip, i.e., $s_j=s_{j+1}=-1$, changes $S(N)$ by $\Delta S=-2(2j-1)-2(2j+1)=-8j$, which preserves $S(N)$ modulo $8$ and allowing to tune $S(N)$ in steps of $8$ indexed by the pair position $j$.
\end{enumerate}
When $\gcd(8,N)=1$ (odd $N$), we set two congruences: first $S(N)\bmod 8 = S_1$ by single-site flips, and then $S(N)\bmod N=k$ by using adjacent-pair flips to realize $S(N)\bmod N = S_1-8t\bmod N = k$. Since $8$ and $N$ are coprime, the Chinese remainder theorem yields a unique solution $S(N)$ modulo $8N$~\cite{gauss1986congruences}. When $\gcd(8,N)>1$ (even $N$), the congruence $8t\bmod N=\Delta$ that encodes the modulo $N$ tuning admits a solution iff $d\coloneqq \gcd(8,N)$ divides $\Delta \equiv S_1 - k \pmod{N}$. In that case, we can define $t\equiv (8/d)^{-1}(\Delta/d)\pmod{N'}$, with $N'\coloneqq N/d$ and $a^{-1}$ the modular inverse of $a$ modulo $N'$, such that $aa^{-1}\equiv 1\pmod{N'}$. 

As an illustration, take $N=9$ and $S(N)\bmod 72 = k = 37$. Initially starting from all $s_j=+1$, $S_0=\sum_{j=1}^9(2j-1)\bmod 72=81\bmod 72=9$, so $S_0\bmod 8=1$. To reach $S(N)\bmod 8=5$, flip $j=1$ and $j=9$: $\Delta S=-2(1)-2(17)=-36$, yielding $S_1=45\bmod 8=5$. For the modulo $9$ condition, solve $8t\bmod 9= S_1-k\bmod 9= 8$, giving $t\equiv 1 \pmod{9}$. Choosing the adjacent pairs $j=4$ and $j=6$, whose total change is $\Delta S=-8\cdot 4 - 8\cdot 6 = -80\implies S(N)=45-80 \bmod 72= 37$, as expected. Finally, a valid set of spins is given by $\bm{s} = (s_1,\dots,s_9) = (-,+,+,-,-,-,-,+,-)$. For even $N$, let $N=10$ and $S(N)\bmod 80 = k=14$. Starting from all $s_j=+1$, $S_0=\sum_{j=1}^{10}(2j-1)\bmod 80=100\bmod 80 = 20$, so $S_0\bmod 8 = 4$. We flip $j=10$ to get $\Delta S=-2(19)=-38$ such that $S_1\bmod 8=62\bmod 8 = 6$. For the congruence $8t\bmod 10= S_1-k \bmod 10 = 62-14\bmod 10 = 8$, we have $d=\gcd(8,10)=2$ and $d=2$ divides $8$, so divide to obtain $4t\bmod 5 = 4$, hence $t\bmod 5 = 1$. Taking the adjacent pair $j=6$ yields $\Delta S=-8\cdot 6=-48$ and $S(N)=62-48\bmod 80=14$, as expected. Finally, a valid spin string is $\bm{s}=(s_1,\dots,s_{10})=(+,+,+,+,+,-,-,+,+,-)$.

Following the same reasoning as for the OBC case, we can extract the eigenphases when PBC are considered. Under these boundary conditions, it can be shown that the monodromy under Majorana operators yields
\begin{equation}\label{eq:monodromy_pbc}
    (U^\dagger(1,0))^N \gamma_a U^N(1,0) = \begin{dcases}
        \gamma_a & \text{if }N\text{ even} \\
        iP\gamma_a & \text{otherwise}
    \end{dcases}.
\end{equation}
Starting from the $N$ even case, we obtain eigenphases in the form of $e^{i\phi N} = +1 \implies \phi = 2\pi k/N$ $(\text{mod }2\pi)$,
which sets the eigenphases lying on the unit circle. Doing a similar block-diagonalization of the Floquet operator and rewriting it as a Gaussian operator [Eqs.~\eqref{eq:block_obc}-\eqref{eq:epsilon_obc}], the resulting eigenphases can be constructed as
\begin{equation}\label{eq:phi_pbc}
    \Phi(N) = \frac{2\pi}{4N}\sum_{k=1}^N s_k k = \frac{2\pi}{4N} \Bigg[ N(N+1) -4\sum_{k\in G_-} k \Bigg].
\end{equation}
Written in this form, we can conclude that the for $N$ even and PBC, the inner bracket of~\eqlabel{eq:phi_pbc} generates all the residues $l\in\mathbb{Z}_{4N}$ such that $l\bmod 4 = N\bmod 4\in\{0,2\}$.

Regarding the $N$ odd case, as the parity operator $P$ in the monodromy of the Majorana operators [\eqlabel{eq:monodromy_pbc}], the eigenvalues of the Floquet operator come from the union of sets of eigenphases coming from the even and odd parity sectors ($\epsilon_+$ and $\epsilon_-$, respectively). We can pick eigenstates of the Floquet operator $U(1,0)$ lying in the even and odd parity sectors ($\ket{\psi_+}$ and $\ket{\psi_-}$, respectively), such that $\braket{\psi_+|\gamma_j|\psi_-}\neq 0$, and exploit that $U(1,0)\ket{\psi_\pm}=e^{-i\epsilon_\pm}\ket{\psi_\pm}$ and $P\ket{\psi_\pm}=\pm\ket{\psi_\pm}$. Therefore, we have that
\begin{equation}
\left.\begin{aligned}
    \braket{\psi_+| (U^\dagger(1,0))^N \gamma_j U^N(1,0) | \psi_-} &= e^{iN(\epsilon_+ - \epsilon_-)} \braket{\psi_+|\gamma_j|\psi_-} \\
    &= \braket{\psi_+| iP\gamma_j | \psi_-} = i\braket{\psi_+|\gamma_j|\psi_-}
\end{aligned}\right\} \implies e^{iN(\epsilon_+ - \epsilon_-)} = i.
\end{equation}
Writing the eigenphases as $\epsilon_\pm = 2\pi m_\pm/4N \implies m_+ - m_- \bmod 4 = 1$. Setting $m_+=0\implies m_-=3$. Therefore, for the PBC case we finally obtain as set of distinct eigenphases
\begin{equation}\label{eq:eigenvalues_pbc}
    \Phi(N) = \begin{dcases}
        \left\{\frac{2\pi k}{N} \,|\, k\in\mathbb{Z}_N \right\} & \text{if }N\bmod 4=0 \\
        \left\{ \frac{2\pi (2k+1)}{2N} \,|\, k\in\mathbb{Z}_N \right\} & \text{if }N\bmod 4=2 \\
        \left\{ \frac{2\pi k}{N} \,|\, k\in\mathbb{Z}_N \right\} \cup \left\{ \frac{2\pi (4k+3)}{4N} \,|\, k\in\mathbb{Z}_N \right\} & \text{if }N\text{ odd}
    \end{dcases},
\end{equation}
where we plot each of these sets in~\figlabel{fig:eigenvalues}(c).

As a side note, we can see that at the self-dual point the many-body Floquet spectrum of~\eqlabel{eq:ref_ham1} is highly degenerated, where the number of distinct eigenvalues is proportional to the system size $N$ regardless of the boundary conditions and parity on the number of spins. As a concluding remark, all these results were verified numerically using exact diagonalization up to $N=16$. 
\begin{figure}[!tb]
    \centering
    \includegraphics[width=\linewidth]{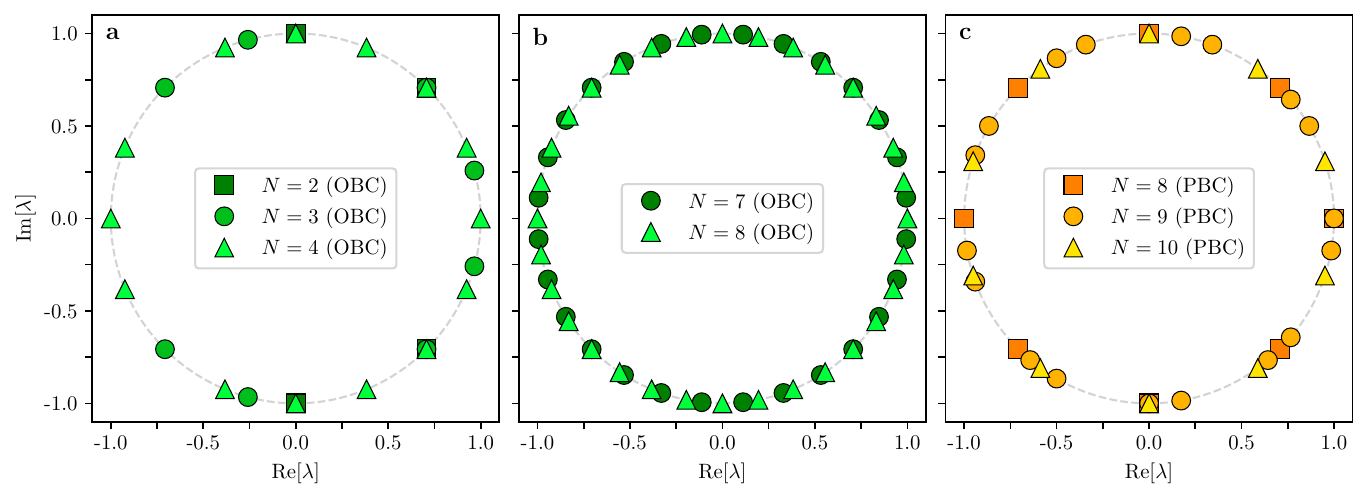}\vspace{-2mm}%
    \caption{Real and imaginary parts of the KIC Floquet operator eigenvalues. (a) Eigenvalues $\lambda=e^{i\Phi(N)}$ under OBC and system sizes $N\in\{2,3,4\}$ [\eqlabel{eq:eigenvalues_obc}]. (b) Same plot but considering OBC for $N\in\{7,8\}$, where all residues are included. (c) Eigenvalues under PBC and $N\in\{8,9,10\}$ [\eqlabel{eq:eigenvalues_pbc}]. The unit circle is drawn as a reference.}\label{fig:eigenvalues}
\end{figure}%

\subsection{Exact solution using Clifford quantum cellular automata}

While Clifford quantum cellular automata (CQCA) have been previously studied in the context of quantum circuits, quantum error correction codes, and dual-unitary dynamics~\cite{feynman1982simulating,watrous1995onedimensional,schumacher2004reversible,schlingemann2008structure,gross2012index,farrelly2020review}, our work identifies and exploits a precise correspondence between the self-dual operator regime~\cite{bertini2019entanglement} and CQCA evolution in the kicked-Ising model. The dynamics of energy injection of our quantum battery (QB) is analyzed by using CQCA under the self-dual operator regime. Specifically, when the Ising and transverse-field strengths are proportional to $\pi/4$, each Floquet layer is a Clifford unitary, enabling full analytic description of the energy dynamics.

\subsubsection{\texorpdfstring{$H^{xx}_1$}{H1xx} charger case}\label{sec:h1xx}

In the following lines, we derive the exact solution of the KIC charging dynamics under the $H^{xx}_1$ charger [\eqlabel{eq:ref_ham1}] for arbitrary system size, number of kicks and boundary conditions. For simplicity and without loss of generality, we use as battery Hamiltonian $H_0=(\omega_0/2)\sum_{i=1}^N\sigma^z_i$ whose ground state is given by $\ket{\psi(0)}=\ket{1}^{\otimes N}$ with $\ket{1}=[0,1]^\text{T}$. Additionally, its entire spectrum can be obtained following Ref.~\cite{romero2024scramblingchargingquantumbatteries}, whose corresponding energy levels are $\epsilon_k=(k-N/2)\omega_0$ (multiplicity $\binom{N}{k}$) $\forall k\in[0,N]$ where here we work without shifting to zero its ground state energy. This particular choice of one- and two-body interaction terms are naturally provided across several platforms, as discussed in the main text, fact that may enable KIC QB implementation on current setups. Hereinafter, we work under the self-dual operator regime, where we consider as parameters $J=\pi/4$ and $b=-\pi/4$.

Equation~\eqref{eq:ref_ham1} satisfies translational invariance under PBC. Let $T$ be the unitary translation operator, which shifts all spins by one site, such that $T\sigma^\alpha_iT^{-1}=\sigma^\alpha_{i+1}$ with $T^N=I$. We can easily see that $TH^{xx}_1T^{-1}=H^{xx}_1 \implies [T,H^{xx}_1(t)]=0$. Therefore, the time-evolved state also preserves translational invariance since $T\ket{\psi(0)}=\ket{\psi(0)}$, thus $T\ket{\psi(t)} = TU(t,0)\ket{\psi(0)} = U(t,0)T\ket{\psi(0)} = \ket{\psi(t)}$. Additionally, for both PBC and OBC, \eqlabel{eq:ref_ham1} preserves fermionic parity along the evolution ($[P,H(t)]=0$), with $P$ the parity operator. As we will discuss later, this implies that if an initial state that is either in the even or odd parity sector is evolved on time, it will continue being part of it along the evolution.

Depending on the boundary conditions considered and the parity on the number of cells, different charging dynamics appear, as seen in Fig. 2 in the main text. Expanding the definition of energy injected, $E_N(m) = \braket{\psi(m) | H_0 | \psi(m)} = \braket{\psi(0) | (U^\dagger_IU^\dagger_K)^m H_0 (U_KU_I)^m | \psi(0)} = \braket{\psi(0) | H_0(m) | \psi(0)}$, with $U_{\{I,K\}} \coloneqq \exp(-iH^{xx}_{\{I,K\}})$ [\eqlabel{eq:ref_ham1}] and $H_0(m) \coloneqq (U^\dagger_IU^\dagger_K)^m H_0 (U_KU_I)^m$ the Heisenberg picture of $H_0$. As we work in the self-dual operator regime, both $U_I$ and $U_K$ are Clifford gates, feature that can be exploited to keep track of how the charging dynamics evolve. Given that for any Pauli operator $\mathcal{P}$ and Clifford unitary $U_C$, $U^\dagger_C\mathcal{P}U_C$ returns another Pauli operator, we can derive how each individual quantum cell operator of $H_0$ transforms after applying an arbitrary number of kicks. Once computed, the energy injected in the chosen quantum cell can be obtained from the resulting operator. Since the initial state is a product of ground states of $\sigma^z$, $E_N(m)$ will only return a nonzero value in those cases where the evolved Pauli operator only contains identity and Pauli-Z operators, since $\braket{1 | \sigma^{\{x,y\}} | 1} = 0$. 

For this purpose, we use as key result that
\begin{equation}\label{eq:clifford}
    e^{i\theta \mathcal{P}}\mathcal{O}e^{-i\theta \mathcal{P}}=\begin{cases}
    \mathcal{O} & \text{if }[\mathcal{O},\mathcal{P}]=0 \\
    \cos(2\theta)\mathcal{O} - \frac{i}{2}\sin(2\theta)[\mathcal{O},\mathcal{P}] & \text{otherwise}
    \end{cases},
\end{equation}
with $\theta$ a real number and $\mathcal{O}$ and $\mathcal{P}$ two Pauli strings, as well as that two Pauli strings commute (anticommute) if the number of spin indices at which their respective Pauli operators do not commute is even (odd).

Without loss of generality, we study as a reference the evolution of the quantum cell located at the center of the chain, $\sigma^z_{\frac{N}{2}}$, assuming that the rest of cells behave similarly. While this holds trivially true when PBC are considered, due to translational invariance, for OBC we recover similar evolutions regardless the spin index accounted. We obtain its Heisenberg picture $\sigma^\alpha_i(m)\coloneqq (U^\dagger_IU^\dagger_K)^m \sigma^\alpha_i (U_KU_I)^m$ ($\alpha\in\{x,y,z\}$) under different boundary conditions and parity on the number of spins, which are required to derive their corresponding energy injected profiles.

\emph{Periodic boundary conditions and even number of spins.}---For the particular case when PBC and an even number of spins $N$ are considered, the system features a periodicity $U(t,0)=U(t+N,0)$. The Pauli operator evolves after $m$ kicks as
\begin{equation}\label{eq:xx_pbc_even}
    \sigma^z_{\frac{N}{2}}(m) = \begin{cases}
    \sigma^z_{\frac{N}{2}} & \text{if }m=0 \\
    -\sigma^x_{\frac{N}{2}-m}\bigotimes_{i=\frac{N}{2}-m+1}^{\frac{N}{2}-1+m}\sigma^z_i\sigma^x_{\frac{N}{2}+m} & \text{if }0<m<\frac{N}{2} \\
    -\bigotimes_{i=1}^{N-1}\sigma^z_i & \text{if }m=\frac{N}{2} \\
    -\sigma^y_{m-\frac{N}{2}}\bigotimes_{i=m-\frac{N}{2}+1}^{\frac{3N}{2}-m-1}\sigma^z_i \sigma^y_{\frac{3N}{2}-m} & \text{otherwise}
    \end{cases},
\end{equation}
results that can be proven by induction. See that the locality of the Pauli operators follow a light-cone-like behavior, where it increases and decreases linearly (sometimes referred as \emph{ballistically}), feature that will be present in all of our derivations due to the commutator present in~\eqlabel{eq:clifford}. The second case of~\eqlabel{eq:xx_pbc_even}, we can start by the base case ($m=1$) to see how the given equation holds, where
\begin{equation}\label{eq:xx_pbc_even_c2_1}
    U^\dagger_IU^\dagger_K \sigma^z_{\frac{N}{2}}(m=0) U_KU_I = U^\dagger_IU^\dagger_K \sigma^z_{\frac{N}{2}} U_KU_I = U^\dagger_I \sigma^z_{\frac{N}{2}} U_I = -\sigma^x_{\frac{N}{2}-1} \sigma^z_{\frac{N}{2}} \sigma^x_{\frac{N}{2}+1} = \sigma^z_{\frac{N}{2}}(m=1).
\end{equation}
Now, assuming that the equation holds true for $0<m<N/2$ kicks, we can compute the Pauli operator for the $(m+1)$th kick as
\begin{equation}\label{eq:xx_pbc_even_c2_2}
\begin{split}
    U^\dagger_IU^\dagger_K \sigma^z_{\frac{N}{2}}(m) U_KU_I &= U^\dagger_IU^\dagger_K \left[ -\sigma^x_{\frac{N}{2}-m}\bigotimes_{i=\frac{N}{2}-m+1}^{\frac{N}{2}-1+m}\sigma^z_i\sigma^x_{\frac{N}{2}+m} \right] U_KU_I = - U^\dagger_I \left[ \sigma^y_{\frac{N}{2}-m}\bigotimes_{i=\frac{N}{2}-m+1}^{\frac{N}{2}-1+m}\sigma^z_i\sigma^y_{\frac{N}{2}+m} \right] U_I \\
    &= -\sigma^x_{\frac{N}{2}-m-1}\bigotimes_{i=\frac{N}{2}-m}^{\frac{N}{2}+m}\sigma^z_i\sigma^x_{\frac{N}{2}+m+1},
\end{split}
\end{equation}
which finalizes the proof. For the $N/2$th kick, region where the injected energy maximizes [Fig. 2a in the main text], we obtain
\begin{equation}
    U^\dagger_IU^\dagger_K \sigma^z_{\frac{N}{2}}\bigg(\frac{N}{2}-1\bigg) U_KU_I = U^\dagger_IU^\dagger_K \left[ -\sigma^x_1 \bigotimes_{i=2}^{N-2}\sigma^z_i\sigma^x_{N-1} \right] U_KU_I = -U^\dagger_I \left[ \sigma^y_1 \bigotimes_{i=2}^{N-2}\sigma^z_i\sigma^y_{N-1} \right] U_I = -\bigotimes_{i=1}^{N-1}\sigma^z_i.
\end{equation}
If we evaluate this quantity with our initial state, we obtain $\braket{\psi(0) | \sigma^z_{\frac{N}{2}}(N/2) | \psi(0)} = -\prod_{i=1}^{N-1} \braket{1|\sigma^z|1} = -(-1)^{N-1} = 1$. Using translational invariance, we can extend this result to all the $N$ cells, such that at kick $N/2$ the energy injected is given by $E_N(t=N/2) = \braket{\psi(N/2) | H_0 | \psi(N/2)} = (N\omega_0/2) \braket{\psi(0) | \sigma^z_{\frac{N}{2}}(N/2) | \psi(0)} = N\omega_0/2$, as expected.

For the latter case of~\eqlabel{eq:xx_pbc_even}, we can start by the base case $m=N/2+1$, where
\begin{equation}\label{eq:xx_pbc_even_c4_1}
    U^\dagger_IU^\dagger_K \sigma^z_{\frac{N}{2}}\bigg(\frac{N}{2}\bigg) U_KU_I = U^\dagger_IU^\dagger_K \left[ -\bigotimes_{i=1}^{N-1}\sigma^z_i \right] U_KU_I = - U^\dagger_I \bigotimes_{i=1}^{N-1}\sigma^z_i U_I = -\sigma^y_1\bigotimes_{i=2}^{N-2}\sigma^z_i\sigma^y_{N-1}
\end{equation}
which is contemplated in~\eqlabel{eq:xx_pbc_even}. Assuming that the given equation holds true when $N/2<m<N$ kicks are applied, for the $(m+1)$th kick only in the endpoints non-commuting Pauli strings appear, returning
\begin{equation}\label{eq:xx_pbc_even_c4_2}
\begin{aligned}
    U^\dagger_IU^\dagger_K \sigma^z_{\frac{N}{2}}(m) U_KU_I &= U^\dagger_IU^\dagger_K \left[ -\sigma^y_{m-\frac{N}{2}}\bigotimes_{i=m-\frac{N}{2}+1}^{\frac{3N}{2}-m-1}\sigma^z_i \sigma^y_{\frac{3N}{2}-m} \right] U_KU_I = -U^\dagger_I \left[ \sigma^x_{m-\frac{N}{2}}\bigotimes_{i=m-\frac{N}{2}+1}^{\frac{3N}{2}-m-1}\sigma^z_i \sigma^x_{\frac{3N}{2}-m} \right] U_I \\
    &=-\sigma^y_{m+1-\frac{N}{2}}\bigotimes_{i=m-\frac{N}{2}+2}^{\frac{3N}{2}-m-2}\sigma^z_i \sigma^y_{\frac{3N}{2}-m-1}
\end{aligned}
\end{equation}
which proofs the formula. To check its consistency, we can compute the Pauli string obtained at the $N$th kick making use of~\eqlabel{eq:xx_pbc_even_c4_2} for $m=N-1$ kicks, where we should recover the initial local operator since the system has periodicity $U(t,0)=U(t+N,0)$. In particular,
\begin{equation}\label{eq:final_kick}
    U^\dagger_IU^\dagger_K \sigma^z_{\frac{N}{2}}(N-1) U_KU_I = U^\dagger_IU^\dagger_K \left[ -\sigma^y_{\frac{N}{2}-1} \sigma^z_{\frac{N}{2}} \sigma^y_{\frac{N}{2}+1} \right] U_KU_I = -U^\dagger_I \sigma^x_{\frac{N}{2}-1} \sigma^z_{\frac{N}{2}} \sigma^x_{\frac{N}{2}+1} U_I = \sigma^z_{\frac{N}{2}}.
\end{equation}

Therefore, following~\eqlabel{eq:xx_pbc_even}, we can conclude that the energy injected is minimal when $m=0$ ($E_N(0)=-N\omega_0/2$), maximal at $m=N/2$ ($E_N(N/2)=N\omega_0/2$) and zero otherwise.

\emph{Periodic boundary conditions and odd number of spins.}---Following the same methodology, it is also possible to derive the Pauli strings obtained after applying an arbitrary number of kicks $m$ under PBC but $N$ odd. The system features again a periodicity $U(t,0)=U(t+N,0)$, where the Pauli operator spreads as
\begin{equation}\label{eq:xx_pbc_odd}
    \sigma^z_{\lceil\frac{N}{2}\rceil}(m) = \begin{cases}
    \sigma^z_{\lceil\frac{N}{2}\rceil} & \text{if }m=0 \\
    -\sigma^x_{\lceil\frac{N}{2}\rceil-m}\bigotimes_{i=\lceil\frac{N}{2}\rceil-m+1}^{\lceil\frac{N}{2}\rceil-1+m}\sigma^z_i\sigma^x_{\lceil\frac{N}{2}\rceil+m} & \text{if }0<m\le\left\lceil\frac{N}{2}\right\rceil \\
    -\sigma^y_{m-\lceil\frac{N}{2}\rceil}\bigotimes_{i=m-\lceil\frac{N}{2}\rceil+2}^{3\lceil\frac{N}{2}\rceil-m-2}\sigma^z_i \sigma^y_{3\lceil\frac{N}{2}\rceil-m} & \text{otherwise}
    \end{cases},
\end{equation}
which looks similar to the expression obtained in~\eqlabel{eq:xx_pbc_even}. In fact, the last two cases can be obtained following the same arguments as the ones used to prove the second and last cases of~\eqlabel{eq:xx_pbc_even} (see Eqs.~\eqref{eq:xx_pbc_even_c2_1}-\eqref{eq:xx_pbc_even_c2_2} and~\eqref{eq:xx_pbc_even_c4_1}-\eqref{eq:xx_pbc_even_c4_2}, respectively). As seen in the main text, the energy injected under these conditions is $-N\omega_0/2$ at $m=0$ and zero otherwise.

\emph{Open boundary conditions and even number of spins.}---When OBC are considered, the translational invariance of the system is broken, which might affect how the Pauli operators evolve. In fact, their spreading slightly differs from the results obtained previously. As we will see, the periodicity of~\eqlabel{eq:ref_ham1} under OBC and $N$ even also changes, which satisfies $U(t,0)=U(t+2N,0)$. For the first $N-1$ kicks, we obtain
\begin{equation}\label{eq:xx_obc_even}
    \sigma^z_{\frac{N}{2}}(m) = \begin{cases}
        \sigma^z_{\frac{N}{2}} & \text{if }m=0 \\
        -\sigma^x_{\frac{N}{2}-m}\bigotimes_{i=\frac{N}{2}-m+1}^{\frac{N}{2}-1+m}\sigma^z_i\sigma^x_{\frac{N}{2}+m} & \text{if }0<m<\frac{N}{2} \\
        \sigma^y_1\bigotimes_{i=2}^{N-1}\sigma^z_i\sigma^x_N & \text{if }m=\frac{N}{2} \\
        -\sigma^y_{m-\frac{N}{2}+1}\bigotimes_{i=m-\frac{N}{2}+2}^{\frac{3N}{2}-m}\sigma^z_i \sigma^y_{\frac{3N}{2}-m+1} & \text{otherwise}
    \end{cases},\qquad (\text{if }m<N)
\end{equation}
The second case can be obtained following again Eqs.~\eqref{eq:xx_pbc_even_c2_1}-\eqref{eq:xx_pbc_even_c2_2}. For $m=N/2$, only the rightmost endpoint has a non-commuting Pauli matrix before applying the Ising contribution, where we get
\begin{equation}\label{eq:xx_obc_even_c3}
    U^\dagger_IU^\dagger_K \sigma^z_{\frac{N}{2}}\bigg(\frac{N}{2}-1\bigg) U_KU_I = U^\dagger_IU^\dagger_K \left[ -\sigma^x_1 \bigotimes_{i=2}^{N-2}\sigma^z_i\sigma^x_{N-1} \right] U_KU_I = -U^\dagger_I \left[ \sigma^y_1 \bigotimes_{i=2}^{N-2}\sigma^z_i\sigma^y_{N-1} \right] U_I = \sigma^y_1 \bigotimes_{i=2}^{N-1}\sigma^z_i\sigma^x_{N}.
\end{equation}

For the subsequent kick, $m=N/2+1$, now the leftmost endpoint have a non-commuting part, where we obtain
\begin{equation}\label{eq:xx_obc_even_c4}
    U^\dagger_IU^\dagger_K \sigma^z_{\frac{N}{2}}\bigg(\frac{N}{2}\bigg) U_KU_I = U^\dagger_IU^\dagger_K \left[ \sigma^y_1 \bigotimes_{i=2}^{N-1}\sigma^z_i\sigma^x_{N} \right] U_KU_I = U^\dagger_I \left[ -\sigma^x_1 \bigotimes_{i=2}^{N-1}\sigma^z_i\sigma^y_{N} \right] U_I = -\sigma^y_2 \bigotimes_{i=3}^{N-2}\sigma^z_i\sigma^y_{N},
\end{equation}
which is the same operator as the one shown in~\eqlabel{eq:xx_pbc_even_c4_1} but with spin indices displaced towards the right one position. To conclude with the demonstration, we can simply follow the same arguments used in~\eqlabel{eq:xx_pbc_even_c4_2}. 

As expected, for the $N$th kick we will not recover the initial Pauli operator, obtaining $\sigma^z_{\frac{N}{2}}(m=N)=\sigma^z_{\frac{N}{2}+1}$ instead. Starting from this operator and following the same procedure, one can show that for $N\le m<2N$ the Pauli operator evolves as~\eqlabel{eq:xx_obc_even} but applying the map $\sigma^z_k\mapsto\sigma^z_{N+1-k}$. Therefore, it is possible to prove that it is actually at kick $m=2N$ when the initial operator is recovered. In particular, this is true for each initial operator, with the $k$th quantum cell operator being displaced at kick $m=N$ from $\sigma^z_k$ to $\sigma^z_{N+1-k}$. Nonetheless, while the periodicity of~\eqlabel{eq:ref_ham1} under OBC and $N$ even is $U(t,0)=U(t+2N,0)$ locally, the actual periodicity of the model is given by $U(t,0)=U(t+N,0)$ globally, as seen in Fig. 2b in the main text. As a final remark, from~\eqlabel{eq:xx_obc_even}, we can trivially see that the energy injected is $-N\omega_0/2$ at $m=0$ and zero otherwise.

\emph{Open boundary conditions and odd number of spins.}---Finally, now we study the system when OBC and $N$ odd are considered, showing a periodicity $U(t,0)=U(t+N,0)$. Under these conditions, the Pauli operator spreads as
\begin{equation}\label{eq:xx_obc_odd}
    \sigma^z_{\lceil\frac{N}{2} \rceil}(m) = \begin{cases}
        \sigma^z_{\lceil\frac{N}{2} \rceil} & \text{if }m=0 \\
        -\sigma^x_{\lceil\frac{N}{2}\rceil-m}\bigotimes_{i=\lceil\frac{N}{2}\rceil-m+1}^{\lceil\frac{N}{2}\rceil-1+m}\sigma^z_i\sigma^x_{\lceil\frac{N}{2}\rceil+m} & \text{if }0<m<\left\lceil\frac{N}{2}\right\rceil \\
        -\sigma^y_{m-\lceil\frac{N}{2}\rceil}\bigotimes_{i=m-\lceil\frac{N}{2}\rceil+3}^{3\lceil\frac{N}{2}\rceil-m-1}\sigma^z_i \sigma^y_{3\lceil\frac{N}{2}\rceil-m+1} & \text{otherwise}
    \end{cases},
\end{equation}
where we can prove these results using again Eqs.~\eqref{eq:xx_pbc_even_c2_1}-\eqref{eq:xx_pbc_even_c2_2} and~\eqref{eq:xx_pbc_even_c4_1}-\eqref{eq:xx_pbc_even_c4_2} for the latter two cases, respectively. We can see that the energy injected is minimal at $m=0$ ($E_N(0)=-N\omega_0/2$) and zero otherwise.

As a summary of our results, we observe the following patterns under the $H^{xx}_1$ charger and using $q\in\mathbb{Z}$: 
\begin{itemize}
\item When PBC are considered:%
    \begin{itemize}%
        \item If $N$ is even (periodicity $U(m+N,0)=U(m,0)$):
        \begin{itemize}
            \item If $m=(q+1/2)N$, cells are maximally charged ($E_N(m)=N\omega_0/2$), most excited energy of $H_0$.
            \item If $m=qN$, cells are maximally discharged ($E_N(m)=-N\omega_0/2$), ground state energy of $H_0$.
            \item Otherwise, $E_N(m)=0$.
        \end{itemize}
        \item If $N$ is odd (periodicity $U(m+N,0)=U(m,0)$):
        \begin{itemize}
            \item If $m=qN$, cells are maximally discharged ($E_N(m)=-N\omega_0/2$), ground state energy of $H_0$.
            \item Otherwise, $E_N(m)=0$.
        \end{itemize}
    \end{itemize}
\item When OBC are considered, regardless of the parity of $N$ (periodicity $U(m+N,0)=U(m,0)$):
    \begin{itemize}
        \item If $m=qN$, cells are maximally discharged ($E_N(m)=-N\omega_0/2$), ground state energy of $H_0$.
        \item Otherwise, $E_N(m)=0$.
    \end{itemize}
\end{itemize}
We can see that the energy injection profiles when OBC are considered, regardless of the parity of the number of spins, are equal to those when PBC are considered and $N$ is odd. As concluding remark, if we shift to zero the ground state energy of $H_0$, i.e., we perform the map $E_N(m)\mapsto E_N(m)+N\omega_0/2$, we recover the energy values reported in the main text.

\subsubsection{\texorpdfstring{$H^{zz}_1$}{H1zz} charger case}\label{sec:h1zz}

Building upon the results obtained in the previous section, now we derive the exact solution of the KIC charging dynamics under the $H^{zz}_1$ charger [Eq.~(3) in the main text] for arbitrary system size and number of kicks. For this case, we use as battery Hamiltonian $H_0=(\omega_0/2)\sum_{i=1}^N\sigma^y_i$, whose ground state is given by $\ket{\psi(0)}=\ket{-i}^{\otimes N}$ with $\ket{\pm i}=(\ket{0} \pm i\ket{1})/\sqrt{2}$. Its energy spectrum has as energy levels $\epsilon_k=(k-N/2)\omega_0$ (multiplicity $\binom{N}{k}$) $\forall k\in[0,N]$~\cite{romero2024scramblingchargingquantumbatteries}, where again we work without shifting to zero its ground state energy for simplicity. Let $H_1^{zz}(t)=H^{zz}_I+H^{zz}_K(t)$ be the charger Hamiltonian, with Ising and kicked-transverse field contributions
\begin{equation}\label{eq:ref_ham_zz1}
    H^{zz}_I = J\sum_{\braket{ij}}\sigma^z_i\sigma^z_j,\qquad H^{zz}_K(t) = b\sum_{j=1}^N\sigma^x_j\sum_{t_i\in\mathbb{Z}}\delta(t-t_i),
\end{equation}
Without loss of generality, we use the same coupling strengths as in the main text, where $J=\pi/4$ and $b=-\pi/4$. As a side note, this model is equivalent to~\eqlabel{eq:ref_ham1} up to a global $\frac{\pi}{2}$-rotation around the Y-axis.

Equation~\eqref{eq:ref_ham_zz1} also satisfies translational invariance under PBC and preserves fermionic parity along the time evolution, as~\eqlabel{eq:ref_ham1} does. As we continue working in the self-dual operator regime, we can apply the same logic as in the previous section to enlighten the charging dynamics behind this model under different boundary conditions and parity in the number of spins. Since the initial state is a product of ground states of $\sigma^y$, $E_N(m)$ will only return a nonzero value in those cases where the evolved Pauli operator only contains identity and Pauli-Y operators, since $\braket{-i | \sigma^{\{x,z\}} | -i} = 0$. 

\emph{Periodic boundary conditions and even number of spins.}---The Pauli operator under PBC and $N$ even showcases a periodicity of $U(t,0)=U(t+N,0)$, where for arbitrary number of kicks it reads as
\begin{equation}\label{eq:zz_pbc_even}
    \sigma^y_{\frac{N}{2}}(m) = \begin{cases}
        \sigma^y_{\frac{N}{2}} & \text{if }m=0 \\
        \sigma^z_{\frac{N}{2}} & \text{if }m=1 \\
        \sigma^z_{\frac{N}{2}-m+1}\bigotimes_{i=1}^{m-1} \sigma^y_{\frac{N}{2}-m+2i}\sigma^z_{\frac{N}{2}-m+2i+1} & \text{if }1<m\le\frac{N}{2} \\
        \sigma^y_{m-\frac{N}{2}}\bigotimes_{i=1}^{N-m} \sigma^z_{2i+m-\frac{N}{2}-1}\sigma^y_{2i+m-\frac{N}{2}} & \text{otherwise}
    \end{cases},
\end{equation}
where we can see that the energy is minimal at $m=0$ ($E_N(0)=-N\omega_0/2$) and zero otherwise. While the second case can be trivially obtained, the latter two ones of can be proven by induction. Starting from the third case of~\eqlabel{eq:zz_pbc_even}, we can check that for its base case ($m=2$) the equation holds. Using $\sigma^y_{\frac{N}{2}}(m=1)=\sigma^z_{\frac{N}{2}}$, we obtain
\begin{equation}\label{eq:zz_pbc_even_c31}
    U^\dagger_IU^\dagger_K \sigma^y_{\frac{N}{2}}(m=1) U_KU_I = U^\dagger_IU^\dagger_K \sigma^z_{\frac{N}{2}} U_KU_I = U^\dagger_I (-\sigma^y_{\frac{N}{2}})U_I = \sigma^z_{\frac{N}{2}-1}\sigma^y_{\frac{N}{2}}\sigma^z_{\frac{N}{2}+1} = \sigma^y_{\frac{N}{2}}(m=2),
\end{equation}
which matches the expected result. For the $(m+1)$th kick, we obtain
\begin{equation}\label{eq:zz_pbc_even_c32}
\begin{split}
    &U^\dagger_IU^\dagger_K \sigma^y_{\frac{N}{2}}(m) U_KU_I \\
    &\qquad= U^\dagger_IU^\dagger_K \bigg[ \sigma^z_{\frac{N}{2}-m+1}\bigotimes_{i=1}^{m-1} \sigma^y_{\frac{N}{2}-m+2i}\sigma^z_{\frac{N}{2}-m+2i+1} \bigg] U_KU_I \\
    &\qquad= U^\dagger_I \bigg[ (-1)^m\sigma^y_{\frac{N}{2}-m+1}\bigotimes_{i=1}^{m-1} \sigma^z_{\frac{N}{2}-m+2i}\sigma^y_{\frac{N}{2}-m+2i+1} \bigg] U_I \\
    &\qquad= (-1)^m \exp\bigg[i\frac{\pi}{4}\sum_{j=1}^{m}\sigma^z_{\frac{N}{2}-m+2j-1}\sigma^z_{\frac{N}{2}-m+2j}\bigg] \sigma^z_{\frac{N}{2}-m} \bigotimes_{i=0}^{m-1}\sigma^x_{\frac{N}{2}-m+2i+1} \exp\bigg[-i\frac{\pi}{4}\sum_{j=1}^{m}\sigma^z_{\frac{N}{2}-m+2j-1}\sigma^z_{\frac{N}{2}-m+2j}\bigg] \\
    &\qquad= (-1)^{2m} \sigma^z_{\frac{N}{2}-m}\bigotimes_{i=1}^{m}\sigma^y_{\frac{N}{2}-m+2i-1}\sigma^z_{\frac{N}{2}-m+2i} = \sigma^z_{\frac{N}{2}-m}\bigotimes_{i=1}^{m}\sigma^y_{\frac{N}{2}-m+2i-1}\sigma^z_{\frac{N}{2}-m+2i},
\end{split}
\end{equation}
which finalizes the proof. Following a similar strategy for the last case of~\eqlabel{eq:zz_pbc_even}, we can start from the base case ($m=N/2+1$), where
\begin{equation}\label{eq:zz_pbc_even_c41}
\begin{split}
    U^\dagger_IU^\dagger_K \sigma^y_{\frac{N}{2}}\!\left(\frac{N}{2}\right) U_KU_I &= U^\dagger_IU^\dagger_K \sigma^z_1 \bigotimes_{i=0}^{\frac{N}{2}-2} \sigma^y_{2(i+1)}\sigma^z_{2(i+1)+1} U_KU_I = U^\dagger_I \bigg[ (-1)^{\frac{N}{2}} \sigma^y_1 \bigotimes_{i=0}^{\frac{N}{2}-2} \sigma^z_{2(i+1)}\sigma^y_{2(i+1)+1} \bigg] U_I \\
    &=(-1)^{\frac{N}{2}} \exp\!\left[i\frac{\pi}{4}\sum_{j=0}^{\frac{N}{2}-1} \sigma^z_{2j+1}\sigma^z_{2(j+1)} \right] \bigotimes_{i=0}^{\frac{N}{2}-1}\sigma^x_{2i+1} \sigma^z_N \exp\!\left[-i\frac{\pi}{4}\sum_{j=0}^{\frac{N}{2}-1} \sigma^z_{2j+1}\sigma^z_{2(j+1)} \right] \\
    &= (-1)^N \sigma^y_1 \bigotimes_{i=0}^{\frac{N}{2}-2} \sigma^z_{2(i+1)}\sigma^y_{2(i+1)+1} = \sigma^y_1 \bigotimes_{i=0}^{\frac{N}{2}-2} \sigma^z_{2(i+1)}\sigma^y_{2(i+1)+1},
\end{split}
\end{equation}
as expected. Assuming that $\sigma^y_{\frac{N}{2}}(m)$ for $N/2<m<N$ holds true, for $m+1$ we obtain
\begin{equation}\label{eq:zz_pbc_even_c42}
\begin{split}
    &U^\dagger_IU^\dagger_K \sigma^y_{\frac{N}{2}}(m) U_KU_I \\
    &\qquad= U^\dagger_IU^\dagger_K \sigma^y_{m-\frac{N}{2}}\bigotimes_{i=0}^{N-m-1}\left( \sigma^z_{2i+m-\frac{N}{2}+1}\sigma^y_{2(i+1)+m-\frac{N}{2}} \right)  U_KU_I \\
    &\qquad= U^\dagger_I \bigg[ (-1)^{N-m} \sigma^z_{m-\frac{N}{2}}\bigotimes_{i=0}^{N-m-1}\left( \sigma^y_{2i+m-\frac{N}{2}+1}\sigma^z_{2(i+1)+m-\frac{N}{2}} \right) \bigg] U_I \\
    &\qquad=(-1)^{N-m} \exp\left[i\frac{\pi}{4}\sum_{j=1}^{N-m} \sigma^z_{2i+m-\frac{N}{2}-1}\sigma^z_{2i+m-\frac{N}{2}} \right] \bigotimes_{i=1}^{N-m}\sigma^x_{2i+m-\frac{N}{2}-1}\sigma^z_{\frac{3N}{2}-m} \exp\left[-i\frac{\pi}{4}\sum_{j=1}^{N-m} \sigma^z_{2i+m-\frac{N}{2}-1}\sigma^z_{2i+m-\frac{N}{2}} \right] \\
    &\qquad=(-1)^{2(N-m)} \sigma^y_{m-\frac{N}{2}+1}\bigotimes_{i=1}^{N-m-1}\left( \sigma^z_{2i+m-\frac{N}{2}}\sigma^y_{2i+m-\frac{N}{2}+1} \right) = \sigma^y_{m-\frac{N}{2}+1}\bigotimes_{i=1}^{N-m-1}\left( \sigma^z_{2i+m-\frac{N}{2}}\sigma^y_{2i+m-\frac{N}{2}+1} \right),
\end{split}
\end{equation}
which completes the proof.

\emph{Periodic boundary conditions and odd number of spins.}---Now, when PBC and $N$ odd are considered, the system showcases a periodicity of $U(t,0)=U(t+4N,0)$, where the Pauli operator evolves for arbitrary number of kicks as
\begin{equation}\label{eq:zz_pbc_odd}
    \sigma^y_{\lceil\frac{N}{2}\rceil}(m) = \begin{cases}
        \sigma^y_{\lceil\frac{N}{2}\rceil} & \text{if }m=0 \\
        \sigma^z_{\lceil\frac{N}{2}\rceil} & \text{if }m=1 \\ 
        \Big[ \bigotimes_{i=1}^{\lfloor \frac{N}{2}\rfloor} \sigma^x_i \Big] \sigma^z_{\lceil\frac{N}{2}\rceil} \Big[ \bigotimes_{i=\lceil \frac{N}{2} \rceil + 1}^N \sigma^x_i \Big] & \text{if }m=N \\ 
        -\Big[ \bigotimes_{i=1}^{\lfloor \frac{N}{2}\rfloor} \sigma^x_i \Big] \sigma^y_{\lceil\frac{N}{2}\rceil} \Big[ \bigotimes_{i=\lceil \frac{N}{2} \rceil + 1}^N \sigma^x_i \Big] & \text{if }m=N+1 \\
        \sigma^z_{\lceil\frac{N}{2}\rceil-m+1} \bigotimes_{i=0}^{m-2} \sigma^y_{\lceil \frac{N}{2}\rceil-m+2i}\sigma^z_{\lceil \frac{N}{2}\rceil-m+2i+1} & \text{if }1<m\le\left\lceil\frac{N}{2}\right\rceil \\
        \Big[ \bigotimes_{i=1}^{m- \lceil \frac{N}{2}\rceil} \sigma^x_i \Big] \sigma^z_{m-\lceil\frac{N}{2}\rceil +1} \bigotimes_{i=1}^{N-m} \sigma^y_{2i+m-\lceil\frac{N}{2}\rceil} \sigma^z_{2i+m-\lceil\frac{N}{2}\rceil+1}  \Big[ \bigotimes_{i=3\lceil\frac{N}{2}\rceil-m}^{N} \sigma^x_i \Big] & \text{if }\left\lceil\frac{N}{2}\right\rceil < m <N \\
        - \Big[ \bigotimes_{i=0}^{3\lceil\frac{N}{2}\rceil-m-1} \sigma^x_i \Big] \sigma^z_{3\lceil\frac{N}{2}\rceil-m} \bigotimes_{i=0}^{m-2} \sigma^y_{3\lceil \frac{N}{2}\rceil-m+2i-1}\sigma^z_{3\lceil \frac{N}{2}\rceil-m+2i} \Big[ \bigotimes_{i=0}^{3\lceil\frac{N}{2}\rceil-m-1} \sigma^x_i \Big] & \text{if }N+1<m< 3\left\lceil \frac{N}{2}\right\rceil  \\
        - \sigma^y_{m-3\lceil \frac{N}{2}\rceil +1} \bigotimes_{i=0}^{N-m-1} \sigma^z_{2(i+1)+m-3\lceil\frac{N}{2}\rceil} \sigma^y_{2(i+1)+m-3\lceil\frac{N}{2}\rceil+1} & \text{if }3\left\lceil\frac{N}{2}\right\rceil \le m< 2N \\
        -\sigma^y_{\lceil\frac{N}{2}\rceil}(m-2N) & \text{otherwise}
    \end{cases}.
\end{equation}
It can be seen the energy injected is minimal at $m=0$ ($E_N(0)=-N\omega_0/2$), maximal at $m=2N$ ($E_N(2N)=N\omega_0/2$) and zero otherwise. The fifth case can be proven using Eqs.~\eqref{eq:zz_pbc_even_c31}-\eqref{eq:zz_pbc_even_c32} and the eighth one using Eqs.~\eqref{eq:zz_pbc_even_c41}-\eqref{eq:zz_pbc_even_c42}. We can start by proving using induction the sixth case stated. We get for the base case ($m=\lceil\frac{N}{2}\rceil + 1$)
\begin{equation}
\begin{split}
    U^\dagger_IU^\dagger_K \sigma^y_{\lceil\frac{N}{2}\rceil}\left(\left\lceil\frac{N}{2}\right\rceil\right) U_KU_I &= U^\dagger_IU^\dagger_K \sigma^z_1 \bigotimes_{i=1}^{\lceil\frac{N}{2}\rceil - 1} \sigma^y_{2i}\sigma^z_{2i+1}  U_KU_I = U^\dagger_I \left[ (-1)^{\lceil\frac{N}{2}\rceil} \sigma^y_1 \bigotimes_{i=1}^{\lceil\frac{N}{2}\rceil -1} \sigma^z_{2i}\sigma^y_{2i+1} \right] U_I \\
    &= (-1)^{2\lceil\frac{N}{2}\rceil-1} \exp\left[i\frac{\pi}{4}\sum_{j=1}^{\lceil\frac{N}{2}\rceil -1} \sigma^z_{2j}\sigma^z_{2j+1} \right] \bigotimes_{i=1}^{\lceil\frac{N}{2}\rceil -1}\sigma^x_{2i-1}\sigma^y_N \exp\left[-i\frac{\pi}{4}\sum_{j=1}^{\lceil\frac{N}{2}\rceil -1} \sigma^z_{2j}\sigma^z_{2j+1} \right] \\
    &=-(-1)\sigma^x_1 \sigma^z_2 \bigotimes_{i=1}^{\lceil\frac{N}{2}\rceil -2}\sigma^y_{2i+1}\sigma^z_{2(i+1)} \sigma^x_N = \sigma^x_1 \sigma^z_2 \bigotimes_{i=1}^{\lceil\frac{N}{2}\rceil -2}\sigma^y_{2i+1}\sigma^z_{2(i+1)} \sigma^x_N.
\end{split}
\end{equation}
Now, assuming that the~\eqlabel{eq:zz_pbc_odd} applies for $\lceil\frac{N}{2}\rceil < m < N$, and making use of $\sigma^{\alpha\beta}_i\coloneqq\sigma^\alpha_i\sigma^\beta_{i+1}$ ($\alpha,\beta\in\{x,y,z\}$ and $\text{e}_\pm[\cdot]\coloneqq\exp[\pm i\frac{\pi}{4}\cdot]$ as more compact notations, for the kick $m+1$ we obtain
\begin{equation}
\begin{split}
    &U^\dagger_IU^\dagger_K \sigma^y_{\lceil\frac{N}{2}\rceil}(m) U_KU_I \\
    &\qquad= U^\dagger_IU^\dagger_K \left[ \bigotimes_{i=1}^{m- \lceil \frac{N}{2}\rceil} \sigma^x_i \right] \sigma^z_{m-\lceil\frac{N}{2}\rceil +1} \bigotimes_{i=1}^{N-m} \sigma^{yz}_{2i+m-\lceil\frac{N}{2}\rceil}  \left[ \bigotimes_{i=3\lceil\frac{N}{2}\rceil-m}^{N} \sigma^x_i \right] U_KU_I \\
    &\qquad= (-1)^{N-m+1} U^\dagger_I \left[ \bigotimes_{i=1}^{m- \lceil \frac{N}{2}\rceil} \sigma^x_i \right] \sigma^y_{m-\lceil\frac{N}{2}\rceil +1} \bigotimes_{i=1}^{N-m} \sigma^{zy}_{2i+m-\lceil\frac{N}{2}\rceil} \left[ \bigotimes_{i=3\lceil\frac{N}{2}\rceil-m}^{N} \sigma^x_i \right] U_I \\
    &\qquad=(-1)^{2(N-m)+1}\text{e}_+\left[\sum_{j=1}^{N-m}\sigma^{zz}_{2j+m-\lceil\frac{N}{2}\rceil}\right] \left[ \bigotimes_{i=1}^{m- \lceil \frac{N}{2}\rceil} \sigma^x_i \right] \bigotimes_{i=1}^{N-m} \sigma^x_{2i+m-\lceil\frac{N}{2}\rceil-1} \sigma^y_{3\lceil\frac{N}{2}\rceil-m-1} \left[ \bigotimes_{i=3\lceil\frac{N}{2}\rceil-m}^{N} \sigma^x_i \right] \text{e}_-\left[\sum_{j=1}^{N-m}\sigma^{zz}_{2j+m-\lceil\frac{N}{2}\rceil}\right] \\
    &\qquad= -(-1) \left[ \bigotimes_{i=1}^{m- \lceil \frac{N}{2}\rceil+1} \sigma^x_i \right] \sigma^z_{m-\lceil\frac{N}{2}\rceil+1} \bigotimes_{i=1}^{N-m-1} \sigma^y_{2i+m-\lceil\frac{N}{2}\rceil-1} \sigma^z_{3\lceil\frac{N}{2}\rceil-m-1} \left[ \bigotimes_{i=3\lceil\frac{N}{2}\rceil-m-1}^{N} \sigma^x_i \right] \\
    &\qquad= \left[ \bigotimes_{i=1}^{m- \lceil \frac{N}{2}\rceil+1} \sigma^x_i \right] \sigma^z_{m-\lceil\frac{N}{2}\rceil+1} \bigotimes_{i=1}^{N-m-1} \sigma^y_{2i+m-\lceil\frac{N}{2}\rceil-1} \sigma^z_{3\lceil\frac{N}{2}\rceil-m-1} \left[ \bigotimes_{i=3\lceil\frac{N}{2}\rceil-m-1}^{N} \sigma^x_i \right],
\end{split}
\end{equation}
which completes the proof. From this point, to prove the cases $m=N$ and $m=N+1$ are trivial and they are left to the reader. 

To conclude with the demonstration, the seventh case of~\eqlabel{eq:zz_pbc_odd} can be also proven by induction. Starting from the base case $m=N+2$, we obtain
\begin{equation}
\begin{split}
    U^\dagger_IU^\dagger_K \sigma^y_{\lceil\frac{N}{2}\rceil}(N+1) U_KU_I &=  - U^\dagger_IU^\dagger_K \left[ \bigotimes_{i=1}^{\lfloor \frac{N}{2}\rfloor} \sigma^x_i \right] \sigma^y_{\lceil\frac{N}{2}\rceil} \left[ \bigotimes_{i=\lceil \frac{N}{2} \rceil + 1}^N \sigma^x_i \right] U_KU_I = - U^\dagger_I \left[ \bigotimes_{i=1}^{\lfloor \frac{N}{2}\rfloor} \sigma^x_i \right] \sigma^z_{\lceil\frac{N}{2}\rceil} \left[ \bigotimes_{i=\lceil \frac{N}{2} \rceil + 1}^N \sigma^x_i \right] U_I \\
    &= -\exp\left[ i\frac{\pi}{4}\sigma^z_{\lceil\frac{N}{2}\rceil} \sigma^z_{\lceil\frac{N}{2}\rceil+1}\right] \left[ \bigotimes_{i=1}^{\lfloor \frac{N}{2}\rfloor -1} \sigma^x_i \right] \sigma^y_{\lceil\frac{N}{2}\rceil-1} \left[ \bigotimes_{i=\lceil \frac{N}{2} \rceil + 1}^N \sigma^x_i \right] \exp\left[ -i\frac{\pi}{4}\sigma^z_{\lceil\frac{N}{2}\rceil} \sigma^z_{\lceil\frac{N}{2}\rceil+1}\right] \\
    &= - \left[ \bigotimes_{i=1}^{\lceil \frac{N}{2}\rceil -2} \sigma^x_i \right] \sigma^y_{\lceil\frac{N}{2}\rceil-1} \sigma^z_{\lceil\frac{N}{2}\rceil} \sigma^y_{\lceil\frac{N}{2}\rceil+1} \left[ \bigotimes_{i=\lceil \frac{N}{2} \rceil + 2}^N \sigma^x_i \right],
\end{split}
\end{equation}
which matches the formula provided in~\eqlabel{eq:zz_pbc_odd}. Assuming that for $N+1 < m < 3\lceil\frac{N}{2}\rceil - 1$ the formula also holds, and making use again of the aforementioned compact notations $\sigma^{\alpha\beta}_i$ and $\text{e}_{\pm}[\cdot]$, for the kick $m+1$
\begin{equation}\label{eq:zz_obc_even_c72}
\begin{split}
    &U^\dagger_IU^\dagger_K \sigma^y_{\lceil\frac{N}{2}\rceil}(m) U_KU_I \\
    &\qquad= - U^\dagger_IU^\dagger_K \left[ \bigotimes_{i=0}^{3\lceil\frac{N}{2}\rceil-m-1} \sigma^x_i \right] \sigma^z_{3\lceil\frac{N}{2}\rceil-m} \bigotimes_{i=0}^{m-2} \sigma^{yz}_{3\lceil \frac{N}{2}\rceil-m+2i-1} \left[ \bigotimes_{i=0}^{3\lceil\frac{N}{2}\rceil-m-1} \sigma^x_i \right] U_KU_I \\
    &\qquad= (-1)^{m+1} U^\dagger_I \left[ \bigotimes_{i=0}^{3\lceil\frac{N}{2}\rceil-m-1} \sigma^x_i \right] \sigma^y_{3\lceil\frac{N}{2}\rceil-m} \bigotimes_{i=0}^{m-2} \sigma^{zy}_{3\lceil \frac{N}{2}\rceil-m+2i-1} \left[ \bigotimes_{i=0}^{3\lceil\frac{N}{2}\rceil-m-1} \sigma^x_i \right] U_I \\
    &\qquad=(-1)^{2m} \text{e}_+\left[\sum_{i=0}^{m-2} \sigma^{zz}_{3\lceil \frac{N}{2}\rceil-m+2i-1} \right] \left[ \bigotimes_{i=0}^{3\lceil\frac{N}{2}\rceil-m} \sigma^x_i \right] \bigotimes_{i=0}^{m-3} \sigma^x_{3\lceil \frac{N}{2}\rceil-m+2i} \sigma^y_{3\lceil \frac{N}{2}\rceil + m+4} \left[ \bigotimes_{i=0}^{3\lceil\frac{N}{2}\rceil-m-1} \sigma^x_i \right] \text{e}_-\left[\sum_{i=0}^{m-2} \sigma^{zz}_{3\lceil \frac{N}{2}\rceil-m+2i-1} \right] \\
    &\qquad= - \left[ \bigotimes_{i=0}^{3\lceil\frac{N}{2}\rceil-m} \sigma^x_i \right] \sigma^z_{3\lceil \frac{N}{2}\rceil - m + 1} \bigotimes_{i=0}^{m-3} \sigma^{yz}_{3\lceil \frac{N}{2}\rceil-m+2i} \left[ \bigotimes_{i=0}^{3\lceil\frac{N}{2}\rceil-m-1} \sigma^x_i \right]
\end{split}
\end{equation}
which completes the proof. To conclude with the demonstration, finally we derive the last case of~\eqlabel{eq:zz_pbc_odd}, where for $m=2N$ we obtain
\begin{equation}
\begin{split}
    U^\dagger_IU^\dagger_K \sigma^y_{\lceil\frac{N}{2}\rceil}(2N-1) U_KU_I &= -U^\dagger_IU^\dagger_K \sigma^y_{\lceil\frac{N}{2}\rceil -1} \sigma^z_{\lceil\frac{N}{2}\rceil} \sigma^y_{\lceil\frac{N}{2}\rceil +1} U_KU_I = -(-1) U^\dagger_I \sigma^z_{\lceil\frac{N}{2}\rceil -1} \sigma^y_{\lceil\frac{N}{2}\rceil} \sigma^z_{\lceil\frac{N}{2}\rceil +1} U_I \\
    &= (-1) \exp\left[i\frac{\pi}{4}\sigma^z_{\lceil\frac{N}{2}\rceil}\sigma^z_{\lceil\frac{N}{2}\rceil+1}\right] \sigma^x_{\lceil\frac{N}{2}\rceil}\sigma^z_{\lceil\frac{N}{2}\rceil+1} \exp\left[-i\frac{\pi}{4}\sigma^z_{\lceil\frac{N}{2}\rceil}\sigma^z_{\lceil\frac{N}{2}\rceil+1}\right] = -\sigma^y_{\lceil\frac{N}{2}\rceil},
\end{split}
\end{equation}
which returns the initial operator but with opposite sign, point where the QB maximally charges. Therefore, for kicks $2N\le m < 4N$ we obtain as Pauli operators $-\sigma^y_{\lceil\frac{N}{2}\rceil}(m-2N)$, which are the same operators as for $m < 2N$, whose forms were already proven by induction, but with the sign flipped.

\emph{Open boundary conditions and even number of spins.}---When OBC and an even number of spins are considered, the system also showcases a periodicity of $U(t,0)=U(t+4N,0)$, where the Pauli operator evolves for arbitrary number of kicks as
\begin{equation}\label{eq:zz_obc_even}
    \sigma^y_{\frac{N}{2}}(m) = \begin{cases}
        \sigma^y_{\frac{N}{2}} & \text{if }m=0 \\
        \sigma^z_{\frac{N}{2}} & \text{if }m=1 \\ 
        \sigma^z_{\frac{N}{2}-m+1}\bigotimes_{i=1}^{m-1}\sigma^y_{\frac{N}{2}-m+2i}\sigma^z_{\frac{N}{2}-m+2i+1} & \text{if }1<m\le\frac{N}{2} \\
        -\sigma^x_1\sigma^z_2\bigotimes_{i=2}^{\frac{N}{2}} \sigma^y_{2i-1}\sigma^z_{2i} & \text{if }m=\frac{N}{2}+1 \\
        -\Big[\bigotimes_{i=1}^{m-\frac{N}{2}} \sigma^x_i \Big] \sigma^z_{m-\frac{N}{2}+1} \bigotimes_{i=1}^{N-m} \sigma^y_{2i+m-\frac{N}{2}} \sigma^z_{2i+1+m-\frac{N}{2}} \Big[\bigotimes_{i=\frac{3N}{2}-m+2}^N \sigma^x_i \Big] & \text{if }\frac{N}{2}+1 < m < N \\
        -\Big[\bigotimes_{i=1}^{\frac{N}{2}} \sigma^x_i \Big] \sigma^z_{\frac{N}{2}+1} \Big[\bigotimes_{i=\frac{N}{2}+2}^N \sigma^x_i \Big] & \text{if }m=N \\
        \Big[\bigotimes_{i=1}^{\frac{N}{2}} \sigma^x_i \Big] \sigma^y_{\frac{N}{2}+1} \Big[\bigotimes_{i=\frac{N}{2}+2}^N \sigma^x_i \Big] & \text{if }m=N+1 \\
        \Big[ \bigotimes_{i=1}^{\frac{3N}{2} -m+1} \sigma^x_i \Big] \sigma^y_{\frac{3N}{2}-m+2} \bigotimes_{i=2}^{N-m}\sigma^z_{2i-1+\frac{3N}{2}-m} \sigma^y_{2i+\frac{3N}{2}-m} \Big[ \bigotimes_{i=m-\frac{N}{2}+1}^{N} \sigma^x_i \Big] & \text{if }N+1 < m < \frac{3N}{2} \\
        \sigma^x_1\sigma^y_2\bigotimes_{i=2}^{\frac{N}{2}} \sigma^z_{2i-1}\sigma^y_{2i} & \text{if }m=\frac{3N}{2} \\
        -\sigma^y_1\bigotimes_{i=1}^{\frac{N}{2}-1}\sigma^z_{2i}\sigma^y_{2i+1} & \text{if }m=\frac{3N}{2} +1 \\
        - \sigma^y_{m-\frac{3N}{2} +1} \bigotimes_{i=1}^{N-m} \sigma^z_{2i+m-\frac{3N}{2}} \sigma^y_{2i+m-\frac{3N}{2}+1} & \text{if }\frac{3N}{2}+1 < m < 2N \\
        -\sigma^y_{\frac{N}{2}}(m-2N) & \text{otherwise}
    \end{cases}.
\end{equation}
Again, it can be seen that the energy injected is minimal at $m=0$ ($E_N(0)=-N\omega_0/2$), maximal at $m=2N$ ($E_N(2N)=N\omega_0/2$) and zero otherwise. 

To prove the fourth case of~\eqlabel{eq:zz_obc_even}, at kick $m=\frac{N}{2}+1$ we obtain
\begin{equation}
\begin{split}
    U^\dagger_IU^\dagger_K \sigma^y_{\frac{N}{2}}\left(\frac{N}{2}\right) U_KU_I &= U^\dagger_IU^\dagger_K \sigma^z_1\bigotimes_{i=1}^{\frac{N}{2}-1}\sigma^y_{2i}\sigma^z_{2i+1} U_KU_I = (-1)^{\frac{N}{2}} U^\dagger_I \sigma^y_1\bigotimes_{i=1}^{\frac{N}{2}-1}\sigma^z_{2i}\sigma^y_{2i+1} U_I \\
    &= (-1)^{\frac{N}{2}} \exp\left[ i\frac{\pi}{4}\sum_{i=1}^{\frac{N}{2}-1} \sigma^z_{2i}\sigma^z_{2i+1} \right] \sigma^x_1 \bigotimes_{i=1}^{\frac{N}{2}-1} \sigma^x_{2i+1}\sigma^z_N \exp\left[ -i\frac{\pi}{4}\sum_{i=1}^{\frac{N}{2}-1} \sigma^z_{2i}\sigma^z_{2i+1} \right] \\
    &= (-1)^{N-1} \sigma^x_1\sigma^z_2\bigotimes_{i=2}^{\frac{N}{2}} \sigma^y_{2i-1}\sigma^z_{2i} = - \sigma^x_1\sigma^z_2\bigotimes_{i=2}^{\frac{N}{2}} \sigma^y_{2i-1}\sigma^z_{2i}.
\end{split}
\end{equation}
To conclude with the demonstration, we can use the steps followed in~\eqlabel{eq:zz_obc_even_c72}.

\begin{equation}
\begin{split}
    U^\dagger_IU^\dagger_K \sigma^y_{\frac{N}{2}}\left( \frac{3N}{2}-1 \right) U_KU_I &= U^\dagger_IU^\dagger_K \sigma^x_1\sigma^x_2 \sigma^y_3\bigotimes_{i=2}^{\frac{N}{2}-1} \sigma^z_{2i}\sigma^y_{2i+1} \sigma^x_N U_KU_I = (-1)^{\frac{N}{2}-2} U^\dagger_I \sigma^x_1\sigma^x_2 \sigma^z_3\bigotimes_{i=2}^{\frac{N}{2}-1} \sigma^y_{2i}\sigma^z_{2i+1}\sigma^x_N U_I \\
    &= (-1)^{\frac{N}{2}-1} \exp\left[ i\frac{\pi}{4}\sum_{i=1}^{\frac{N}{2}-1} \sigma^z_{2i}\sigma^z_{2i+1} \right] \sigma^x_1\sigma^x_2 \bigotimes_{i=2}^{\frac{N}{2}-1} \sigma^x_{2i} \sigma^y_N \exp\left[ -i\frac{\pi}{4}\sum_{i=1}^{\frac{N}{2}-1} \sigma^z_{2i}\sigma^z_{2i+1} \right] \\
    &= (-1)^{N-2} \sigma^x_1\sigma^y_2\bigotimes_{i=2}^{\frac{N}{2}} \sigma^z_{2i-1}\sigma^y_{2i} =  \sigma^x_1\sigma^y_2\bigotimes_{i=2}^{\frac{N}{2}} \sigma^z_{2i-1}\sigma^y_{2i}.
\end{split}
\end{equation}
For the subsequent kick, $m=\frac{3N}{2}+1$, we obtain
\begin{equation}
\begin{split}
    U^\dagger_IU^\dagger_K \sigma^y_{\frac{N}{2}}\left( \frac{3N}{2} \right) U_KU_I &= U^\dagger_IU^\dagger_K \sigma^x_1\sigma^y_2\bigotimes_{i=2}^{\frac{N}{2}} \sigma^z_{2i-1}\sigma^y_{2i} U_KU_I = (-1)^{\frac{N}{2}-1} U^\dagger_I \sigma^x_1\sigma^z_2\bigotimes_{i=2}^{\frac{N}{2}} \sigma^y_{2i-1}\sigma^z_{2i} U_I \\
    &= (-1)^{\frac{N}{2}} \exp\left[ i\frac{\pi}{4}\sum_{i=1}^{\frac{N}{2}-1} \sigma^z_{2i}\sigma^z_{2i+1} \right] \sigma^y_1 \bigotimes_{i=2}^{\frac{N}{2}} \sigma^x_{2i-1} \exp\left[ -i\frac{\pi}{4}\sum_{i=1}^{\frac{N}{2}-1} \sigma^z_{2i}\sigma^z_{2i+1} \right] \\
    &= (-1)^{N-1} \sigma^y_1\bigotimes_{i=1}^{\frac{N}{2}-1} \sigma^z_{2i}\sigma^y_{2i+1} = - \sigma^y_1\bigotimes_{i=1}^{\frac{N}{2}-1} \sigma^z_{2i}\sigma^y_{2i+1}.
\end{split}
\end{equation}

\emph{Open boundary conditions and odd number of spins.}---For the case when OBC and an odd number of spins are considered, the system also showcases a periodicity of $U(t,0)=U(t+4N,0)$, where the Pauli operator evolves for arbitrary number of kicks as
\begin{equation}\label{eq:zz_obc_odd}
    \sigma^y_{\lceil\frac{N}{2}\rceil}(m) = \begin{cases}
        \sigma^y_{\lceil\frac{N}{2}\rceil} & \text{if }m=0 \\
        \sigma^z_{\lceil\frac{N}{2}\rceil} & \text{if }m=1 \\ 
        \Big[ \bigotimes_{i=1}^{\lfloor \frac{N}{2}\rfloor} \sigma^x_i \Big] \sigma^z_{\lceil\frac{N}{2}\rceil} \Big[ \bigotimes_{i=\lceil \frac{N}{2} \rceil + 1}^N \sigma^x_i \Big] & \text{if }m=N \\ 
        -\Big[ \bigotimes_{i=1}^{\lfloor \frac{N}{2}\rfloor} \sigma^x_i \Big] \sigma^y_{\lceil\frac{N}{2}\rceil} \Big[ \bigotimes_{i=\lceil \frac{N}{2} \rceil + 1}^N \sigma^x_i \Big] & \text{if }m=N+1 \\
        \sigma^z_{\lceil\frac{N}{2}\rceil-m+1} \bigotimes_{i=0}^{m-2} \sigma^y_{\lceil \frac{N}{2}\rceil-m+2i}\sigma^z_{\lceil \frac{N}{2}\rceil-m+2i+1} & \text{if }1<m\le\left\lceil\frac{N}{2}\right\rceil \\
        \Big[ \bigotimes_{i=1}^{m- \lceil \frac{N}{2}\rceil} \sigma^x_i \Big] \sigma^z_{m-\lceil\frac{N}{2}\rceil +1} \bigotimes_{i=1}^{N-m} \sigma^y_{2i+m-\lceil\frac{N}{2}\rceil} \sigma^z_{2i+m-\lceil\frac{N}{2}\rceil+1}  \Big[ \bigotimes_{i=}^{N} \sigma^x_i \Big] & \text{if }\left\lceil\frac{N}{2}\right\rceil < m <N \\
        - \Big[ \bigotimes_{i=0}^{3\lceil\frac{N}{2}\rceil-m-1} \sigma^x_i \Big] \sigma^z_{3\lceil\frac{N}{2}\rceil-m} \bigotimes_{i=0}^{m-2} \sigma^y_{3\lceil \frac{N}{2}\rceil-m+2i-1}\sigma^z_{3\lceil \frac{N}{2}\rceil-m+2i} \Big[ \bigotimes_{i=0}^{3\lceil\frac{N}{2}\rceil-m-1} \sigma^x_i \Big] & \text{if }N+1<m< 3\left\lceil \frac{N}{2}\right\rceil  \\
        - \sigma^y_{m-3\lceil \frac{N}{2}\rceil +1} \bigotimes_{i=0}^{N-m-1} \sigma^z_{2(i+1)+m-3\lceil\frac{N}{2}\rceil} \sigma^y_{2(i+1)+m-3\lceil\frac{N}{2}\rceil+1} & \text{if }3\left\lceil\frac{N}{2}\right\rceil \le m< 2N \\
        -\sigma^y_{\lceil\frac{N}{2}\rceil}(m-2N) & \text{otherwise}
    \end{cases},
\end{equation}
results that look similar to those obtained in~\eqlabel{eq:zz_pbc_odd}. In fact, all these cases can be proven by induction replicating the arguments used to demonstrate the validity of~\eqlabel{eq:zz_pbc_odd}, but now taking into account that OBC are considered. It can be seen the energy injected is minimal at $m=0$ ($E_N(0)=-N\omega_0/2$), maximal at $m=2N$ ($E_N(2N)=N\omega_0/2$) and zero otherwise.

As a summary of our results, when we $H^{zz}_1$ as charger Hamiltonian now we observe the following patterns:
\begin{itemize}
    \item When PBC are considered:
    \begin{itemize}
        \item If $N$ is even (periodicity $U(m+N,0)=U(m,0)$):
        \begin{itemize}
            \item If $m=qN$, cells are maximally discharged ($E_N(m)=-N\omega_0/2$), ground state energy of $H_0$.
            \item Otherwise, $E_N(m)=0$.
        \end{itemize}
        \item If $N$ is odd (periodicity $U(m+4N,0)=U(m,0)$):
        \begin{itemize}
            \item If $m=2(1+2q)N$, cells are maximally charged ($E_N(m)=N\omega_0/2$), most excited energy of $H_0$.
            \item If $m=4qN$, cells are maximally discharged ($E_N(m)=-N\omega_0/2$), ground state energy of $H_0$.
            \item Otherwise, $E_N(m)=0$.
        \end{itemize}
    \end{itemize}
    \item When OBC are considered, regardless of the parity of $N$ (periodicity $U(m+4N,0)=U(m,0)$):
    \begin{itemize}
            \item If $m=2(1+2q)N$, cells are maximally charged ($E_N(m)=N\omega_0/2$), most excited energy of $H_0$.
            \item If $m=4qN$, cells are maximally discharged ($E_N(m)=-N\omega_0/2$), ground state energy of $H_0$.
            \item Otherwise, $E_N(m)=0$.
        \end{itemize}
\end{itemize}
We can see that the energy injection profiles when OBC are considered, regardless of the parity of the number of spins, are equal to those when PBC are considered and $N$ is odd. This feature is also present in the $H^{xx}_1$ charger [\seclabel{sec:h1xx}]. As concluding remark, if we shift to zero the ground state energy of $H_0$, i.e., we perform the map $E_N(m)\mapsto E_N(m)+N\omega_0/2$, we recover again the energy values reported in the main text.

\subsection{Exact solution using momentum space and Cayley-Hamilton theorem}

Despite we have solved the exact charging dynamics of the KIC model (Eqs.~\eqref{eq:ref_ham1} and~\eqref{eq:ref_ham_zz1}, respectively) under Clifford quantum cellular automata, this method is only valid when the angle $\theta$ in~\eqlabel{eq:clifford} is a multiple of $\pi/4$ in absolute value, turning $e^{\pm i\theta P}$ into a Clifford gate. However, we might be interested on obtaining the exact dynamics for arbitrary coupling strengths. In the following lines, we derive the charging dynamics from a momentum space perspective, which can be used for arbitrary coupling strengths of the Ising and transverse field terms ($J$ and $b$, respectively). After obtaining the Floquet operator in momentum space, we make use of the Cayley-Hamilton theorem to derive its $m$th power, resulting operator after applying $m$ kicks. Notice that this method can only be applied when PBC are considered.

\subsubsection{\texorpdfstring{$H^{xx}_1$}{H1xx} charger case}

Our starting point follows~\seclabel{sec:h1xx}, where we consider~\eqlabel{eq:ref_ham1} as charger Hamiltonian $H^{xx}_1(t)$, with the same battery Hamiltonian $H_0=(\omega_0/2)\sum_{i=1}^N\sigma^z_i$ and initial state $\ket{\psi(0)} = \ket{1}^{\otimes N}$, ground state of $H_0$. Additionally, we define for convenience $b(t)\coloneqq b\sum_{t_i\in\mathbb{Z}}\delta(t-t_i)$, which encapsulates the periodically-kicked transverse field coupling strength. 

To begin with the exact diagonalization of this model, we apply the Jordan-Wigner transformation $\sigma^x_i = -(c_i + c^\dagger_i)\prod_{j<i} (1-2c^\dagger_j c_j)$ and $\sigma^z_i = 1-2c^\dagger_i c_i$~\cite{jordan1928uber}, with $c^\dagger_i$ ($c_i$) a spinless creation (annihilation) fermion at the $i$th site, following the anticommutation rules $\{c_i,c^\dagger_j\}=\delta_{ij}$ and $\{c_i,c_j\}=0$. Under the spinless fermion representation, the charger Hamiltonian reads as
\begin{equation}\label{eq:reduced_hamiltonian}
    H(t) = J\sum_{\braket{ij}} (c^\dagger_ic_j -c_ic_j +\hc) +b(t)\sum_{i=1}^N (1-2c^\dagger_ic_i),
\end{equation}
which resembles a Kitaev chain with a periodically-kicked chemical potential~\cite{kitaev2001unpaired}. This Hamiltonian can be split into its even and odd parity sectors~\cite{dziarmaga2005dynamics,damski2014exact}, $H^+$ and $H^-$ respectively, as $H=P^+H^+P^+ + P^-H^-P^-$, where $P^\pm\coloneqq (1\pm P)/2$ are the projectors on the subspaces with even and odd number of $c$-quasiparticles $\mathcal{N}=\sum_{i=1}^N c^\dagger_ic_i$, respectively. Here, $H^\pm$ are given by~\eqlabel{eq:reduced_hamiltonian} but following antiperiodic (periodic) boundary conditions $c_{N+1}=-c_1$ ($c_{N+1}=c_1$) for the even (odd) parity sector. If we apply the parity operator in our initial state, we obtain $P\ket{\psi(0)}=(-1)^N\ket{\psi(0)}$. So, depending on the parity of $N$, we should work only with the even parity sector ($N$ even) or the odd one ($N$ odd), thus consider either antiperiodic or periodic boundary conditions, respectively. Since we want to exactly diagonalize~\eqlabel{eq:ref_ham1} for arbitrary $N$, we start by solving the $N$ even case and later solve the $N$ odd case, whose procedure will be similar.

Next step is to diagonalize~\eqlabel{eq:reduced_hamiltonian} using the Fourier transform followed by the Bogoliubov transformation~\cite{lieb1961two,katsura1962statistical}, where we use as Fourier transform 
\begin{equation}\label{eq:fourier}
    c_l=\frac{e^{-i\pi/4}}{\sqrt{N}}\sum_k c_ke^{ik(la)}
\end{equation}
with $a$ the lattice spacing and taking as pseudomomenta
\begin{equation}\label{eq:pseudomomenta}
    k\in\left\{ \pm\frac{\pi}{Na}, \pm\frac{3\pi}{Na}, \dots, \pm\frac{(N-1)\pi}{Na} \right\},
\end{equation}
which is consistent with antiperiodic boundary conditions. The total Hilbert space is now a product of two-dimensional spaces spanned by the states $\ket{0_k,0_{-k}}$ and $\ket{1_k,1_{-k}}$ for each $k$ mode. After this transform, the Hamiltonian reads as
\begin{equation}
    H(t) = \sum_k \left\{ 2[J\cos(ka) - b(t)] c^\dagger_kc_k - J\sin(ka)(c^\dagger_kc^\dagger_{-k} + c_{-k}c_k) + b(t) \right\}.
\end{equation}

Diagonalization is completed by the Bogoliubov transformation $c_k=u_k(t)\gamma_k + v^*_{-k}(t)\gamma^\dagger_{-k}$ with $\gamma_k=u^*_kc_k + v_{-k}c^\dagger_{-k}$ a quasiparticle operator, where the Bogoliubov modes $[u_k(t),v_k(t)]$ must satisfy the Heisenberg equation $i\text{d}c_k/\text{d}t=[c_k,H(t)]$ with the constraint $\text{d}\gamma_k/\text{d}t=0$. Therefore, they are eigenstates of the dynamical Bogoliubov-de Gennes equations
\begin{equation}\label{eq:bog_degennes}
    i\frac{\text{d}}{\text{d}t}u_k = 2[J\cos(ka)-b(t)]u_k-2J\sin(ka)v_k, \qquad
    i\frac{\text{d}}{\text{d}t}v_k = -2[J\cos(ka)-b(t)]v_k-2J\sin(ka)u_k,
\end{equation}
with the normalization condition $|u_k|^2 + |v_k|^2 = 1$ $\forall t$. These equations can be written in a more convenient form by defining for the $k$th mode
\begin{equation}\label{eq:matricial_bdg}
    H_k(t)\coloneqq 2[J\cos(ka)-b(t)]\tau^z -2J\sin(ka)\tau^x= 2\begin{bmatrix}
        J\cos(ka)-b(t) & -J\sin(ka) \\
        -J\sin(ka) & b(t)-J\cos(ka)
    \end{bmatrix},
\end{equation}
with $\tau^{\{x,y,z\}}$ the Pauli matrices, and $\ket{\psi_k(t)}\coloneqq [u_k(t),v_k(t)]^\text{T}$, satisfying the time-dependent Schr\"odinger-like equation $i\text{d}\ket{\psi_k}/\text{d}t=H_k\ket{\psi_k}$. The Floquet operator after one kick is given by
\begin{equation}\label{eq:unitary_k}
\begin{split}
    U_k(1,0) &= \mathfrak{T}\exp\left[ -i\int_{0}^{1} \text{d}tH_k(t) \right] = \exp(2ib\tau^z)\exp(-2iJ[\cos(ka)\tau^z-\sin(ka)\tau^x]) \\
    &= \begin{bmatrix}
    e^{2ib}[\cos(2J) - i\sin(2J)\cos(ka)] & ie^{2ib}\sin(2J)\sin(ka) \\
    ie^{-2ib}\sin(2J)\sin(ka) & e^{-2ib}[\cos(2J) + i\sin(2J)\cos(ka)]
    \end{bmatrix}=\begin{bmatrix}
        \alpha_k & -\beta^*_k \\
        \beta_k & \alpha^*_k
    \end{bmatrix}.
\end{split}
\end{equation}

After $m$ kicks, the state of the $k$th mode will be given by $\ket{\psi_k(m)}=U^m_k(1,0)\ket{\psi_k(0)}$. Relying on the Cayley-Hamilton theorem, where every square matrix satisfies its own characteristic equation, we can compute the $m$th power of the Floquet operator by means of the Chebyshev polynomials of the second kind of degree $m$ in $x$~\cite{griffiths2001waves,das2024insights}, $\mathcal{U}_m(x)$, as
\begin{equation}
    U^m_k(1,0) = U_k(1,0)\mathcal{U}_{m-1}(\xi_k) - \mathcal{U}_{m-2}(\xi_k),\qquad \xi_k=\frac{1}{2}\trace{U_k(1,0)}=\text{Re}[\alpha_k].
\end{equation}
Therefore, the state after $m$ kicks can be expressed in terms of the initial state as
\begin{equation}
    \begin{bmatrix}
        u_k(m) \\
        v_k(m)
    \end{bmatrix}
    = \begin{bmatrix}
        \alpha_k\mathcal{U}_{m-1}(\xi_k)-\mathcal{U}_{m-2}(\xi_k) & -\beta_k^*\mathcal{U}_{m-1}(\xi_k) \\
        \beta_k\mathcal{U}_{m-1}(\xi_k) & \alpha^*_k\mathcal{U}_{m-1}(\xi_k)-\mathcal{U}_{m-2}(\xi_k)
    \end{bmatrix}
    \begin{bmatrix}
        u_k(0) \\
        v_k(0)
    \end{bmatrix},
\end{equation}
where the state of the whole system can be recovered as a product state $\ket{\psi(t)}=\prod_k \ket{\psi_k(t)}$. We take as initial state $\ket{\psi(0)} = \prod_k \ket{1}$, returning
\begin{align}\label{eq:uk}
    u_k(m) &= -\beta_k^*\mathcal{U}_{m-1}(\xi_k) = ie^{2ib}\sin(2J)\sin(ka)\frac{\sin m\theta_k}{\sin\theta_k}, \\ \label{eq:vk}
    v_k(m) &= \alpha^*_k\mathcal{U}_{m-1}(\xi_k)-\mathcal{U}_{m-2}(\xi_k) = e^{-2ib}[\cos(2J) + i\sin(2J)\cos(ka)]\frac{\sin m\theta_k}{\sin\theta_k} - \frac{\sin (m-1)\theta_k}{\sin\theta_k},
\end{align}
where we used that $\mathcal{U}_m(\cos\theta)=\sin(m+1)\theta/\sin\theta$ with
\begin{equation}
    \theta_k=\cos^{-1}\xi_k = \cos^{-1}[\cos(2b)\cos(2J) + \sin(2b)\sin(2J)\cos(ka)].
\end{equation}

The energy injected in the battery Hamiltonian $H_0$ after $m$ kicks can be computed as
\begin{equation}\label{eq:energy}
    E_N(m)=\frac{\omega_0}{2}\sum_k \braket{\psi_k(m)|\sigma^z|\psi_k(m)} = \frac{\omega_0}{2}\sum_k(2|u_k(m)|^2-1) = \omega_0\sin^2(2J)\sum_k \sin^2(ka)\frac{\sin^2 m\theta_k}{\sin^2 \theta_k} - \frac{N\omega_0}{2}.
\end{equation}
The even parity of $E_N(m)$ in terms of the pseudomomenta $k$ indicates that the pairs of modes $(k,-k)$ are equally charged along the time evolution. Setting $a=1$ for simplicity and working under the self-dual operator regime $|J|=|b|=\pi/4$, the energy injected simplifies to $E_N(m) = \omega_0\sum_k \sin^2 mk - N\omega_0/2$. In this form and following~\eqlabel{eq:pseudomomenta}, we can identify different regions of interest for the injected energies $E_N$ depending on the number of kicks. Therefore, for $N$ even and defining $q\in\mathbb{Z}$:
\begin{itemize}
    \item If $m=(q+1/2)N$ all modes are maximally charged, returning $E_N(m)=N\omega_0/2$, most excited energy of $H_0$.
    \item If $m=qN$ all modes are maximally discharged, returning $E_N(m)=-N\omega_0/2$, ground state energy of $H_0$.
    \item Otherwise, since $\sum_k \sin^2 mk = (N - \sum_k \cos2mk)/2 = N/2$ (where we use the Lagrange trigonometric identity $\sum_{i=1}^n \cos (2i -1)\alpha = \sin 2n \alpha/2 \sin \alpha$, which is zero for $\alpha=2q\pi/n$), $E_N(m)=0$.
\end{itemize}
In the limit of $N\to\infty$, we can turn the sum into an integral $\sum_k \mapsto (N/2\pi)\int^\pi_{-\pi} \text{d}k$. Under this limit, the energy injected becomes $\lim_{N\to\infty} E_N(m)/N = (\omega_0/2\pi)\int^\pi_{-\pi} \text{d}k\sin^2mk -\omega_0/2 = -\omega_0\sin2\pi m/4\pi m = 0$ since $m\in\mathbb{Z}$, result that is consistent with our findings.

For the case where $N$ is odd, the parity operator $P$ acting on the initial state returns an odd parity, since $P\ket{\psi(0)}=(-1)^N\ket{\psi(0)}$. Therefore, we can proceed with the same steps as for the $N$-even case but now considering periodic boundary conditions $c_{N+1}=c_1$ and taking as pseudomomenta
\begin{equation}\label{eq:pseudomomenta_odd}
    k\in\left\{ 0, \pm\frac{2\pi}{Na}, \pm\frac{4\pi}{Na}, \dots, \pm\frac{2(N-1)\pi}{Na} \right\},
\end{equation}
which is consistent with periodic boundary conditions.

After applying the Fourier transform [\eqlabel{eq:fourier}], our new Hamiltonian in the spinless fermion basis becomes
\begin{equation}
    H(t) = \sum_{k-\{0\}} \left\{ 2[J\cos(ka) - b(t)] c^\dagger_kc_k - J\sin(ka)(c^\dagger_kc^\dagger_{-k} + c_{-k}c_k) + b(t) \right\} + 2[J-b(t)](c^\dagger_0c_0 - c_0c^\dagger_0).
\end{equation}
While the sum over nonzero pseudomomenta returns the same dynamical Bogoliubov-de Gennes equations as~\eqlabel{eq:bog_degennes}, thus their solutions are equal as the $N$-even case, for the zero pseudomomenta-related term we obtain
\begin{equation}
    i\frac{\text{d}}{\text{d}t}u_0 = 2[J-b(t)]u_0, \qquad
    i\frac{\text{d}}{\text{d}t}v_0 = -2[J-b(t)]v_0,
\end{equation}
with $|u_0|^2 + |v_0|^2=1$ $\forall t$. This set of equations that can be rewritten as $H_0(t)\coloneqq 2[J-b(t)]\tau^z$, whose corresponding Floquet operator after one kick returns
\begin{equation}
    U_0(1,0) = \mathfrak{T}\exp\left[ -i\int_{0}^{1} \text{d}t H_0(t) \right] = \exp(-2i[J-b]\tau^z)
    = \begin{bmatrix}
    e^{-2i(J-b)} & 0 \\
    0 & e^{2i(J-b)}
    \end{bmatrix}=\begin{bmatrix}
        \alpha_0 & -\beta^*_0 \\
        \beta_0 & \alpha^*_0
    \end{bmatrix}.
\end{equation}

Following Eqs.~\eqref{eq:uk} and~\eqref{eq:vk}, after $m$ kicks we obtain
\begin{align}
    u_0(m) &= -\beta_0^*\mathcal{U}_{m-1}(\xi_0) = 0, \\
    v_0(m) &= \alpha^*_0\mathcal{U}_{m-1}(\xi_0)-\mathcal{U}_{m-2}(\xi_0) = e^{2i(J-b)}\frac{\sin 2m(J-b)}{\sin 2(J-b)} - \frac{\sin 2(m-1)(J-b)}{\sin 2(J-b)}=1,
\end{align}
with $\xi_0=\text{Re}[\alpha_0]=\cos 2(J-b)$. Since $u_0(m)=0$ $\forall m$, following~\eqlabel{eq:energy}, the contribution to the injected energy of $k=0$ is given by 
\begin{equation}
    \frac{\omega_0}{2}\braket{\psi_0(m)|\sigma^z|\psi_0(m)}=\frac{\omega_0}{2}(2|u_0(m)|^2-1)=-\frac{\omega_0}{2}.
\end{equation} 
Therefore, summing over all the pseudomomenta, for $N$ odd the injected energy reads as~\eqlabel{eq:energy} where again, under the self-dual operator regime $|J|=|b|=\pi/4$ and setting $a=1$ for simplicity, we recover the same expression $E_N(m)=\omega_0\sum_k \sin^2mk-N\omega_0/2$ but now with a sum running over the pseudomomenta given by~\eqlabel{eq:pseudomomenta_odd}. For $N$ odd and using $q\in\mathbb{Z}$, the charging dynamics obey:
\begin{itemize}
    \item If $m=qN$ all modes are maximally discharged, returning $E_N(m)=-N\omega_0/2$, ground state energy of $H_0$.
    \item Otherwise, since $\sum_k \sin^2 mk = (N - \sum_k \cos2mk)/2 = N/2$ (where we use the Lagrange trigonometric identity $\sum_{i=1}^n \cos 2i\alpha = \sin n \alpha\cos(n-1)\alpha/\sin \alpha$, which is zero for $\alpha=2q\pi/n$), $E_N(m)=0$.
\end{itemize}

As expected, for both even and odd $N$ the results obtained are in agreement for those obtained under CQCA perspective when PBC are considered. We can also see that for both scenarios the periodicity of the injected energy curves under this model is $U(t,0)=U(t+N,0)$, where after $m=qN$ kicks the energy injected drops to the ground state energy and the pattern starts again.

Finally, it is possible to still use the same framework when kicks are not applied uniformly. In this case, the Floquet operator changes per kick, thus the unitary operator after $m$ kicks does not reduce to the $m$th power of the Floquet operator of a single kick. If we parametrize an arbitrary unitary matrix like~\eqlabel{eq:unitary_k} as
\begin{equation}
U(\alpha,\beta)\coloneqq
\begin{bmatrix}
\alpha & -\beta^*\\
\beta & \alpha^*
\end{bmatrix},
\qquad
\text{with }|\alpha|^2+|\beta|^2=1,
\end{equation}
we can compute the ordered product after $m$ kicks as $U_{\text{tot}}\coloneqq\prod_{k=1}^m U(\alpha_k,\beta_k)$. Let $z\coloneqq\beta/\alpha$ and $q\coloneqq\alpha^*/\alpha$, with the pair $(z,q)$ carrying the degrees of freedom that parametrizes SU$(2)$. For two factors, we can observe that $U(\alpha_{12}, \beta_{12}) = U(\alpha_1,\beta_1)U(\alpha_2,\beta_2)$, with $\alpha_{12}\coloneqq\alpha_1\alpha_2-\beta_1^*\beta_2$ and $\beta_{12}\coloneqq\beta_1\alpha_2+\alpha_1^*\beta_2$. Writing $\beta_i\coloneqq z_i\alpha_i$ and $\alpha_i^*\coloneqq q_i\alpha_i$ yields
\begin{equation}
\alpha_{12}=\alpha_1\alpha_2(1-q_1 z_1^*z_2),
\quad
z_{12}=\frac{z_1+q_1 z_2}{1-q_1 z_1^*z_2},
\quad
q_{12}=\frac{\alpha_{12}^*}{\alpha_{12}}.
\end{equation}
This identity leads to a linear-time algorithm for $m$ factors, which is a M\"obius-transformation representation of SU$(2)$~\cite{beardon1983mobius}. If we initialize
$\alpha=\alpha_1$, $z=\beta_1/\alpha_1$, and $q=\alpha^*/\alpha$, for
$k=2,\ldots,N$ we update
\begin{equation}
\alpha\leftarrow \alpha\alpha_k (1-qz^*z_k),\quad
z\leftarrow \frac{z+qz_k}{1-qz^*z_k},\quad
q\leftarrow \frac{\alpha^*}{\alpha}.
\end{equation}
After the final step $\beta=\alpha z$, and $U_{\text{tot}}=U(\alpha,\beta)$. 

As concluding remarks, in general for open boundary conditions (OBC) it is not possible to exact diagonalize~\eqlabel{eq:ref_ham1}, where the final Hamiltonian in the spinless fermionic basis has the form $H=\sum_{ij} [ c_i^\dagger A_{ij} c_j + \tfrac{1}{2}(c_i^\dagger B_{ij} c_j^\dagger + \hc) ]$, with $A_{ij}$ and $B_{ij}$ the hopping and pairing matrices, respectively~\cite{lieb1961two}. This method requires the exact diagonalization of a $2N\times 2N$ matrix, which might not be computationally possible for large systems. However, we verified via tensor network simulations that, regardless the parity on the number of spins, the behavior of the energy injected for OBC evolves as the $N$-odd case for PBC. 

\subsubsection{\texorpdfstring{$H^{zz}_1$}{H1zz} charger case}

For this case, now our starting point follows~\seclabel{sec:h1zz}, where we consider~\eqlabel{eq:ref_ham_zz1} as charger Hamiltonian $H^{zz}_1(t)$, using the same battery Hamiltonian $H_0=(\omega_0/2)\sum_{i=1}^N\sigma^y_i$ and initial state $\ket{\psi(0)}=\ket{-i}^{\otimes N}$, ground state of $H_0$. Before starting with the diagonalization, first notice that this model is equivalent to~\eqlabel{eq:ref_ham1} up to a global $\frac{\pi}{2}$-rotation around the Y-axis. Let $\text{RY}(\frac{\pi}{2})\coloneqq \exp(-i\frac{\pi}{4}\sigma^y)$ and $R\coloneqq\prod_{j=1}^N \text{RY}_j(\frac{\pi}{2})$, with $j$ the qubit index. Since $\text{RY}(\frac{\pi}{2})\sigma^{\{x,z\}}\text{RY}^\dagger(\frac{\pi}{2})=\mp\sigma^{\{z,x\}}$ [\eqlabel{eq:clifford}], the rotated Hamiltonian reads as $H_1^{xx'}(t)=H^{xx'}_I + H^{xx'}_K(t)$ with
\begin{equation}
    H^{xx'}_I = J\sum_{\braket{ij}}\sigma^x_i\sigma^x_j,\qquad H^{xx'}_K(t) = -b(t)\sum_{i=1}^N\sigma^z_i,
\end{equation}
which only differs by a sign-flip from the kicked-transverse field contribution of~\eqlabel{eq:ref_ham1}. After applying the Jordan-Wigner transformation, the charger Hamiltonian in the spinless fermion representation now reads as 
\begin{equation}\label{eq:reduced_hamiltonian_xx}
    H(t) = J\sum_{\braket{ij}} (c^\dagger_ic_j -c_ic_j +\hc) -b(t)\sum_{i=1}^N (1-2c^\dagger_ic_i).
\end{equation}

In this case, if we apply the parity operator over the initial state we obtain that $P\ket{\psi(0)}=P\ket{-i}^{\otimes N} = \ket{i}^{\otimes N}$, thus the initial state is not an eigenstate of $P$. In such cases, we need to split the state into its corresponding even and odd sectors and solve each of them independently. We can do that by defining $\ket{\psi_{\{e,o\}}(0)}\coloneqq (\ket{-i}^{\otimes N} \pm \ket{i}^{\otimes N})/2$, such that $\ket{\psi(0)} = \ket{\psi_e(0)} + \ket{\psi_o(0)}$, as the even and odd initial state relatives, respectively. To prove that, let $w(s)$ be the Hamming weight of the bitstring $s$, integer value that accounts the number of ones in $s$. See that
\begin{equation}
    \ket{\pm i}^{\otimes N} = \left[ \frac{\ket{0} \pm i\ket{1}}{\sqrt{2}} \right]^{\otimes N} = \frac{1}{2^{N/2}}\sum_{s\in \{0,1\}^N} (\pm i)^{w(s)} \ket{s}.
\end{equation}
Therefore,
\begin{align}
    \ket{\psi_e(0)} &= \frac{1}{2^{N/2}} \sum_{s\in \{0,1\}^N} \frac{i^{w(s)} + (-i)^{w(s)}}{2} \ket{s} = \frac{1}{2^{N/2}} \sum_{\substack{s\in \{0,1\}^N\\w(s)\text{ even}}} (-1)^{w(s)/2} \ket{s},  \\
    \ket{\psi_o(0)} &= -\frac{1}{2^{N/2}} \sum_{s\in \{0,1\}^N} \frac{i^{w(s)} - (-i)^{w(s)}}{2} \ket{s} = -\frac{i}{2^{N/2}} \sum_{\substack{s\in \{0,1\}^N\\w(s)\text{ odd}}} (-1)^{(w(s)-1)/2} \ket{s}.
\end{align}
From both expressions, we can trivially see that $P\ket{\psi_{\{e,o\}}(0)} = \pm \ket{\psi_{\{e,o\}}(0)}$ regardless the parity of $N$. However, both states do not feature pairwise correlations over $(k,-k)$ fermion modes, thus nontrivial entanglement is present across multiple $k$-modes, containing higher-order products of the creation and annihilation operators. This fact leads to a lack of fermionic parity conservation within the Bogoliubov basis. Therefore, despite it is possible to split~\eqlabel{eq:reduced_hamiltonian_xx} as a sum of $2\times 2$ matrices (as we did in the previous section), the initial state cannot be written as a product of initial states at each mode, requiring a real space treatment. Nonetheless, the charging dynamics for this case under the self-dual operator regime were already obtained under the CQCA perspective in~\seclabel{sec:h1zz}. Nonetheless, we derive the remaining modes for completeness.

We start by considering the $N$ even case. While it was already derived for the even parity sector in the previous subsection, now we also need to obtain the exact same results but for the odd parity sector, thus we have to consider periodic boundary conditions $c_{N+1}=c_1$ in~\eqlabel{eq:reduced_hamiltonian_xx} and now take as pseudomomenta
\begin{equation}\label{eq:pseudomomenta_odd_h1zz}
    k\in\left\{ 0, \pm\frac{2\pi}{Na}, \pm\frac{4\pi}{Na}, \dots, \pm\frac{2(N-1)\pi}{Na}, \frac{\pi}{a} \right\},
\end{equation}
which is consistent with periodic boundary conditions.

After applying the Fourier transform [\eqlabel{eq:fourier}], our new Hamiltonian in the spinless fermion basis becomes
\begin{equation}
    \begin{split}
        H(t) &= \sum_{k-\{0, \pi/a\}} \left\{ 2[J\cos(ka) + b(t)] c^\dagger_kc_k - J\sin(ka)(c^\dagger_kc^\dagger_{-k} + c_{-k}c_k) - b(t) \right\} \\
        &+ 2[J+b(t)](c^\dagger_0c_0 - c_0c^\dagger_0) + 2[J-b(t)](c^\dagger_{\pi/a}c_{\pi/a} - c_{\pi/a}c^\dagger_{\pi/a}).
    \end{split}
\end{equation}
While the sum over $k\notin\{0,\pi/a\}$ returns similar dynamical Bogoliubov-de Gennes equations as~\eqlabel{eq:bog_degennes}, thus their solutions are similar to the $N$-even case. For the pseudomomenta-related terms with $k\in\{0,\pi/a\}$ we obtain
\begin{equation}
    i\frac{\text{d}}{\text{d}t}u_{\{0,\pi/a\}} = 2[J\pm b(t)]u_{\{0,\pi/a\}}, \qquad
    i\frac{\text{d}}{\text{d}t}v_{\{0,\pi/a\}} = -2[J\pm b(t)]v_{\{0,\pi/a\}},
\end{equation}
with $|u_{\{0,\pi/a\}}|^2 + |v_{\{0,\pi/a\}}|^2=1$ $\forall t$, set of equations that can be rewritten as $H_{\{0,\pi/a\}}(t)\coloneqq 2[J\pm b(t)]\tau^z$, whose corresponding Floquet operator after one kick returns
\begin{equation}
    \begin{split}
    U_{\{0,\pi/a\}}(1,0) &= \mathfrak{T}\exp\left[ -i\int_{0}^{1} \text{d}t H_{\{0,\pi/a\}}(t) \right] = \exp(-2i[J\pm b]\tau^z) \\
    &= \begin{bmatrix}
    e^{-2i(J\pm b)} & 0 \\
    0 & e^{2i(J\pm b)}
    \end{bmatrix}=\begin{bmatrix}
        \alpha_{\{0,\pi/a\}} & -\beta^*_{\{0,\pi/a\}} \\
        \beta_{\{0,\pi/a\}} & \alpha^*_{\{0,\pi/a\}}
    \end{bmatrix}.
\end{split}
\end{equation}

Following Eqs.~\eqref{eq:uk} and~\eqref{eq:vk}, after $m$ kicks we obtain
\begin{align}
    u_{\{0,\pi/a\}}(m) &= -\beta_{\{0,\pi/a\}}^*\mathcal{U}_{m-1}(\xi_{\{0,\pi/a\}}) = 0, \\
    v_{\{0,\pi/a\}}(m) &= \alpha^*_{\{0,\pi/a\}}\mathcal{U}_{m-1}(\xi_{\{0,\pi/a\}})-\mathcal{U}_{m-2}(\xi_{\{0,\pi/a\}}) = e^{2i(J\pm b)}\frac{\sin 2m(J\pm b)}{\sin 2(J\pm b)} - \frac{\sin 2(m-1)(J\pm b)}{\sin 2(J\pm b)}=1,
\end{align}
with $\xi_{\{0,\pi/a\}}=\text{Re}[\alpha_{\{0,\pi/a\}}]=\cos 2(J\pm b)$. 

Again, for open boundary conditions (OBC) it is not possible to exact diagonalize~\eqlabel{eq:ref_ham_zz1}, where the final Hamiltonian in the spinless fermionic basis has the form $H=\sum_{ij} [ c_i^\dagger A_{ij} c_j + \tfrac{1}{2}(c_i^\dagger B_{ij} c_j^\dagger + \hc) ]$, with $A_{ij}$ and $B_{ij}$ the hopping and pairing matrices, respectively, which requires the exact diagonalization of a $2N\times 2N$ matrix, which might not be computationally possible for large systems~\cite{lieb1961two}. Similarly to~\eqlabel{eq:ref_ham1}, this model under the OBC case behaves as the PBC-case for odd $N$ regardless the parity of the number of spins, where we verified our findings using tensor network simulations.

\subsubsection{Final remarks: exact diagonalization of the transverse-field Ising model}

To compare our results with the performance of regular spin chains~\cite{le2018spinchain}, as in Fig. 4 in the main text, now we diagonalize the transverse-field Ising model (TFIM) as charger Hamiltonian, which reads as
\begin{equation}
    H^\text{TFIM}_1 = J\sum_{\braket{ij}} \sigma^x_i\sigma^x_j + b\sum_{i=1}^N \sigma^z_i,
\end{equation}
model that is equivalent to~\eqlabel{eq:ref_ham1} where now the transverse-field contribution is applied continuously. Its exact diagonalization and posterior analytical time-evolution will be helpful to compute the energy injected up to time $t$, saturation value that is reached after applying enough kicks when the KIC model is considered within a time window [Eq.~(5) in the main text]. For simplicity and without loss of generality, we consider PBC and $N$ even as well as $H_0=(\omega_0/2)\sum_{i=1}^N \sigma^z_i$ as battery Hamiltonian. Under these conditions, we can proceed as in Eqs.~\eqref{eq:reduced_hamiltonian}-\eqref{eq:matricial_bdg} but performing $b(t)\mapsto b$, where we have as reduced time-independent Hamiltonian for the $k$th mode
\begin{equation}
    H^\text{TFIM}_k\coloneqq 2[J\cos(ka)-b]\tau^z -2J\sin(ka)\tau^x= 2\begin{bmatrix}
        J\cos(ka)-b & -J\sin(ka) \\
        -J\sin(ka) & b-J\cos(ka)
    \end{bmatrix}.
\end{equation}
Solving for its corresponding time-dependent Schr\"odinger-like equation, $i\text{d}\ket{\psi_k}/\text{d}t=H^\text{TFIM}_k\ket{\psi_k}$, we obtain
\begin{equation}
    \begin{split}
        U_k(t,0) &= \mathfrak{T}\exp\left[ -i\int_{0}^{t} \text{d}t'H^\text{TFIM}_k \right] = \exp(-2it[(J\cos(ka)-b)\tau^z-J\sin(ka)\tau^x]) \\
        &= \frac{1}{\Delta_k} \begin{bmatrix}
            \Delta_k\cos(2t\Delta_k) - i \sin(2t\Delta_k) (J\cos ka - b)
            &
            - i \sin(2t\Delta_k) J \sin ka \\
            - i \sin(2t\Delta_k) J \sin ka
            &
            \Delta_k\cos(2t\Delta_k) + i \sin(2t\Delta_k) (J\cos ka - b)
        \end{bmatrix} \\
        &=\begin{bmatrix}
            \alpha_k(t) & -\beta^*_k(t) \\
            \beta_k(t) & \alpha^*_k(t)
        \end{bmatrix},
    \end{split}
\end{equation}
with $\Delta_k = \sqrt{J^2-2Jb\cos ka + b^2}$. As the state of the whole system can be recovered as a product state $\ket{\psi(t)}=\prod_k \ket{\psi_k(t)} = \prod_k U_k(t,0)\ket{\psi_k(0)}$. We take as initial state $\ket{\psi(0)} = \prod_k \ket{1}$, returning
\begin{align}
    u_k(t) &= -\beta_k^*(t) = - \frac{i}{\Delta_k} \sin(2t\Delta_k) J \sin ka, \\
    v_k(t) &= \alpha^*_k(t) = \cos(2t\Delta_k) + \frac{i}{\Delta_k} \sin(2t\Delta_k) (J\cos ka - b).
\end{align}

As we derived in~\eqlabel{eq:energy}, the energy injected in the battery Hamiltonian $H_0$ can be computed as
\begin{equation}
    E_N(t)=\frac{\omega_0}{2}\sum_k \braket{\psi_k(t)|\sigma^z|\psi_k(t)} = \frac{\omega_0}{2}\sum_k(2|u_k(t)|^2-1) = \omega_0J^2\sum_k \frac{1}{\Delta_k^2} \sin^2(2t\Delta_k) \sin^2 ka - \frac{N\omega_0}{2}.
\end{equation}

Setting $a=1$ for simplicity and working under the self-dual operator regime with $|J|=|b|=\pi/4$, finally the injected energy simplifies to $E_N(t) = \omega_0\sum_k \sin^2(\pi t\cos k/2)\sin^2 k/2 - N\omega_0/2$, where we use that $\cos^2x = (1+\cos 2x)/2$ and $\sin 2x = 2\sin x\cos x$. In the limit of $N\to\infty$, we can turn the sum into an integral $\sum_k \mapsto (N/2\pi)\int^\pi_{-\pi} \text{d}k$. Under this limit, the energy injected becomes $\lim_{N\to\infty} E_N(t)/N = (\omega_0/2\pi) \int^\pi_{-\pi} \text{d}k \sin^2(\pi t\cos k/2)\sin^2 k/2 - \omega_0/2 = (2\omega_0/\pi) \int^{\pi/2}_{0} \text{d}\theta \sin^2(\pi t\cos\theta)\sin^2\theta - \omega_0/2$, where for the last equality we substitute $\theta = k/2$ and use the even parity of the integrand. To the best of our knowledge, there is no known closed form of the resulting integral. However, efficient numerical integration methods can be used, as the Clenshaw-Curtis quadrature~\cite{clenshaw1960method}. As a concluding note, both expressions can be helpful to set an appropriate time window for the KIC QB model with non-uniform kick schedulings by looking at which time $t$ the injected energy is maximized, energy value that can be approached after applying enough kicks, as Fig. 4 in the main text shows.

\section{Performance under different sources of imperfections}

\subsection{Impact of disorder in the parameters}

Deviations in the parameters naturally arise in realistic experimental scenarios and may degrade performance. Following Ref.~\cite{romero2024optimizing}, we can set 
\begin{equation}\label{eq:dis_ham1}
    H^{xx}_I(\sigma_J) = J\sum_{\braket{ij}}(1+\delta J_i)\sigma^x_i\sigma^x_j,\qquad H^{xx}_K(t;\sigma_b) = b\sum_{j=1}^N(1+\delta b_j)\sigma^z_j\sum_{t_i\in\mathbb{Z}}\delta(t-t_i).
\end{equation}
where we have included disorder as $J_{ij}=J(1+\delta J_i)$ and $b_i=b(1+\delta b_i)$ in~\eqlabel{eq:ref_ham1} with randomly sampled $\delta J_i \in [-\sigma_J, \sigma_J]$ and $\delta b_i \in [-\sigma_b, \sigma_b]$ $\forall i$ with $\sigma_{\{J,b\}}$ an upper bound of the disorder introduced. Applying disorder on $J_{ij}$ while remaining $b$ untouched is equivalent to conduct the opposite thing. Therefore, we include disorder on $J_{ij}$ and set $b=-\pi/4$.  
The results for $N=20$ are shown in~\figlabel{fig:disorder} for PBC after averaging $n_d=100$ disorder realizations using $H^{xx}_1$ as charger. As a result, our protocol demonstrates remarkable robustness to disorder under the self-dual operator regime. For moderate and intermediate disorder ($\sigma_J\le 0.2$), the charging performance remains similar to the disorder-free case. For stronger disorder, the normalized mean injected energy accumulates around $E_N(\tau)/N\sim 0.5$, independent of the number of kicks. Similar trends are found for different system sizes, and boundary conditions when using the $H_1^{zz}$ charger, according to their corresponding energy injection profiles. 
\begin{figure}[!b]
    \centering
    \includegraphics[width=.55\linewidth]{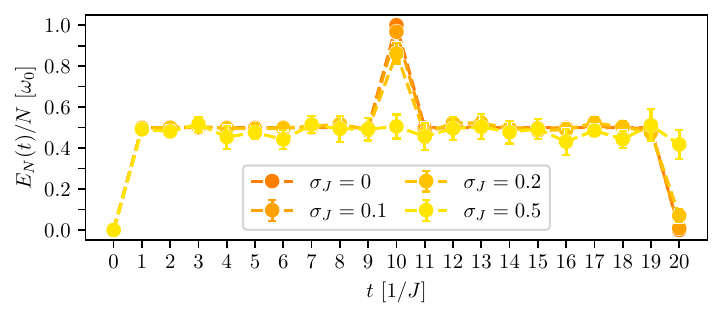}\vspace{-2mm}%
    \caption{Charging performance including disorder. Using $H_1^{xx}$ [\eqlabel{eq:dis_ham1}], normalized injected energies with $N=20$ for PBC using as disorder ratios $\sigma_J$ zero (no disorder), $0.1$, $0.2$ and $0.5$ (darker to lighter colors).}\label{fig:disorder}%
\end{figure}%

\subsection{Long-range interactions}

Throughout our discussion in the main text, we do not consider any contribution beyond nearest-neighbor interactions in the charger Hamiltonians, contributions that are naturally present in several quantum platforms (e.g., neutral atoms on optical lattices), which might have an impact in the performance of our QB. To study the influence of these contributions in our models, we can consider~\eqlabel{eq:ref_ham_zz1} for the charger Hamiltonian $H^{zz}_1$ but now including long-range power-law-decaying interactions (parametrized by $\alpha>0$) for the Ising contribution. The new charger Hamiltonian reads as
\begin{equation}\label{eq:lr_kic}
    H^{zz}_1(\alpha) = J\sum_{i<j} \frac{1}{|i-j|^\alpha} \sigma^z_i\sigma^z_j + b(t)\sum_{i=1}^N \sigma^x_i,
\end{equation}
where we recover the KIC of~\eqlabel{eq:ref_ham_zz1} when $\alpha\to\infty$.

Under the self-dual operator regime with $J=\pi/4$ and $b=-\pi/4$~\cite{Akila_2016,bertini2019entanglement} and setting OBC, in~\figlabel{fig:longrange}(a) we analyze the charging dynamics of a KIC of $N=20$ quantum cells with OBC using~\eqlabel{eq:lr_kic} as charger Hamiltonian and compare its performance with $\alpha\in\{1,3,6\}$, where the latter two are typical values for neutral-atom platforms, against the expected performance for a regular KIC ($\alpha\to\infty$), thus without long-range interactions. We use $H_0=(\omega_0/2)\sum_{i=1}^N \sigma^y_i$ as battery Hamiltonian, after shifting to zero its lowest energy, and its ground state as initial state. For completeness, we compute the power-lay decaying amplitudes with respect $\alpha$, showcasing how fast long-range interactions are suppressed across a KIC chain. While for $\alpha=6$ it is still possible to keep the desired regular KIC profile, strengthening the robustness and versatility of our QB proposal, for $\alpha\in\{1,3\}$ the normalized injected energies accumulate around $E_N/N\sim 0.5$ regardless the number of kicks applied, which reminds to the expected performance under the presence of a strong source of disorder [\figlabel{fig:disorder}]. Moreover, the stability of our protocol is also shown again. Similar results are obtained under different boundary conditions and using $H^{xx}_1$ as charger Hamiltonian [\eqlabel{eq:ref_ham1}] with power-law decaying long-range interactions as in~\eqlabel{eq:lr_kic}.
\begin{figure}[!tb]
    \centering
    \includegraphics[width=\linewidth]{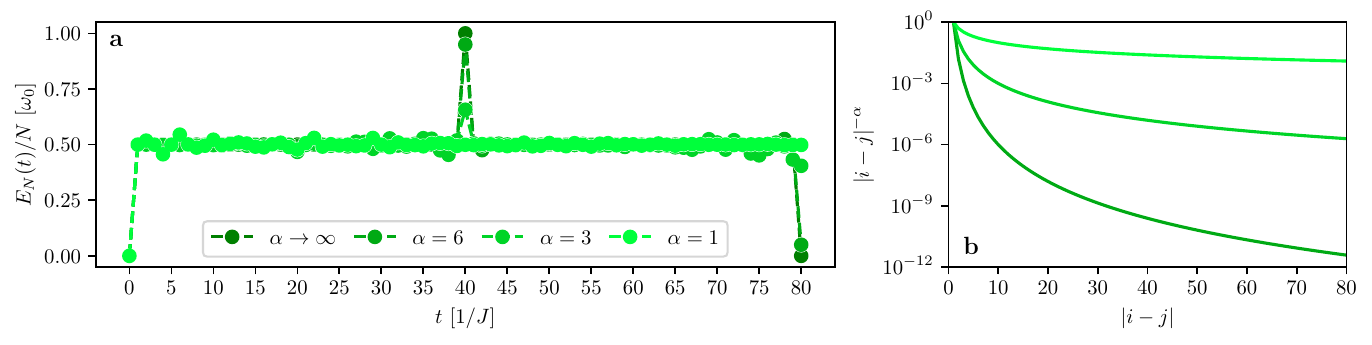}\vspace{-1mm}%
    \caption{Normalized injected energies for a KIC with long-range interactions with $N=20$. (a) For different power-law decays $\alpha$ (where the darker the color, the larger its value), we apply up to $m=4N=80$ kicks using $H^{zz}_1(\alpha)$ as charger Hamiltonian [\eqlabel{eq:lr_kic}] under OBC. In the limit of $\alpha\to\infty$, long-range interactions are completely suppressed so we recover the regular KIC. (b) Relative amplitudes of the long-range interactions using the same values of $\alpha$.}\label{fig:longrange}%
\end{figure}

\subsection{Kick duration: impact of quasikicks}

Another feature that might compromise the performance of our QB is the time interval between switching on and off the kicks. While KIC model requires the implementation of transverse fields during an infinitesimal amount of time, which is far from being practical, it is possible to keep the KIC nature when these fields are applied along a narrow time window, which we define as a \emph{quasikick}. Without loss of generality, we can consider a Blackman window~\cite{blackman1958measurement} of the type
\begin{equation}\label{eq:blackman}
    \Phi(t) = \frac{21}{50} + \frac{1}{2}\cos\frac{2\pi t}{\delta t} + \frac{2}{25}\cos\frac{4\pi t}{\delta t},
\end{equation}
with $\delta t$ its corresponding width, whose area under the curve returns $\mathcal{A}=\int_{-\delta t/2}^{\delta t/2}\text{d}t\Phi(t)=21\delta t/50$, which will be used to normalize the quasikick. With that, we introduce the quasikicked Ising chain (QKIC) model as
\begin{equation}\label{eq:quasi}
    H_Q(t) = H_I + H_K\sum_{m\in\mathbb{N}}\phi(t-m),
\end{equation}
with $\phi(t) \coloneqq \Phi(t)/\mathcal{A}$ $\forall t\in[-\delta t/2, \delta t/2]$ and zero otherwise. Defined in this manner, $\phi(t)$ follows a function of the form $f_\alpha(x) \coloneqq \alpha^{-1}f(x/\alpha)$, with $\alpha>0$ and $\int_\mathbb{R} \text{d}x f_\alpha(x)=1$. Under these conditions, it can be proven that $f_\alpha(x)$ converges to the Dirac delta function when $\alpha\to 0$, i.e., $\lim_{\alpha\to 0}f_\alpha(x)=\delta(x)$~\cite{kanwal1998additional}.

Using the QKIC model, the unitary time-evolved operator between quasikicks is given by
\begin{equation}
    U\!\left(1+\frac{\delta t}{2}; \frac{\delta t}{2}\right) = U_Q = \mathfrak{T}\exp\!\left[ -i\int_{\delta t/2}^{1 + \delta t/2} \text{d}t H_Q(t) \right] = \mathfrak{T}\exp\!\left[ -i\int_{1-\delta t/2}^{1 + \delta t/2} \text{d}t \left( H_I + H_K\phi(t-1) \right) \right] e^{-i(1-\delta t)H_I},
\end{equation}
which converges to the KIC Floquet operator when $\delta t\to 0$, as expected. Moreover, the time-evolved unitary operator after $m$ kicks is also $U_Q^m$. As a side note, both results hold true for any normalized function that is symmetrical around the center with a $\delta t$ width.
\begin{figure}[!tb]
    \centering
    \includegraphics[width=\linewidth]{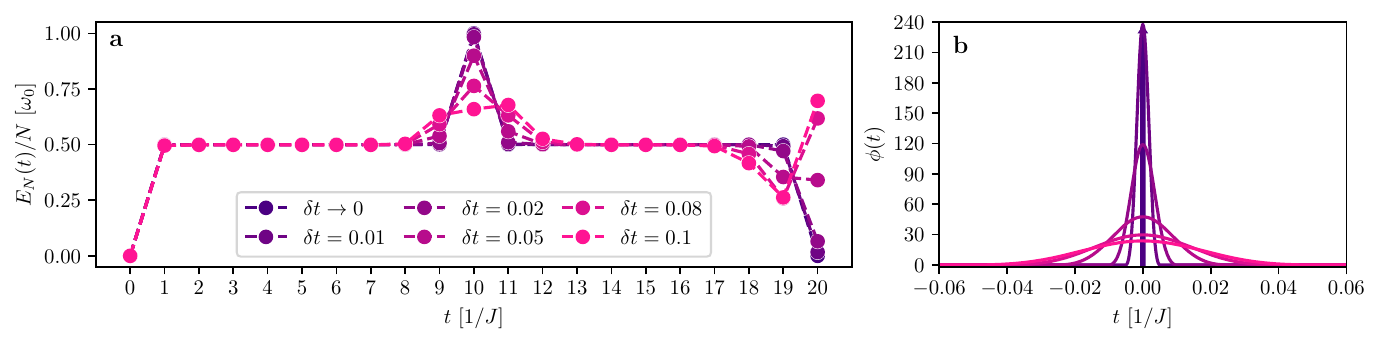}\vspace{-1mm}%
    \caption{Normalized injected energies for a QKIC using different widths with $N=20$. (a) For different widths $\delta t$ (where the lighter the color, the larger its value), we apply up to $m=N=20$ kicks using $H^{xx}_1$ as charger Hamiltonian [\eqlabel{eq:ref_ham1}] under PBC. In the limit of $\delta t\to 0$, we recover the regular KIC. (b) Amplitudes $\phi(t)$ of the quasikicks for different $\delta t$ values. The Dirac delta function ($\delta t\to0$ case) is represented by a vertical arrow.}\label{fig:quasikick}%
\end{figure}

Considering $H^{xx}_1$ [\eqlabel{eq:ref_ham1}] as charger Hamiltonian under PBC, $N=20$ quantum cells and setting again $J=\pi/4$ and $b=-\pi/4$, in~\figlabel{fig:quasikick}(a) we compare the charging dynamics of the regular KIC QB ($\delta t\to0)$ versus the QKIC model of~\eqlabel{eq:quasi} setting different $\delta t$ values. We use as battery Hamiltonian $H_0=(\omega_0/2)\sum_{i=1}^N \sigma^z_i$ after shifting to zero its lowest energy and its corresponding ground state as initial state. While for moderate values of the quasikick width (up to $\delta t\sim0.08$) it is possible to capture the regular KIC nature, for higher values no, where the injected energies accumulate around $E_N/N\sim0.5$, as for very disordered systems [\figlabel{fig:disorder}]. The amplitudes $\phi(t)$ of the quasikicks considered are plotted in~\figlabel{fig:quasikick}(b) [\eqlabel{eq:blackman}]. Again, similar results are obtained under different boundary conditions and using $H^{zz}_1$ as charger Hamiltonian [\eqlabel{eq:ref_ham_zz1}].

\subsection{Effect of slower quenches}

Finally, our last study focuses on the impact of how fast the charger Hamiltonian is introduced within our charging protocol. The system Hamiltonian is described by $H(t) = H_0 + \lambda(t)[H_1(t) - H_0]$, with $\lambda(t)$ an scheduling function that switches on and off the battery and charging Hamiltonians. For simplicity, in the main text we use a step function, which toggles both Hamiltonians infinitely fast, enabling a neater analytical treatment of our models. However, sudden quenches might be unfeasible in realistic setups, where such nonzero time interval might affect performance. With that in mind and without loss of generality, we propose a polynomial scheduling function of the type
\begin{equation}\label{eq:lambda}
    \lambda(t) = \begin{dcases}
        0 & \text{if }t\le -\delta t \\
        3\left(\frac{t + \delta t}{\delta t}\right)^2 - 2\left(\frac{t + \delta t}{\delta t}\right)^3 & \text{if }-\delta t<t<0 \\
        1 & \text{otherwise}
    \end{dcases},
\end{equation}
whose time interval is monitored by the value $\delta t$. If we consider now a kicking scheduling at natural times, i.e., $b(t)=\sum_{t_i\in\mathbb{N}}\delta(t-t_i)$, the corresponding time-evolved unitary operator after one kick reads as
\begin{equation}
    U(1,-\delta t) = \mathfrak{T}\exp\left[-i\int_{-\delta t}^1 \text{d}t H(t)\right] = e^{-iH_K}e^{-iH_I} \mathfrak{T}\exp\left[-i\int_{-\delta t}^0 \text{d}t \left[H_0 + \lambda(t)(H_I - H_0)\right]\right] = U_KU_IU_\lambda(\delta t),
\end{equation}
with $U_\lambda(\delta t)\coloneqq \mathfrak{T}\exp\big[-i\int_{-\delta t}^0 \text{d}t \left[H_0 + \lambda(t)(H_I - H_0)\right]\big]$ the unitary time-evolved operator capturing the initial quench, which converges to the identity when $\delta t\to 0$, as expected. After $m$ kicks, the time evolution would be driven by $U(m,-\delta t) = (U_KU_I)^mU_\lambda(\delta t)$.

Considering $H^{xx}_1$ [\eqlabel{eq:ref_ham1}] as charger Hamiltonian under PBC, $N=20$ quantum cells and setting again $J=\pi/4$ and $b=-\pi/4$, in~\figlabel{fig:quasistep}(a) we compare the charging dynamics of the regular KIC QB ($\delta t\to0)$ against different $\delta t$ values of the polynomial scheduling function $\lambda(t)$ in~\eqlabel{eq:lambda}. We use again as battery Hamiltonian $H_0=(\omega_0/2)\sum_{i=1}^N \sigma^z_i$ after shifting to zero its lowest energy and its corresponding ground state as initial state. From the results drawn, we can conclude that our model is remarkably robust against slow quenches, where even for notably large time intervals ($\delta t> 0.5$) it is still possible to capture the regular KIC nature, despite the maximal energy injected peaks at slightly lower values with increasing $\delta t$. The amplitudes $\lambda(t)$ of the polynomial scheduling function of~\eqlabel{eq:lambda} along their evolution are plotted in~\figlabel{fig:quasikick}(b). As a concluding remark, similar results can be reported using different boundary conditions and $H^{zz}_1$ as charger Hamiltonian [\eqlabel{eq:ref_ham_zz1}].
\begin{figure}[!tb]
    \centering
    \includegraphics[width=\linewidth]{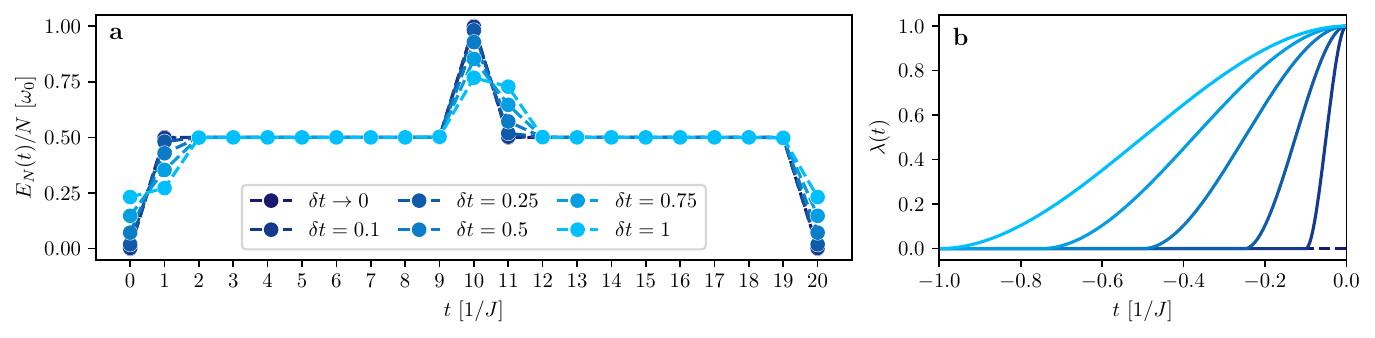}\vspace{-1mm}%
    \caption{Normalized injected energies for a KIC using for slow quenches with $N=20$. (a) For different widths $\delta t$ (where the lighter the color, the larger its value), we apply up to $m=N=20$ kicks using $H^{xx}_1$ as charger Hamiltonian [\eqlabel{eq:ref_ham1}] under PBC. In the limit of $\delta t\to 0$, we recover the regular KIC. (b) Amplitudes $\lambda(t)$ of the scheduling functions for different $\delta t$ values [\eqlabel{eq:lambda}]. The step function ($\delta t\to0$ case) is represented by a horizontal dashed line.}\label{fig:quasistep}
\end{figure}

\section{Entanglement entropy evolution across bonds for the kicked-Ising chain}

The entanglement between two subsystems, in the case of a composite pure state, can be quantified by the von Neumann entropy of the reduced density matrix of either subsystem. In a matrix product state (MPS) representation of the wavefunction, the entanglement entropy for any bipartition is given by the singular values corresponding to the bond that connects the two parts~\cite{Schollwock2011}. For bond index $i$, we denote the $j$th-largest singular value by $\lambda_{ij}$. Here, $j$ has values $j \in \{ 1, 2, \dots, \chi_i \}$ up to the bond dimension $\chi_i$ and the bonds between qubits (tensors in the MPS) are indexed by $i \in \{ 1, \dots, N_b \}$ with $N_b$ the number of bonds and the $i$th bond connecting spins $(i,i+1)$. If PBC are considered, the $N$th bond couples spins $(N,1)$. Therefore, for a linear chain with OBC (PBC), the number of bonds is given by $N_b=N-1$ ($N_b=N$). The entanglement entropy per site at the $i$th bond reads as
\begin{equation}\label{eq:entropy}
    S_i^\text{vN} = -\sum_{j=1}^{\chi_i}\lambda^2_{ij}\log_2\lambda^2_{ij}.
\end{equation}
This definition of entropy is often used to estimate the computational resources required to simulate a system, thus how computationally expensive is to encode with MPS an input state.

Previous works have studied how the Rényi entropy, being von Neumann entropy a particular definition of its family, across partitions evolve under a KIC in the self-dual operator regime when starting from a product state~\cite{bertini2019entanglement}, featuring a linear growth in time up to a saturation point. Similar findings have been addressed in different physical settings, as for example in conformal invariant systems~\cite{calabrese2005evolution,liu2014entanglement}, nonintegrable closed systems~\cite{kim2013ballistic}, from Clifford and random unitary dynamics~\cite{nahum2017quantum,vonkeyserlingk2018operator,sommers2023crystalline} and periodically driven systems~\cite{mishra2015protocol,chan2018solution,bertini2019entanglement}. In the following lines we study how the entanglement entropy evolves with the number of kicks applied using $H^{zz}_1$ [\eqlabel{eq:ref_ham_zz1}] as charger Hamiltonian and an even number of spins $N$, where similar results can be obtained considering different setups. Recall that for this particular charger, its periodicity, thus the length of each Floquet cycle, is given by $U(t,0)=U(t+N,0)$ and $U(t,0)=U(t+4N,0)$ for both PBC and OBC, respectively. In this case and under our definition of entropy [\eqlabel{eq:entropy}], which is more convenient for analyzing the tractability of our MPS simulations, the obtained patterns are given by
\begin{align}\label{eq:vn_obc}
    S_{i, \text{OBC}}^\text{vN}(t) &= \min\left[ \min\left( \left\lfloor t-\frac{1}{2} \right\rfloor, N - \left\lfloor t+\frac{1}{2} \right\rfloor \right), \min(i, N-i) \right], &&\forall t\in[0,N] \\ \label{eq:vn_pbc}
    S_{i, \text{PBC}}^\text{vN}(t) &= \min\left[ \min\left( \left\lfloor 2t-\frac{1}{2} \right\rfloor, N - \left\lfloor 2t+\frac{1}{2} \right\rfloor \right), \min(i, N-i) \right], &&\forall t\in[0,N]
\end{align}
for PBC and OBC, respectively. As for previously reported works, the entanglement entropy per site features a piecewise linear increasing, up to a saturation value given by the bond index, and decreasing with the number of kicks $t$, pattern that is repeated per each Floquet cycle. In~\figlabel{fig:vn_entropy}(a)-(b) the entanglement entropy per bond for a KIC with $N=20$, using both PBC and OBC, is compared with the expected results given by Eqs.~\eqref{eq:vn_obc}-\eqref{eq:vn_pbc}, which are in accordance. 

Building on~\eqlabel{eq:entropy}, we can also define as entanglement entropy $S^\text{vN}\coloneqq \sum_{i=1}^{N_b}S^\text{vN}_i$, which sums the entanglement entropy per site over all bonds, as well as its average over bonds $s^\text{vN}\coloneqq S^\text{vN}/N_b$. Using MPS, in~\figlabel{fig:vn_entropy}(c)-(d) we see how the entanglement entropy $S^\text{vN}(t)$ evolves with $t$ as double-arched parabolas which are symmetrical with respect $N/2$ and $N$ for PBC and OBC, respectively. In other words, they are symmetrical with respect quarter of their corresponding Floquet cycles. The parabolic behavior comes from the linear profile that the entanglement entropy per bond features, whose sum over bonds returns a quadratic shape (recall the triangular sum formula).
\begin{figure}[!tb]
    \centering
    \includegraphics[width=\linewidth]{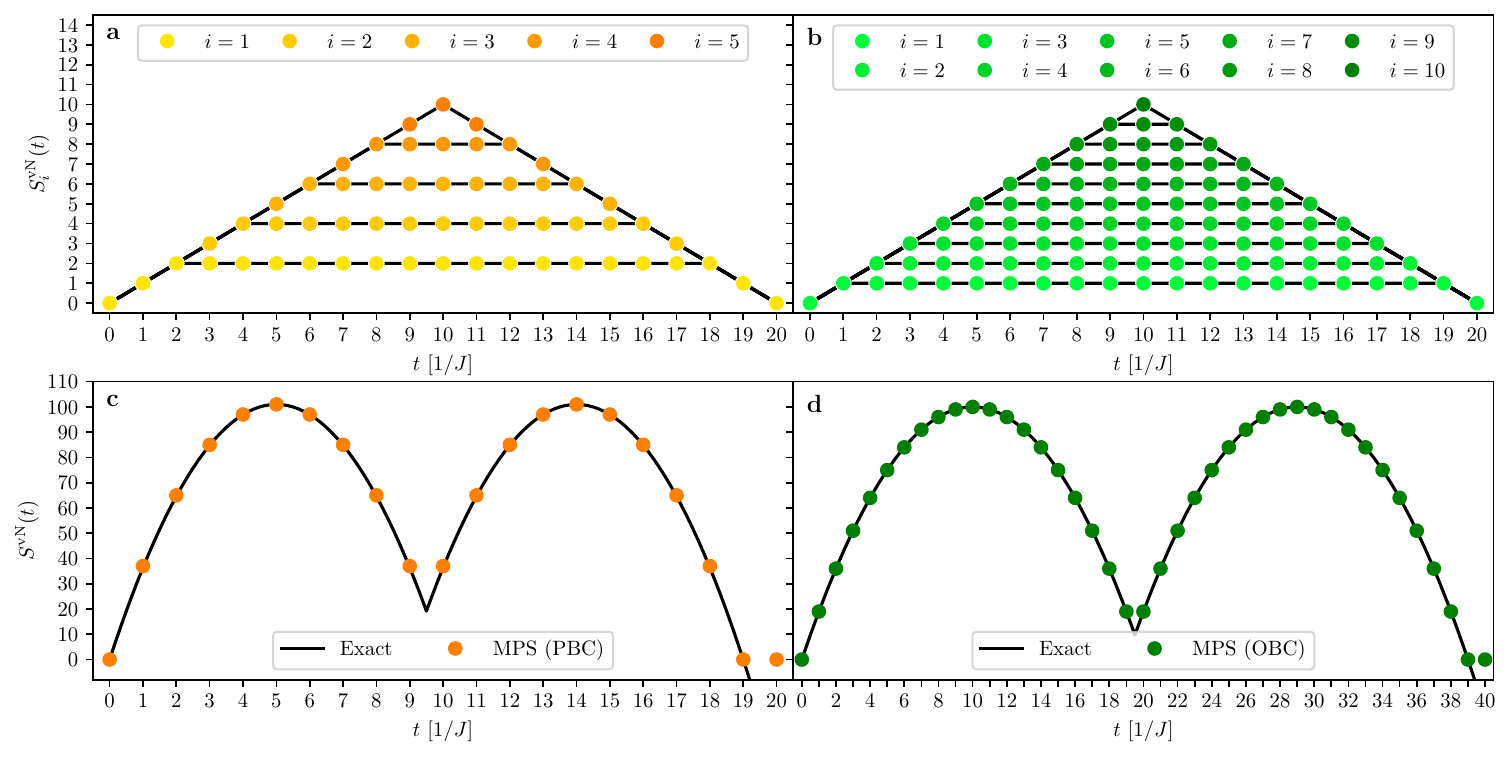}
    \caption{Entanglement entropy evolution for a KIC with $N=20$. (a)-(b) Entanglement entropy evolution per site [\eqlabel{eq:entropy}] for both PBC (yellow to orange) and OBC (light to dark green) using MPS, where the more intense the color, the larger its corresponding bond index. (c)-(d) Entanglement entropy summed over bonds evolution for both PBC (orange) and OBC (green) using MPS, showcasing the expected double-arched parabolic profile [Eqs.~\eqref{eq:vn_obc}-\eqref{eq:vn_pbc}]. For (a) and (d), we plot for visualization purposes data until half of its Floquet cycle ($t=2N=40$), where the most excited state is prepared (product state), since the exact same curves are returned for $t\in[2N,4N]$. In all plots, black lines show the expected results.}\label{fig:vn_entropy}
\end{figure}

To build their corresponding functions, we can make use as ansätze a quadratic function of the form $a(t-b)^2+c$, where $a$, $b$ and $c$ are in general system-size dependent parameters to be obtained. We start by the OBC case, where the entanglement entropy peaks at $t=N/2$, thus returning $b=N/2$, with a value given by
\begin{equation}
    c = S_\text{OBC}^\text{vN}\left( t=\frac{N}{2} \right) = 2\sum_{i=1}^{N/2-1}i + \frac{N}{2} = 2\binom{N/2}{2} + \frac{N}{2} = \frac{N^2}{4},
\end{equation}
where we apply the triangular sum formula. For the remaining parameter, since $S^\text{vN}(t=0)=0\implies a=-c/b^2 = -1$. A similar reasoning can be applied when considering PBC, where we obtain $b=N/4$, $c=1+2\lfloor N^2/8\rfloor$ and $a=-c/b^2 = -16(1+2\lfloor N^2/8\rfloor)/N^2$. Finally, we obtain the following double-arched parabolas,
\begin{align}
    S^\text{vN}_\text{OBC}(t) &= -\min\left( t-\frac{N}{2}, \frac{3N}{2}-1-t \right)^2 + \frac{N^2}{4}, &&\forall t\in[0,2N]\\
    S^\text{vN}_\text{PBC}(t) &= -16\frac{1+2\lfloor N^2/8\rfloor}{N^2}\min\left( t-\frac{N}{4}, \frac{3N}{4} - 1 - t \right)^2 + 1+2\left\lfloor\frac{N^2}{8}\right\rfloor, &&\forall t\in\left[0,N\right]
\end{align}
which are reminiscent of the iconic ``Golden Arches'' logo. In~\figlabel{fig:vn_entropy}(c)-(d) we plot both curves and compare them with the data obtained from MPS simulation of a KIC with $N=20$, where both are in accordance again.

\section{Data analysis of the hardware experiments}

As discussed in the main text, despite KIC model permits an exact implementation on digital quantum hardware, reason for which is often mentioned that it captures gate-like time dynamics~\cite{brown2023quantum}, errors and noise are present and might remarkably degrade the quality of the obtained results. They can appear due to gate and readout errors and short coherence times, among other factors. The charging dynamics provided by the $H^{\{xx,zz\}}_1$ chargers, with flat regions accompanied by abrupt sudden changes after applying a precise number of kicks, offers a simple way to assess how reliable the platform considered performed. Apart from comparing the injected energies, based on Ref.~\cite{visuri2025digitizedcounterdiabaticquantumcritical}, from a set of samples we can also see how the probability of getting a zero per qubit evolves, which is used to quantify the impact that these errors have. While for an ideal implementation we would expect a uniform distribution, i.e., same probability of sampling either a zero or a one, qubits with shorter coherence times and bitflip errors may return different results. For the $i$th qubit, we denote this quantity as $P_{0,i}(t)$.
\begin{figure}[!b]
    \centering
    \includegraphics[width=\linewidth]{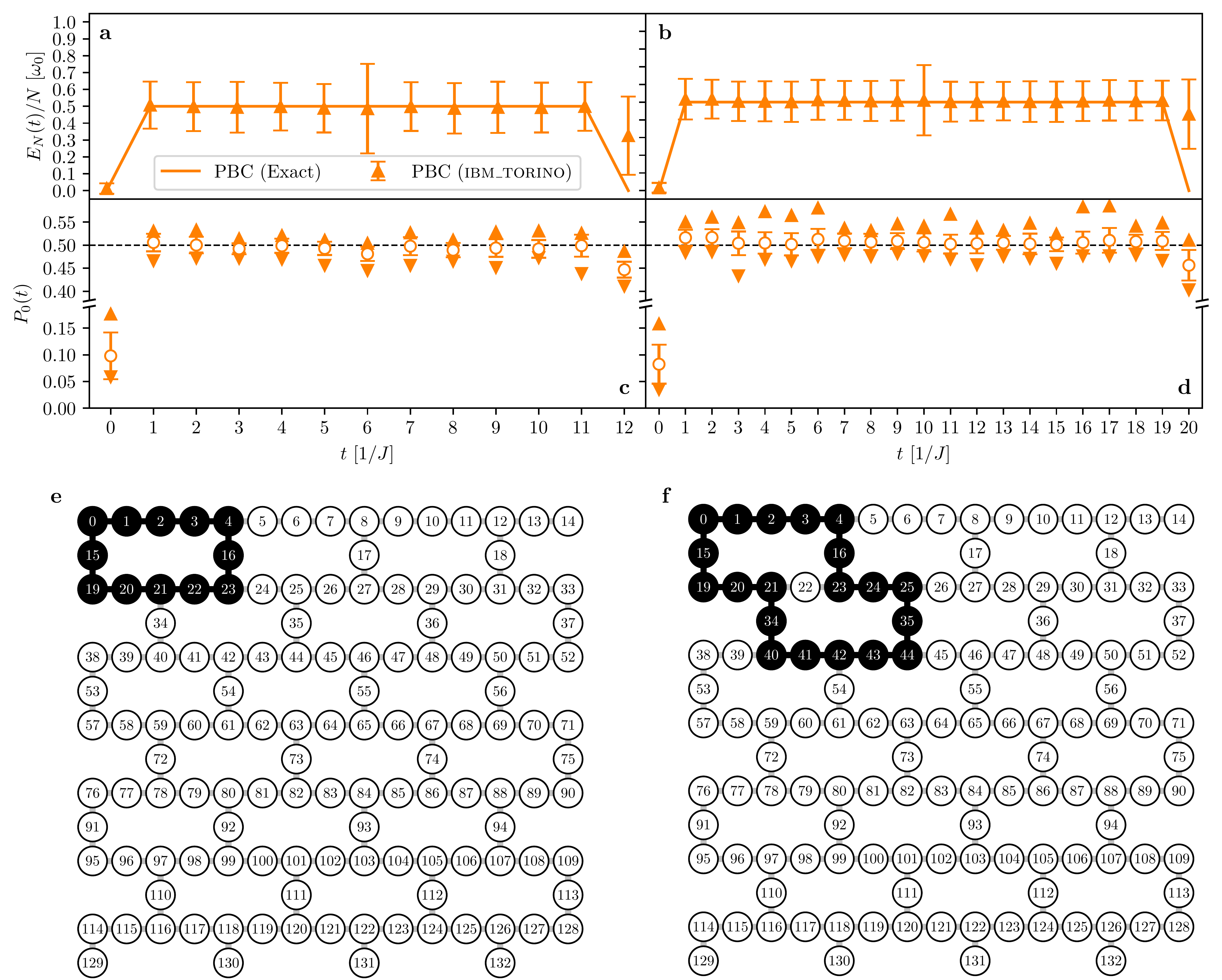}%
    \caption{Analysis of the experimental data obtained from \textsc{ibm\_torino}. (a) Normalized injected energies for KIC with $N=12$ quantum cells using $H_1^{zz}$ as charger [\eqlabel{eq:ref_ham_zz1}], with exact and experimental results up to $m=N=12$ kicks under PBC using \textsc{ibm\_torino}. Error bars correspond to the standard deviation after $1000$ measurements (see Methods in the main text). Exact (experimental) results are represented by solid lines (triangular marks). (b) Same plot as (a) but considering $N=20$ quantum cells and $m=N=20$ kicks. (c) From the experiments shown in (a), evolution of the mean probability of returning a zero $P_0(t)$ (circles), as well as the minimum and maximum values $P^{\{\min,\max\}}_0(t)$ obtained (upwards and downwards triangular markers, respectively). Dashed horizontal line is included as a guidance. (d) Same plot as (c) but for the experiments realized in (b). (e) Heavy-hex coupling map of \textsc{ibm\_torino}, highlighting in black the qubits used for the experiments run for the KIC QB of $N=12$ quantum cells with PBC. (f) Same schematic as (e) but for the $N=20$ experiments. The vertical axes of (c) and (d) are broken for visualization purposes.}\label{fig:experiment}
\end{figure}%

In~\figlabel{fig:experiment} we conduct both tests considering a KIC under PBC of $N=12$ and $N=20$, which are the two smallest closed chains that can be prepared on IBM, using $H^{zz}_1$ [\eqlabel{eq:ref_ham_zz1}] as charger. As the periodicity under these conditions is governed by $U(t,0)=U(t+N,0)$, we apply for both cases up to $m=N$ kicks. In~\figlabel{fig:experiment}(a)-(b) we compare the exact normalized injected energies with experimental results obtained in \textsc{ibm\_torino}, using $100000$ samples, where we observe:
\begin{itemize}
    \item Despite circuit depths and numbers of gates used are moderate (see table I in Methods), both implementations fail to prepare their corresponding ground states at the $N$th kick, with worse results for the largest implementation, where $E_N(t=N)/N\lesssim0.5$. As seen in~\figlabel{fig:disorder}, this is the expected behavior when a large amount of disorder is present in the implementation.
    \item As expected, the standard deviation of the normalized injected energy at the $(N/2)$th kick is larger, region where the entanglement entropy maximizes [\figlabel{fig:vn_entropy}(a)].
\end{itemize}%
As the depth of the circuit implementation of the KIC model is constant with the system size $N$ and led by our results, we apply throughout our studies up to $m=12$ kicks on hardware, where we assume and expect that the simulation of KIC chains with $m>12$ will only return normalized injected energies accumulating at $E_N/N\sim 0.5$, as aforementioned. However, to apply diverse error mitigation and suppression techniques might be favorable to slightly extend this upper bound and obtain more accurate results~\cite{miessen2024benchmarking}.

In~\figlabel{fig:experiment}(c)-(d), for both KIC chains we analyze per kick the probability of sampling a zero per qubit $P_{0,i}(t)$, where we indicate: the mean values $P_0(t)\coloneqq\sum_{i=1}^N P_{0,i}(t)$ (marked with empty circles), their corresponding standard deviations and their maximum and minimum values $P^{\max}_0(t)\coloneqq\max_i P_{0,i}(t)$ and $P^{\min}_0(t)\coloneqq\min_i P_{0,i}(t)$, respectively (marked with upwards and downwards triangles). We observe:
\begin{itemize}
    \item Once the ground state is prepared ($t=0$), ideally we should only obtain samples with all ones after rotating into the Pauli-Y basis, thus $P_0(t=0)=0$. Despite this value is not obtained, it is relatively low, with less than $10\%$ of probability of sampling excited states, which might come from bitflip or readout errors.
    \item For the $N$th kick, a Floquet cycle is completed, thus ideally we should again be in the ground state. Nonetheless, and in accordance with the results obtained in~\figlabel{fig:experiment}(a)-(b), we obtain $P_0(t=N)\sim0.5$, value that we would expect for systems with a strong presence of disorder [\figlabel{fig:disorder}].
    \item For intermediate times ($0<t<N$), experimental data behave as expected, with stable $P^{\{\min, \max\}}_0(t)$ values accumulating around $0.5$. Since few kicks are implemented, thus circuit running times are quite below of current IBM hardware coherence times~\cite{ibm}, we assume that thermalization effects are negligible in the regions studied but bitflip and readout errors are present.
    \item As a concluding remark: if we continue applying kicks, we expect a competition between thermalization effects, where individual qubits relax from their excited toward their ground states due to increased circuit depth, and bitflip errors arising from the larger number of operations. As each qubit in \textsc{ibm\_torino} has coherence times ranging from $1-\SI{100}{\micro\second}$, we expect larger differences between $P^{\{\min, \max\}}_0(t)$, with the qubits with lowest coherence times relaxing faster than the rest.
\end{itemize}
In~\figlabel{fig:experiment}(e)-(f) we can see an schematic of the \textsc{ibm\_torino} platform with the qubits used for our experiments highlighted.

Another way to evaluate the performance our IBM results is to analyze the covariance matrix of the data obtained and compare it with the ideal one. For example, we can use again the same data used in~\figlabel{fig:experiment}, where we find this particular state after $m=N/2$ kicks
\begin{equation}\label{eq:m_n_2}
\begin{split}
    \ket{\psi(t=N/2)} &= \frac{(-1)^{\lfloor N/2 - 1\rfloor}}{2}\left[ \ket{+i}^{\otimes N} - \left( \ket{+i,-i}^{\otimes N/2} + \ket{-i,+i}^{\otimes N/2} \right) - (-1)^{N/2}\ket{-i}^{\otimes N} \right] \\
    &=\frac{(-1)^{\lfloor N/2 - 1\rfloor}}{2}\left[ \ket{0}_\text{Y}^{\otimes N} - \left( \ket{0,1}_\text{Y}^{\otimes N/2} + \ket{1,0}_\text{Y}^{\otimes N/2} \right) - (-1)^{N/2}\ket{1}_\text{Y}^{\otimes N} \right],
\end{split}
\end{equation}
where $\ket{\cdot}_\text{Y}$ indicates that the state is described under the Pauli-Y basis. Ignoring global and relative phases, this state looks like an equal superposition of a Greenberger-Horne-Zeilinger (GHZ) and a Néel state under the Pauli-Y basis. We can also see that after rotating into the Pauli-Y basis, basis where all of our measurements on IBM hardware were performed, only four types of sampled bitstrings are going to be obtained with same probability: $\{00\dots0, 0101\dots1, 1010\dots0, 11\dots1\}$. When expressed as Ising variables, with binary-to-Ising map $s_i=1-2b_i$ $\forall i\in[1,N]$, we get $\bm{s}^{(1)} = [+1, +1, \dots, +1]$, $\bm{s}^{(2)} = [+1, -1, \dots, -1]$, $\bm{s}^{(3)} = [-1, +1, \dots, +1)]$ and $\bm{s}^{(4)} = [-1, -1, \dots, -1]$. 

To compute the covariance matrix between two variables, $\cov(s_i,s_j)=\mathbb{E}[s_is_j] - \mathbb{E}[s_i]\mathbb{E}[s_j]$. We can trivially see that $\mathbb{E}[s_i] = (1/4)\sum_{k=1}^4 s_i^{(k)} = (1 + 1 -1 -1)/4 = 0$. For the second-raw moment $\mathbb{E}[s_is_j]$, we can distinguish two particular cases depending on the relative parity of the bitstring indices:
\begin{itemize}
    \item If $i\equiv j \text{ mod }2$ (same parity): since the four samples $\bm{s}^{(k)}$ feature a two-site translational invariance, under these conditions we can see that $s_i^{(k)}s_j^{(k)}=1$, leading to $\mathbb{E}[s_is_j]=(1/4)\sum_{k=1}^4 s^{(k)}_is^{(k)}_j = (1+1+1+1)/4=1$.
    \item If $i\not\equiv j \text{ mod }2$ (opposite parity): for this case, the signs of $s_i^{\{(2),(3)\}}$ and $s_j^{\{(2),(3)\}}$ differ, thus $s_i^{\{(2),(3)\}}s_j^{\{(2),(3)\}}=-1$. Therefore, $\mathbb{E}[s_is_j]=(1/4)\sum_{k=1}^4 s^{(k)}_is^{(k)}_j = (1-1-1+1)/4=0$.
\end{itemize}
Therefore, the covariance matrix $\cov(s_i,s_j)$ showcases a checkerboard-like pattern. 

In~\figlabel{fig:checkerboard}(a)-(b), we compute the covariance matrix of the samples studied in~\figlabel{fig:experiment} for $N=12$ at kicks $m=N/2=6$ and $m=N=12$, where we should recover ideally the checkerboard pattern aforementioned and the zero matrix, respectively. Although at $m=N/2$ the covariance values for indices with opposite signs return negligible values, as expected, for those with same parity the values returned are smaller than one, reducing with their relative distance $|i-j|$. Moreover, the covariance matrix at $m=N$ yields nonzero values, which are in accordance with the bad energy estimations shown at $m=N$ in~\figlabel{fig:experiment}. They indicate that in our experiments the ground state was far from being prepared correctly despite the small system sizes considered. We perform the same analyses for $N=20$ quantum cells in~\figlabel{fig:checkerboard}(c)-(d), where the results reported are similar but slightly worse in comparison. 

Ideally, at kick $m=N/2$ the uniform probability of sampling each state of~\eqlabel{eq:m_n_2} should return a $1/4$ probability for $\bm{s}^{(k)}$ and zero otherwise in the limit of a large number of samples, while at $m=N$ only the state $\ket{\psi(t=N)}=\ket{1}^{\otimes N}$, ground state of $H_0$, would be sampled. Building on that, in~\figlabel{fig:checkerboard}(e)-(f) we compare the probabilities of the samples obtained, which are far from the expected results, in accordance with the previous discussion. 
\begin{figure}[!tb]
    \centering
    \includegraphics[width=.992\linewidth]{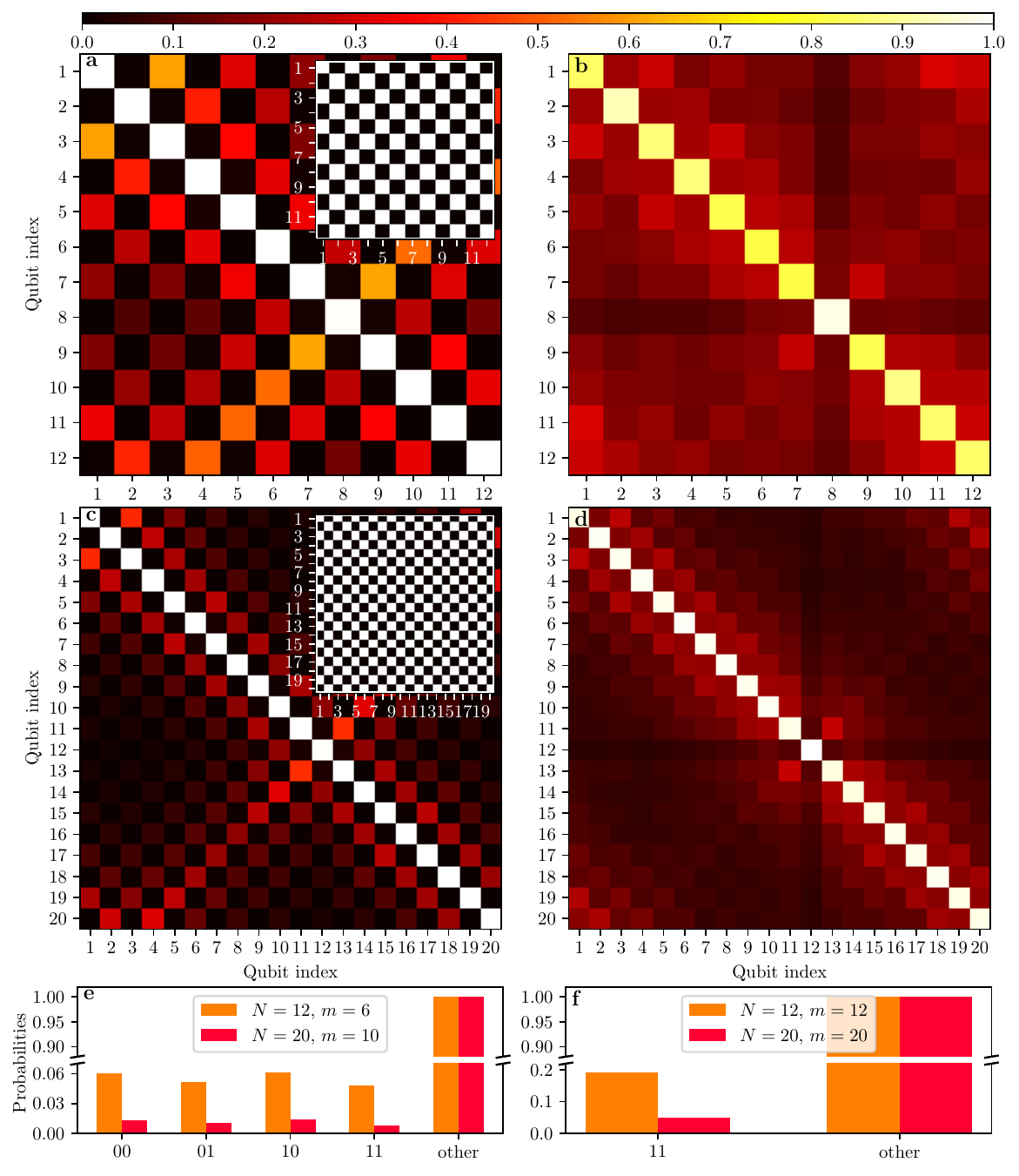}\vspace{-3mm}%
    \caption{Covariance matrices for the experimental results obtained from \textsc{ibm\_torino}. (a)-(b) Using $H_1^{zz}$ as charger [\eqlabel{eq:ref_ham_zz1}] with $N=12$ under PBC, absolute value of the covariance matrix after $m=N/2$ and $m=N$ kicks, known to retrieve~\eqlabel{eq:m_n_2} and the ground state of the battery Hamiltonian, respectively. (c)-(d) Same results as for (a)-(b) but considering $N=20$. (e) Probability of sampling the bitstrings $\bm{s}^{(1)}$, $\bm{s}^{(2)}$, $\bm{s}^{(3)}$, $\bm{s}^{(4)}$ and the remaining states (labeled as ``00'', ``01'', ``10'', ``11'' and ``other'', respectively) for $N=12$ (orange) and $N=20$ (red) after applying $m=N/2$ kicks. (f) Same results as (e) but applying $m=N$ kicks, indicating the probability of sampling the ground state and other states (``11'' and ``other'', respectively). Both (a) and (c) insets show the ideal covariance matrix for comparison. Vertical axes of (e) and (f) are broken for visualization purposes.}\label{fig:checkerboard}
\end{figure}%

\bibliography{bibfile}